\documentclass[12pt,a4paper]{amsart}
\usepackage{diagrams}
\usepackage{amsmath,amstext,amssymb,amsthm,amsfonts}

\newcommand{\nc}{\newcommand}

\nc{\one}{\mbox{\bf 1}}
\nc{\invtensor}{\underset{\leftarrow}{\otimes}}
\nc{\const}{\operatorname{const}}

\nc{\ad}{\operatorname{ad}}
\nc{\tr}{\operatorname{tr}}
\nc{\tp}{\operatorname{top}}
\nc{\rank}{\operatorname{rank}}
\nc{\corank}{\operatorname{corank}}
\nc{\codim}{\operatorname{codim}}
\nc{\sdim}{\operatorname{sdim}}

\nc{\spn}{\operatorname{span}}
\nc{\Sym}{\operatorname{Sym}}
\nc{\sym}{\operatorname{sym}}
\nc{\id}{\operatorname{id}}
\nc{\Id}{\operatorname{Id}}

\nc{\htt}{\operatorname{ht}}
\nc{\Ker}{\operatorname{Ker}}
\nc{\rker}{\operatorname{rKer}}
\nc{\im}{\operatorname{Im}}
\nc{\osp}{\mathfrak{osp}}
\nc{\sgn}{\operatorname{sgn}}
\nc{\F}{\operatorname{F}}
\nc{\Mod}{\operatorname{Mod}}
\nc{\Mat}{\operatorname{Mat}}
\nc{\Soc}{\operatorname{Soc}}
\nc{\Inj}{\operatorname{Inj}}
\nc{\Hom}{\operatorname{Hom}}
\nc{\End}{\operatorname{End}}
\nc{\supp}{\operatorname{supp}}
\nc{\Card}{\operatorname{Card}}
\nc{\Ann}{\operatorname{Ann}}
\nc{\Ind}{\operatorname{Ind}}
\nc{\Coind}{\operatorname{Coind}}
\nc{\wt}{\operatorname{wt}}
\nc{\ch}{\operatorname{ch}}
\nc{\Stab}{\operatorname{Stab}}
\nc{\Sch}{{\mathcal S}\mbox{\em ch}}
\nc{\Irr}{\operatorname{Irr}}
\nc{\Spec}{\operatorname{Spec}}
\nc{\Prim}{\operatorname{Prim}}
\nc{\Aut}{\operatorname{Aut}}
\nc{\Fract}{\operatorname{Fract}}
\nc{\gr}{\operatorname{gr}}
\nc{\deff}{\operatorname{def}}
\nc{\HC}{\operatorname{HC}}

\nc{\wdchi}{\widetilde{\chi}}
\nc{\wdH}{\widetilde{H}}
\nc{\wdN}{\widetilde{N}}
\nc{\wdM}{\widetilde{M}}
\nc{\wdO}{\widetilde{O}}
\nc{\wdR}{\widetilde{R}}
\nc{\wdS}{\widetilde{S}}
\nc{\wdV}{\widetilde{V}}

\nc{\wdC}{\widetilde{C}}

\nc{\Ob}{\operatorname{\mathcal Ob}}
\nc{\Dglie}{\operatorname{{\mathcal D}glie}}
\nc{\Fin}{\operatorname{{\mathcal F}in}}

\nc{\Sg}{{\cS(\fg)}}
\nc{\Shg}{{\cS(\fhg)}}
\nc{\Ug}{{\cU(\fg)}}
\nc{\Uhg}{{\cU(\fhg)}}
\nc{\Sh}{{\cS(\fh)}}
\nc{\Uh}{{\cU(\fh)}}
\nc{\Uhh}{{\cU(\fhh)}}
\nc{\Zg}{{{\mathcal{Z}}(\fg)}}

\nc{\Vir}{{\mathcal{V}ir}}
\nc{\NS}{{\mathcal{NS}}}

\nc{\tZg}{{\widetilde{\mathcal Z}({\mathfrak g})}}
\nc{\Zk}{{\mathcal Z}({\mathfrak k})}

\nc{\Up}{{\mathcal U}({\mathfrak p})}
\nc{\Ah}{{\mathcal A}({\mathfrak h})}
\nc{\Ag}{{\mathcal A}({\mathfrak g})}
\nc{\Ap}{{\mathcal A}({\mathfrak p})}
\nc{\Zp}{{\mathcal Z}({\mathfrak p})}
\nc{\cZ}{\mathcal Z}
\nc{\cS}{\mathcal S}
\nc{\cP}{\mathcal P}
\nc{\cA}{\mathcal A}
\nc{\cU}{\mathcal U}
\nc{\cH}{\mathcal H}
\nc{\cM}{\mathcal M}
\nc{\cL}{\mathcal L}
\nc{\cF}{\mathcal F}
\nc{\fg}{\mathfrak g}

\nc{\fo}{\mathfrak o}

\nc{\CO}{\mathcal O}
\nc{\Cl}{\mathcal {C}\ell}

\nc{\zq}{\mathpzc q}

\nc{\fl}{\mathfrak l}
\nc{\fn}{\mathfrak n}
\nc{\fm}{\mathfrak m}
\nc{\fp}{\mathfrak p}
\nc{\fh}{\mathfrak h}
\nc{\ft}{\mathfrak t}
\nc{\fk}{\mathfrak k}
\nc{\fb}{\mathfrak b}
\nc{\fs}{\mathfrak s}
\nc{\fB}{\mathfrak B}

\nc{\vareps}{\varepsilon}
\nc{\varesp}{\varepsilon}
\nc{\veps}{\varepsilon}

\nc{\fsl}{\mathfrak{sl}}
\nc{\fgl}{\mathfrak{gl}}
\nc{\fso}{\mathfrak{so}}

\nc{\fpq}{\mathfrak{pq}}
\nc{\fq}{\mathfrak q}
\nc{\fsq}{\mathfrak{sq}}
\nc{\fpsq}{\mathfrak{psq}}

\nc{\fhg}{\hat{\fg}}
\nc{\fhn}{\hat{\fn}}
\nc{\fhh}{\hat{\fh}}
\nc{\fhb}{\hat{\fb}}
\nc{\hrho}{\hat{\rho}}

\nc{\hsl}{\hat{\fsl}}
\nc{\fpo}{\mathfrak{po}}
\nc{\dirlim}{\underset{\rightarrow}{\lim}\,}
\nc{\nen}{\newenvironment}
\nc{\ol}{\overline}
\nc{\ul}{\underline}
\nc{\ra}{\rightarrow}
\nc{\lra}{\longrightarrow}
\nc{\Lra}{\Longrightarrow}

\nc{\Lla}{\Longleftarrow}

\nc{\Llra}{\Longleftrightarrow}

\nc{\thla}{\twoheadleftarrow}

\nc{\hra}{\hookrightarrow}

\nc{\iso}{\overset{\sim}{\lra}}

\nc{\ssubset}{\underset{\not=}{\subset}}

\nc{\vac}{|0;c\rangle}

\newarrow{Dline}=====

\nc{\Thm}[1]{Theorem~\ref{#1}}
\nc{\Prop}[1]{Proposition~\ref{#1}}
\nc{\Lem}[1]{Lemma~\ref{#1}}
\nc{\Cor}[1]{Corollary~\ref{#1}}
\nc{\Conj}[1]{Conjecture~\ref{#1}}
\nc{\Claim}[1]{Claim~\ref{#1}}
\nc{\Defn}[1]{Definition~\ref{#1}}
\nc{\Exa}[1]{Example~\ref{#1}}
\nc{\Rem}[1]{Remark~\ref{#1}}
\nc{\Note}[1]{Note~\ref{#1}}
\nc{\Quest}[1]{Question~\ref{#1}}
\nc{\Hyp}[1]{Hypoth\`ese~\ref{#1}}
\nen{thm}[1]{\label{#1}{\bf Theorem.\ } \em}{}
\nen{prop}[1]{\label{#1}{\bf Proposition.\ } \em}{}
\nen{lem}[1]{\label{#1}{\bf Lemma.\ } \em}{}
\nen{cor}[1]{\label{#1}{\bf Corollary.\ } \em}{}
\nen{conj}[1]{\label{#1}{\bf Conjecture.\ } \em}{}

\nen{claim}[1]{\label{#1}{\bf Claim.\ } \em}{}

\nen{defn}[1]{\label{#1}{\bf Definition.\ } }{}
\nen{exa}[1]{\label{#1}{\bf Example.\ } }{}
\nen{rem}[1]{\label{#1}{\em Remark.\ } }{}
\nen{note}[1]{\label{#1}{\em Note.\ } }{}
\nen{exer}[1]{\label{#1}{\em Exercise.\ } }{}
\nen{sket}[1]{\label{#1}{\em Sketch of proof.\ } }{}
\nen{quest}[1]{\label{#1}{\bf Question.\ } \em}{}

\nen{hyp}[1]{\label{#1}{\bf Hypoth\`ese.\ } \em}{}
\begin{document}
\setcounter{section}{-1}

\title{On simplicity of vacuum modules}
\author{Maria Gorelik~$^\dag$}

\address{Dept. of Mathematics, The Weizmann Institute of Science,
Rehovot 76100, Israel}
\email{maria.gorelik@weizmann.ac.il}
\thanks{$^\dag$
Incumbent of the Frances and Max Hersh career development chair}

\author{Victor Kac~$^\ddag$}

\address{Dept. of Mathematics, 2-178, Massachusetts Institute of Technology,
Cambridge, MA 02139-4307, USA}
\email{kac@math.mit.edu}
\thanks{$^\ddag$ Supported in part by NSF Grant  DMS-0501395}

\begin{abstract} We find necessary and sufficient conditions of 
irreducibility of vacuum modules over affine Lie algebras and superalgebras. 
From this we derive conditions of simplicity of minimal $W$-algebras.
Moreover, in the case of the Virasoro and Neveu-Schwarz algebras 
we obtain explicit formulas 
for the vacuum determinants. 
\end{abstract}

\maketitle

\section{Introduction}
\subsection{}\label{sub01}
One of the aims of the present paper is to find conditions of
irreducibility of vacuum modules over the
affine Lie superalgebra 
$$\begin{array}{l}
\hat{\fg}=\fg [t,t^{-1}]+\mathbb{C}K,\\
\ [at^m,bt^n]=[a,b]t^{m+n}+m\delta_{m,-n}B(a|b)K,\ \ [at^m,K]=0,
\end{array}$$
associated to a simple finite dimensional Lie superalgebra $\fg$
with a non-degenerate even invariant bilinear form $B(.|.)$.
Recall that the {\em vacuum module}  is the induced module
$$V^k=\Ind^{\hat{\fg}}_{\fg[t]+\mathbb{C}K} \mathbb{C}_k$$ 
from
the $1$-dimensional module $\mathbb{C}_k$ with trivial action
of $\fg[t]$ and $K=k\in\mathbb{C}$.
\subsection{}
In order to state the result, let $2h^{\vee}_{B}$ be the eigenvalue
of the Casimir operator $\sum_i a_ia^i$ in the adjoint
representation of $\fg$, where $\{a_i\}$ and $\{a^i\}$
are dual bases of $\fg$ i.e. $B(a^i|a_j)=\delta_{ij}$.
The numbers $h^{\vee}_{B}$ and $k$
depends on the normalization of the bilinear form
$B$: if $B$ is multiplied by a non-zero number $\gamma$,
then both  $h^{\vee}_{B}$ and $k$ get multiplied by $\gamma^{-1}$.

For a simple Lie algebra $\fg$ the standard normalization is
$B(\alpha|\alpha)=2$ for a long root $\alpha$. In this case,
$h^{\vee}_B$ is called the dual Coxeter number; it is a positive
integer, denoted by $h^{\vee}$ (these integers are listed, 
e.g. in~\cite{Kbook}). For simple Lie superalgebras a ``standard''
normalization of $B$ was introduced,
and the values of $h^{\vee}$ listed, in~\cite{KWn}.

For a non-isotropic root $\alpha$ introduce 
$$k_{\alpha}:=\frac{k+h^{\vee}_{B}}{B(\alpha|\alpha)}.$$
Note that this number is independent on the normalization of $B$.
\subsubsection{}
\begin{thm}{thm01}
Let $\fg$ be a simple finite-dimensional Lie algebra. 
The vacuum 
$\hat{\fg}$
-module $V^k$ is not irreducible if and only if
$k_{\alpha}\in\mathbb{Q}_{\geq 0}\setminus\{\frac{1}{2m}\}_{m=1}^{\infty}$
for  a short root $\alpha$ of $\fg$
(equivalently, if
and only if $l(k+h^{\vee})$ is a non-negative rational number which is not
the inverse of an integer, where $l$ is the ratio of the lengths squared of 
a long and a short root of $\fg$).
\end{thm}

\subsubsection{}
\begin{thm}{thm02}
Let $\fg$ be a simple Lie superalgebra $\osp(1,2n)$. 
The vacuum $\hat{\fg}$-module $V^k$ is not irreducible
if and only if $k_{\alpha}\in\mathbb{Q}_{\geq 0}\setminus
\{\frac{1}{2m+1}\}_{m=0}^{\infty}$ where $\alpha$
is an odd root of $\fg$ (equivalently, if
and only if $k+2n+1$ is a non-negative rational number which is not
the inverse of an odd integer, if $B(\alpha|\alpha)=1$ for an odd root 
$\alpha$ of $\fg$).
\end{thm}

\subsubsection{}
\begin{conj}{conj03}
Let $\fg$ be an (almost) simple finite-dimensional 
Lie superalgebra of positive defect~\cite{KWn}, i.e.
one of the Lie superalgebras $\fsl(m,n)\ (m,n\geq 1), 
\osp(m,2n)\ (m\geq 2,n\geq 1), D(2,1,a), F(4)$ or $G(3)$.
Then the $\hat{\fg}$-module $V^k$ is not irreducible if and only if
\begin{equation}\label{02}
k_{\alpha}\in\mathbb{Q}_{\geq 0}\ 
\text{ for some even root } \alpha \text{ of }  \fg.
\end{equation}
\end{conj}

Note that  $V^k$
is always reducible at the {\em critical} level $k=-h^{\vee}_B$.

\subsubsection{}
\begin{thm}{thm04}
\Conj{conj03} holds for simple Lie superalgebras of defect $1$, i.e. 
$\fg=\fsl(1,n), \osp(2,n), \osp(n,2), \osp(3,n)$ with $n\geq 2$,  
$D(2,1,a), F(4), G(3)$, and for $\fg=\fgl(2,2)$.
\end{thm}

More explicitly, in the standard normalization (see~\ref{stnorm})
for the Lie superalgebras $\fg=\fsl(1,n), \osp(2,2n)$,
the module $V^k$ is not irreducible if and only if
$k+n-1$ is a non-negative rational number. For the Lie superalgebras
$\fg=\osp(3,n), \osp(n,2)$ 
with $n>2$,  $F(4), G(3), D(2,1,a)$ with $a\in\mathbb{Q}$, 
the module $V^k$ is not irreducible iff
$k+h^{\vee}$ is a  rational number, where $h^{\vee}$
is given in the table in~\ref{stnorm}. 
For $D(2,1,a), a\not\in\mathbb{Q}$, the vacuum module
$V^k$ is not irreducible iff 
$k\in\mathbb{Q}_{\geq 0}\cup\mathbb{Q}_{>0}a\cup\mathbb{Q}_{>0}(-1-a)$.
For $\fg=\fgl(2,2)$ 
the standard normalization is $B(\alpha|\alpha)=2$ for
an even root $\alpha$; the module $V^k$ is not irreducible iff
$k$ is a rational number.

\subsubsection{}
In order to prove these results, we derive a formula for the determinant
of the Shapovalov form on any generalized Verma module, 
induced from a $1$-dimensional representation of a parabolic subalgebra
of an arbitrary symmetrizable contragredient Lie superalgebra, using methods
of~\cite{Jan} and~\cite{KK}. Unfortunately, unlike
in the Verma module case~\cite{KK}, the exponents of the factors
of the determinant are rather complicated alternating sums, and it is a 
non-trivial problem to find when these sums are positive.
It is a very interesting problem to find a determinant formula
for a vacuum module over $\hat{\fg}$ with manifestly positive
exponents.

\subsection{}
We were unable to find such a formula 
affine Lie superalgebras, but we did succeed in the case of the Virasoro
algebra $\Vir$ and the Neveu-Schwarz superalgebra $\NS$. 

Recall that $\Vir$ is a Lie algebra with a basis 
$\{L_n (n\in\mathbb{Z}), C\}$ and commutation relations
\begin{equation}\label{03}
\ [L_m,L_n]=(m-n)L_{m+n}+\frac{m^3-m}{12}\delta_{m,-n}C,\ \ \ [C,L_m]=0.
\end{equation}
Given $c\in\mathbb{C}$, a {\em vacuum module} over $\Vir$ is the induced module
$$V^c=\Ind^{\Vir}_{\Vir_+}\mathbb{C}_c,$$
where $\Vir_+=\mathbb{C}C+\sum_{n\geq -1} \mathbb{C}L_n$ and $\mathbb{C}_c$
is the $1$-dimensional $\Vir_+$-module with trivial action of $L_n$'s,
and $C=c$. The problem is to compute the determinant of the Shapovalov form,
restricted to the $N$-th eigenspace of $L_0$ in $V^c$, 
$N\in\mathbb{Z}_{\geq 0}$. This is a polynomial
in $c$, which we denote by ${\det}_N'(c)$. (It is defined up to a non-zero
constant factor, depending on a basis of the eigenspace.)

\subsubsection{}
In the case of a Verma module over $\Vir$ the answer is given by the Kac
determinant formula~\cite{K3},\cite{KR}:
$${\det}_{N+h}(h,c)=\const\prod_{r,s\in\mathbb{Z}_{\geq 1}}
\varphi_{r,s}(h,c)^{p_{cl}(N-rs)},$$
where $h$ is the eigenvalue of $L_0$ on the highest weight vector  
$|h,c\rangle$,  $\varphi_{r,s}(h,c)$ are some (explicitly known) polynomials
of $c$ and $h$ of degree $\leq 2$,
and $p_{cl}$ is the classical partition function. From this,
using that a Verma module over $\Vir$ has no subsingular 
vectors (see~\cite{Ast} for a simple proof of this fact), one obtains
immediately the roots of ${\det}_{N}'(c)$, 
but it is a non-trivial problem to compute exponents.
We obtain the following formula (via checking a simple combinatorial
identity):
\begin{equation}\label{04}
{\det}'_N(c)=\const\prod_{\scriptstyle{p>q\geq 2,\atop 
 (p,q)=1}}
\bigl(c-(1-\frac{6(p-q)^2}{pq})\bigr)^{\dim L^{p,q}_N},
\end{equation}
where $p,q\in\mathbb{Z}_{\geq 2}$,  
$L^{p,q}=L((p-1)(q-1);1-\frac{6(p-q)^2}{pq})$ is the irreducible
highest weight $\Vir$-module with the lowest eigenvalue of $L_0$ 
equal $(p-1)(q-1)$ and $c=1-\frac{6(p-q)^2}{pq}$, and $L^{p,q}_N$
is the $N$-th eigenspace of $L_0$ in $L^{p,q}$. The dimensions of these
eigenspaces are known explicitly~\cite{FF}:
\begin{equation}\label{05}
\dim L^{p,q}_N=\sum_{j\in\mathbb{Z}\setminus\{0\}} 
\bigl(p_{cl}(N-(jp+1)(jq+1))-
p_{cl}(N-(jp+1)(jq-1)-1)\bigr).
\end{equation}

Next, we prove the following fact
(which can be deduced from ~\cite{FF}, but our proof is simpler
and can be extended to other cases).

\subsubsection{}
\begin{thm}{thm05}
Let $v\in V^c$ be an eigenvector of $L_0$, killed by all $L_n$ with $n>0$,
and not proportional to the highest weight vector $|0,c\rangle$.
Then $L_0v=2Nv$ for some positive integer $N$, and in the decomposition
$$v=\sum_{\scriptstyle{j_1\geq j_2\geq \ldots j_s\geq 2\atop
j_1+j_2+\ldots +j_s=2N}} c_{j_1j_2\ldots j_s}L_{-j_1}L_{-j_2}\ldots L_{-j_s}
|0,c\rangle$$
the coefficient of $L_{-2}^N|0,c\rangle$ is non-zero.
\end{thm}

The following corollary of  formula~(\ref{04}) and~\Thm{thm05} is
well-known.

\subsubsection{}
\begin{cor}{cor06}
The following conditions on the $\Vir$-module $V^c$ are equivalent:
\begin{enumerate}
\item
$V^c$ is not irreducible;
\item
$c=1-\frac{6(p-q)^2}{pq}$ for some relatively prime integers
$p,q\in\mathbb{Z}_{\geq 2}$;
\item 
 $\spn\{L_{-n}v|\ n>2,v\in V_c\}$, where $V_c$ is the irreducible factor
module of the $\Vir$-module
$V^c$, has finite codimension in $V_c$.
\end{enumerate}
\end{cor}

We also obtain results, analogous to formula~(\ref{04}),~\Thm{thm05}, 
and~\Cor{cor06}, for the Neveu-Schwarz algebra, the simplest super extension
of the Virasoro algebra.

\subsection{}
Recall (see e.g.~\cite{Kv}) that the $\hat{\fg}$-module $V^k$ (resp., 
$\Vir$-module $V^c$) carries a canonical structure of a vertex algebra,
and the irreducibility of these modules is equivalent to the simplicity
of the associated vertex algebras, i.e. to the isomorphism $V_k\cong V^k$
(resp., $V_c\cong V^c$). An important problem, coming 
from conformal field theory, is when a vertex algebra satisfies 
Zhu's $C_2$ condition~\cite{Z}. In the case of the vertex algebra $V_c$, 
$C_2$ condition is property (iii) of~\Cor{cor06}; thus, this corollary
says that $V_c$ satisfies $C_2$ condition if and only if $V^c$ is not simple.
We also show that the same property holds for the Neveu-Schwarz algebra
(but it does not hold for $N>1$ superconformal algebras).

It is easy to see that the vertex algebras $V^k$ and 
the vertex algebras $W^k(\fg,f)$, obtained from $V^k$ by quantum Hamiltonian
reduction (where $f$ is a nilpotent even element
of $\fg$) \cite{KWR},\cite{KW} never satisfy the $C_2$ condition.
It is also not difficult to show that among their quotients only the simple
ones have a chance to satisfy the $C_2$ condition, and a simple affine vertex
algebra $V_k$ satisfies the $C_2$ condition if and only if
$\fg$ is either a simple Lie algebra, or $\fg=\osp(1,2n)$ (i.e. $\fg$ has
defect zero), and the $\hat{\fg}$-module $V_k$ is integrable.

Much more non-trivial is the problem for the simple quotients $W_k(\fg,f)$
of the vertex algebra $W^k(\fg,f)$, which includes the Virasoro, Neveu-Schwarz,
and other superconformal algebras. It has been proved in many
cases~\cite{Ar} that the image of a simple $V^k$-module under the 
quantum Hamiltonian reduction is either a simple $W^k(\fg,f)$-module, or
$0$. Using this, ~\Thm{thm01} and the Kazhdan-Lusztig theory, 
we were able to find the
necessary and sufficient conditions on $k$ for which $W^k(\fg,f)$
is simple in the case when $\fg$ is a simple Lie algebra and $f$ is 
a minimal nilpotent element. 
Namely, for $\fg\not=\mathfrak{sl}_2$, the $k$ for which $W^k(\fg,f)$ is 
simple are given by~\Thm{thm01} (since $W^k(\mathfrak{sl}_2,f)$
is the Virasoro vertex algebra, in this case  the answer is 
given by~\Cor{cor06}). Consequently, for these values of $k$ the vertex
algebra $W_k(\fg,f)$ does not satisfy the $C_2$ condition.

\section{Preliminaries}
Our base field is $\mathbb{C}$.
We set $\mathbb{Z}_{\geq n}:=\{m\in\mathbb{Z}|\ m\geq n\}$.
If $V$ is a superspace, we denote by $p(v)$ the parity of a vector $v\in V$.
For a Lie superalgebra $\fg$ considered in this paper,
any root space $\fg_{\gamma}$ is either pure even or pure odd; 
we denote by $p(\gamma)\in\mathbb{Z}/2\mathbb{Z}$
the parity of $\fg_{\gamma}$ and let $s(\gamma)=(-1)^{p(\gamma)}$.
 For a Lie (super)algebra $\fg$
we denote by $\cU(\fg)$ its universal enveloping (super)algebra.

\subsection{Contragredient Lie superalgebras}

Let $J$ be a finite index set, and let $p: J\to \mathbb{Z}/ 2\mathbb{Z}=
\{\ol{0},\ol{1}\}$ be a map called the {\em parity map}.
Consider a triple $\cA=(\fh,\Pi,\Pi^{\vee})$,
 where $\fh$ is a finite-dimensional vector space over $\mathbb{C}$,
$\Pi=\{\alpha_i\}_{j\in J}$ is a linearly independent subset of $\fh^*$,
 and $\Pi^{\vee}=\{h_i\}_{j\in J}$  is a linearly independent set of 
vectors of $\fh$. One associates to the data $(\cA,p)$ the
{\em contragredient Lie superalgebra} $\fg(\cA,p)$ as follows~\cite{Kadv},
\cite{Kbook}.

\subsubsection{}\label{partord}
First, introduce an auxiliary  Lie superalgebra $\tilde{\fg}(\cA,p)$
with the generators $e_j,f_j\ (j\in J)$ and $\fh$, the parity defined by
$p(e_j)=p(f_j)=p(j),\ p(\fh)=\ol{0}$, and the following defining relations:
$$\begin{array}{ll}
\ [e_i,f_j]=\delta_{ij}h_i\  \ (i,j\in J), & [h,h']=0\ \ (h,h'\in\fh),\\
\ [h,e_j]=\alpha_j(h)e_j,\  & [h,f_j]=-\alpha_j(h)f_j\ \ (h\in\fh, j\in J).
\end{array}$$

The free abelian group $Q$ on generators $\{\alpha_i\}_{j\in J}$
is called the {\em root lattice}. Denote by $Q^+$ 
the subset of $Q$, consisting of linear combinations of $\alpha_j$ with
non-negative coefficients. Define the standard partial ordering on 
$\fh^*$: $\alpha\geq \beta$ for $\alpha-\beta\in Q^+$.
Letting 
$$\deg e_j=\alpha_j=-\deg f_j, \ \deg \fh=0$$
defines a $Q$-grading of the Lie superalgebra $\tilde{\fg}(\cA,p)$:
$$\tilde{\fg}(\cA,p)=\displaystyle\oplus_{\alpha\in Q} \tilde{\fg}_{\alpha}.$$

It is clear that each $\tilde{\fg}_{\alpha}$ has parity $p(\alpha)$,
where $p: Q\to \mathbb{Z}/ 2\mathbb{Z}$ is defined by additively 
extending $p: J\to \mathbb{Z}/ 2\mathbb{Z}$.
One has the triangular decomposition 
$$\tilde{\fg}(\cA,p)=\tilde{\fn}_-\oplus\fh\oplus\tilde{\fn}_+,$$
where $\tilde{\fn}_-$ (resp., $\tilde{\fn}_+$) is a subalgebra
of $\tilde{\fg}(\cA,p)$ generated by the $f_j$'s (resp., $e_j$'s).
Consequently, $\tilde{\fg}_0=\fh,\ \ 
\tilde{\fn}_{\pm}=\oplus_{\alpha\in Q^+} \tilde{\fg}_{\pm\alpha}$.

\subsubsection{}\label{tildede}
Let $I(\cA,p)$ be the sum of all $Q$-graded ideals of $\tilde{\fg}(\cA,p)$,
which have zero intersection with the subalgebra $\fh$, and let
$$\fg(\cA,p):=\tilde{\fg}(\cA,p)/I(\cA,p).$$
The Lie superalgebra $\fg(\cA,p)$ carries the induced root space decomposition
$$\fg(\cA,p)=\displaystyle\oplus_{\alpha\in Q} \fg_{\alpha},$$
and the induced triangular decomposition 
$$\fg(\cA,p)=\fn_-\oplus\fh\oplus\fn_+,\ \text{ where } \fg_0=\fh,\ 
\fn_{\pm}=\displaystyle\oplus_{\alpha\in Q^+} {\fg}_{\pm\alpha}.$$

An element $\alpha\in Q$ is called a {\em root} (resp., a {\em
positive root}) if $\dim\fg_{\alpha}\not=0$ and $\alpha\not=0$
(resp., $\alpha\in Q^+$).
Denote by $\Delta$ the set  of all roots $\alpha$  and by
$\Delta^+=\Delta\cap Q^+$ the set of all positive roots. 

The Lie superalgebra  $\fg(\cA,p)$ carries an anti-involution
$\sigma$ (i.e. $\sigma([a,b])=[\sigma(b),\sigma(a)]$ and $\sigma^2=\id$)
 defined on the generators by
$$\sigma(e_j)=f_j,\ \sigma(f_j)=e_j,\ \sigma|_{\fh}=\id_{\fh}.$$

\subsubsection{}
The matrix $A:=\bigl(\alpha_j(h_i) \bigr)_{i,j\in J}$ 
is called the {\em Cartan matrix}
of the data $\cA$ (one can show that $A$ and $\dim\fh$ uniquely determine
$\cA$, and that, given $A$ the triple
 $\cA$ exists iff $\dim\fh\geq |J|+\corank A$).
The matrix $A$ is called {\em symmetrizable} if there exists
an invertible diagonal matrix $D=diag (d_j)_{j\in J}$, such that the
matrix $DA=(b_{ij})$ is symmetric.
It is easy to see that if
$A$ is symmetrizable then there exists a non-degenerate symmetric bilinear
form $(.|.)$ on $\fh$ such that
\begin{equation}\label{3}
d_j(h_j|h)=\alpha_j(h)\ \text{ for all } j\in J, h\in\fh.
\end{equation}
This bilinear form induces an isomorphism $\nu:\fh\to\fh^*$, defined
by $\nu(h)(h')=(h,h'),\ h,h'\in\fh$, and we have:
$$\alpha_j=d_j\nu(h_j),\ j\in J,$$ 
and, for the induced bilinear form $(.|.)$ on $\fh^*$ we have:
$$(\alpha_i|\alpha_j)=b_{ij},\  i,j\in J.$$

\subsubsection{}
The following proposition is proved as in~\cite{Kbook}.

\begin{prop}{propDA}
Suppose that the Cartan matrix $A$ is symmetrizable, and let
$(.|.)_{\fh}$ be a non-degenerate symmetric bilinear
form on $\fh$. Then $\fg(\cA,p)$ carries a unique invariant bilinear form
$(.|.)$ (i.e $([a,b]|c) =(a|[b,c])$), whose restriction to $\fh$ 
is the bilinear form $(.|.)_{\fh}$ if and only if
$(.|.)_{\fh}$ satisfies~(\ref{3}) for some non-zero $d_j$'s such that
the matrix $DA$ is symmetric. Moreover, this bilinear form
has the following properties
\begin{enumerate}
\item
$(\fg_{\alpha}|\fg_{\beta})=0\ \text{ if } \alpha+\beta\not=0,\ \
(.|.)_{\fg_{\alpha}+\fg_{-\alpha}} \text{ is non-degenerate for }
\alpha\in\Delta$;
\item
$ [a,b]=(a|b)\nu^{-1}(\alpha),\ \text{ if } a\in\fg_{\alpha},
 b\in\fg_{-\alpha},\alpha\in\Delta$.
\item $(.|.)$ is supersymmetric.
\end{enumerate}
\end{prop}

\subsubsection{}\label{Casimir}
Choose $\rho\in\fh^*$ in such a way that
$$\rho(h_j)=\alpha_j(h_j)/2=a_{jj}/2\ \text{ for any } j\in J.$$
One has $(\rho|\alpha_j)=(\alpha_j|\alpha_j)/2$ if $A$ is symmetrizable.
(Note that if $\det A=0$ then $\rho$ is not uniquely defined.)

Assume that $A$ is symmetrizable.
For each $\alpha\in\Delta^+\cup\{0\}$ choose a basis $\{e^i_{\alpha}\}$
of $\fg_{\alpha}$ and the dual basis $\{e_{-\alpha,i}\}$
of $\fg_{-\alpha}$, i.e. $(e^i_{\alpha}|e_{-\alpha,j})=\delta_{ij}$, and
define the generalized Casimir operator
$$\Omega:=2\nu^{-1}(\rho)+\sum_i e_{0,i}e_0^i+2\sum_{\alpha\in\Delta^+}\sum_i
e_{-\alpha,i}e^i_{\alpha}.$$
This operator is well defined in any restricted $\fg(\cA,p)$-module $V$, i.e.
a module $V$ such that for any $v\in V$, $\fg_{\alpha}v=0$ for all
but finitely many $\alpha\in\Delta^+$. The following proposition is proved
as in~\cite{Kbook}:

\begin{prop}{propCasimir}
\begin{enumerate}
\item
The operator $\Omega$ commutes with $\fg(\cA,p)$ 
in any restricted $\fg(\cA,p)$-module.
\item
If $N$ is a $\fg(\cA,p)$-module and $v\in N$ is such that
$e_jv=0$ for all $j\in J$, and for some $\lambda\in\fh^*$ one has
$hv=\lambda(h)v$ for all $h\in\fh$, then
$\Omega(v)=(\lambda+2\rho|\lambda)v$.
Moreover, if $v$ generates the module $N$ then $N$ is restricted and
$\Omega=(\lambda+2\rho|\lambda)\Id_N$.
\end{enumerate}
\end{prop}

\subsubsection{}
Let $s(\alpha)=(-1)^{p(\alpha)}$ for $\alpha\in Q$, and introduce the
following (in general infinite) product:
$$R:=\displaystyle\prod_{\alpha\in\Delta^+}
(1-s(\alpha)e^{-\alpha})^{s(\alpha)\dim\fg_{\alpha}}.$$
Using the geometric series, we can expand the inverse of this product:
$$R^{-1}=\sum_{\alpha\in Q^+}K(\alpha)e^{-\alpha},\ \text{ where }
K(\alpha)\in\mathbb{Z}_{\geq 0}.$$
Set $K(\mu)=0$ for $\mu\in Q\setminus Q^+$.
Note that $K(\alpha)$ (the Kostant partition function) is 
the number of partition of $\alpha$ into a sum of positive roots 
(counting multiplicities), where odd roots
appear at most once.

\subsection{Generalized Verma modules}
Write $\fg(\cA,p)$ as $\fg$.
Given $I\subset J$, let $Q_I$ be the $\mathbb{Z}$-span of 
$\{\alpha_i\}_{i\in I}$. Set 
$$\fn_{\pm,I}:=\displaystyle\oplus_{\alpha\in Q_I}\fg_{\pm\alpha}, \ \ 
\fh_I:=\sum_{i\in I}\mathbb{C}h_i,\ \ 
\fh^{\perp}_I:=\{\lambda\in \fh^*|\ \lambda(h_i)=0\ \forall i\in I\}.$$
Note that $\fh_I=[\fn_{+},\fn_{-,I}]\cap\fh=[\fn_{+,I},\fn_{-,I}]\cap\fh$.

\subsubsection{}
Introduce the following $(\fn_++\fh+\fn_{-,I})$-module
structure on the symmetric algebra $S(\fh/\fh_I)$:  the action of
$(\fn_++\fh_I+\fn_{-,I})$ is trivial and
$h\in\fh$ acts by the multiplication by the image $\ol{h}$
of the map $\ol{ }:\fh\to \fh/\fh_I$. Introduce the following $\fg$-module
$$M_I:=\Ind_{\fn_++\fh+\fn_{-,I}}^{\fg}S(\fh/\fh_I).$$ 
Note that $M_I$ is a $\fg$--$\fh$ bimodule.

\subsubsection{}
For $\lambda\in\fh^{\perp}_I$ denote by $\Ker_{\lambda}$ the kernel
of $\lambda$ in $S(\fh/\fh_I)$.
The {\em generalized Verma module } $M_I(\lambda)$
is the evaluation of $M_I$ at $\lambda$, that is
$$M_I(\lambda):=M_I/ M_I\Ker_{\lambda}=
\Ind_{\fn_++\fh+\fn_{-,I}}^{\fg}\mathbb{C}_{\lambda},$$
where  $\mathbb{C}_{\lambda}$ is an even one-dimensional space, 
$(\fn_++\fh_I+\fh_{-,I})$ acts trivially on $\mathbb{C}_{\lambda}$,
and $\fh$ acts via the character $\lambda$.

\subsubsection{}
If $I$ is empty then $\fh^{\perp}_I=\fh,\ \fn_{-,I}=0$
and $M_{\emptyset}(\lambda)=M(\lambda)$ is the usual Verma module
with the highest weight $\lambda$; denote 
by $L(\lambda)$ its unique simple quotient.
Clearly, for any $I\subset J$ the generalized Verma module
$M_I(\lambda)$ is a quotient of $M(\lambda)$.

\subsubsection{}\label{HCpr}
Identify the universal enveloping algebra $U(\fh)$
 with the symmetric algebra $S(\fh)$.
The triangular decomposition $\fg=\fn_-\oplus\fh\oplus\fn_+$
induces the  following decomposition of the  universal enveloping superalgebra:
$U(\fg)=S(\fh)\oplus (\fn_-U(\fg)+U(\fg)\fn_+)$; the corresponding
projection $\HC:U(\fg)\to S(\fh)$ is called
the {\em Harish-Chandra projection}. Let  $\HC_I:U(\fg)\to S(\fh/\fh_I)$
be the composition  of $\HC$ and  the canonical map 
$S(\fh)\to S(\fh/\fh_I)=S(\fh)/S(\fh)\fh_I$.
Define the bilinear form $S(.,.)$, called the {\em Shapovalov form}, 
on $M_I$ as follows:
$$S(u.1,u'.1)=\HC_I(\sigma(u)u')\ \text{ for }u,u'\in U(\fg),$$
where dot denotes the action and
$1$ stands for the canonical generator of $M_I$.

It is easy to see that  the bilinear form
$S: M_I\otimes M_I\to S(\fh/\fh_I)$ satisfies the following properties,
which determine it uniquely:
\begin{equation}\label{defformS}
\begin{array}{ll}
S(1,1)=1,\\
S(uv,v')=S(v,\sigma(u)v')\ \text{ for } u\in\fg,\ 
v,v'\in M_I,\\
S(v,v'h)=S(vh,v')=S(v,v')h\ \text{ for } h\in\fh,\ 
v,v'\in M_I .
\end{array}
\end{equation}
One easily deduces from the uniqueness that this bilinear form is symmetric.

\subsubsection{}
The module $M_I$ is graded by $Q^+$: 
$M_I=\oplus_{\nu\in Q^+} M_{I,\nu}$, where
$$M_{I,\nu}=\{v\in M_I|\ hv-vh=-\nu(h)v\ \text{ for } h\in\fh\}.$$
The image of $M_{I,\nu}$ in $M_I(\lambda)$
is the weight space $M_{I,\nu}(\lambda)$ of weight $\lambda-\nu$.

It is easy to see that $S(M_{I,\nu},M_{I,\mu})=0$ for $\nu\not=\mu$.
Let $S_{\nu}$ be the restriction of $S$
to $M_{I,\nu}$. Each component $M_{I,\nu}$ is a free
$S(\fh/\fh_I)$-module of rank not greater than $K(\nu)$.
Therefore $\det S_{\nu}$ is an element of $S(\fh/\fh_I)$,
defined up to a  non-zero constant factor, 
depending on the basis of $M_{I,\nu}$. 
Clearly, $\det S_{\nu}=1$ for $\nu=0$.

\subsubsection{}
For $\lambda\in\fh_I^{\perp}$ 
the evaluation of $S$ at $\lambda$ gives a bilinear form 
$S(\lambda):M_I(\lambda)\otimes M_I(\lambda)\to \mathbb{C}$,
whose restriction to $M_{I,\nu}(\lambda)$ is $S_{\nu}(\lambda)$.
It is easy to show that the kernel of $S(\lambda)$
 coincides with the maximal proper submodule
of  $M_I(\lambda)$. As a consequence, 
$$M_I(\lambda)\ \text{ is simple }\ \Longleftrightarrow\ 
\det S_{\nu}(\lambda)\not=0\ \forall\nu\in Q^+.$$

\subsubsection{}
Introduce the following linear function 
$\phi_{\alpha}(\lambda)$ on $\fh^*$ for each $\alpha\in Q$:
$$\phi_{\alpha}(\lambda)=(\lambda+\rho|\alpha)-\frac{1}{2}(\alpha|\alpha).$$
Set $\Delta_I^+:=\Delta\cap Q_I$ and introduce the following product:
$$R_I:=\displaystyle\prod_{\alpha\in\Delta^+_I}
(1-s(\alpha)e^{-\alpha})^{s(\alpha)\dim\fg_{\alpha}}.$$
Using the geometric series, we can expand this product:
$$R_I=\sum_{\alpha\in Q^+_I}k_I(\alpha)e^{-\alpha},\ \text{ where }
k_I(\alpha)\in\mathbb{Z};$$
set $k_I(\alpha)=0$ for $\alpha\in Q\setminus Q^+_I$.

\subsubsection{}
In Sect.~\ref{SectSh} we will prove the following theorem

\begin{thm}{thmsh}
Let $\fg(\cA,p)$ be the contragredient
Lie superalgebra, attached to the data $(\cA,p)$
with a symmetrizable Cartan matrix, and let $I\subset J$. Then 
one has for $\lambda\in\fh^{\perp}_I$ (up to a non-zero constant factor
depending on the basis of $M_{I,\nu}(\lambda)$):
$$\det S_{\nu}(\lambda)=\prod_{r=1}^{\infty}
\prod_{\gamma\in \Delta^+\setminus Q^+_I}
\prod_{\alpha\in Q^+_I}\phi_{r\gamma+\alpha}(\lambda)^{
(-1)^{(r-1)p(\gamma)}k_I(\alpha)K(\nu-r\gamma-\alpha)\dim\fg_{\gamma}}
.$$
\end{thm}

\subsubsection{}
\begin{rem}{}
For $I=\emptyset$, i.e. ordinary Verma modules, $Q^+_I=\{0\}$, 
$k_I(\alpha)=\delta_{\alpha,0}$,
and we recover the determinant formula from~\cite{KK} in the non-super case, 
and from~\cite{K3} in the super case.
\end{rem}

\subsubsection{}
\begin{rem}{rem11}
Let $\fh'$ be a subspace of $\fh$, contaning $\Pi^{\vee}$. Then
$\fg'(\cA,p)=\fn_-\oplus\fh'\oplus\fn_+$ is a subalgebra
of $\fg(\cA,p)$, and any generalized Verma module over $\fg'(\cA,p)$
extends (non-uniquely) to that over $\fg(\cA,p)$ by extending 
$\lambda\in (\fh')^*$ to a linear function on $\fh$.
Defining the weight spaces of the former as that of the latter, it is clear
that, by restriction,~\Thm{thmsh} still holds for the 
 generalized Verma module $M_I(\lambda)$ over $\fg'(\cA,p)$, and
the formula for $\det S_{\nu}(\lambda)$ is independent of the extension of
$\lambda$ from $\fh'$ to $\fh$.
\end{rem}

\subsubsection{}
\begin{exa}{exa12}
The simple finite-dimensional Lie algebras carry, of course a unique, up to
an isomorphism, structure of a contragredient Lie algebra. Simple
Lie superalgebras $\fsl(m,n)$ for $m\not=n$, $\osp(m,n)$,
$D(2,1,a)$, $F(4)$ and $G(3)$ carry a
structure of a contragredient Lie superalgebra as well, 
in fact, several non-isomorphic such structures (which depend on the choice
of the set of positive roots)~\cite{Kadv}.

The Lie superalgebra $\fsl(m,m)$ is not quite a contragredient 
Lie superalgebra, but $\fgl(m,m)$ is.
Hence, by~\Rem{rem11},~\Thm{thmsh} holds for 
$\fsl(m,m)$ and 
$\fsl(m,m)/\mathbb{C} I_{2m}$ 
as well. 
\end{exa}

\subsubsection{}
\begin{exa}{exa13}
If $\fg$ is one of the simple Lie superalgebras from~\Exa{exa12}, then the
affine Lie superalgebra 
$\fhg=\fg[t,t^{-1}]\oplus \mathbb{C}K$,
described in~\ref{sub01} is not quite a contragredient Lie superalgebra,
but $\mathbb{C}D\ltimes\hat{\fg}$, where
$D=t\frac{d}{dt}$ on $\fg[t,t^{-1}]$ and [$D,K]=0$, is~\cite{Kbook}.

Recall that the Cartan subalgebra of $\mathbb{C}D\ltimes\hat{\fg}$
is chosen to be 
$$\hat{\fh}=\fh\oplus\mathbb{C}K\oplus\mathbb{C}D,$$
where $\fh$ is a Cartan subalgebra of $\fg$.
Define the linear function $\delta$ on $\fhh$ by 
$\delta|_{\fh\oplus\mathbb{C}K}=0$, $\delta(D)=1$.
Let $\theta\in\Delta^+$ be the highest root, and let 
$e_{\theta}\in\fg_{\theta},\ f_{\theta}\in\fg_{-\theta}$
be such that $(f_{\theta}| e_{\theta})=1$.

Recall that if $\fg=\fg(\cA,p)$, where $\cA=(\fh,\Pi,\Pi^{\vee})$
is a structure of a contragredient Lie superalgebra
on $\fg$, with generators $e_j,f_j$ ($j\in J$) and $\fh$, then
$\mathbb{C}D\ltimes\hat{\fg}=\fg(\hat{\cA},\hat{p})$, where
$\hat{\cA}=(\fhh,\hat{\Pi},\hat{\Pi^{\vee}})$ with
$$\hat{\Pi}=\Pi\cup\{\alpha_0=\delta-\theta\},\ 
\hat{\Pi^{\vee}}={\Pi^{\vee}}\cup\{h_0=K-\nu^{-1}(\theta)\},$$
and the generators
$$e_0=f_{\theta}t,\ e_j\ (j\in J);\ \ f_0=e_{\theta}t^{-1}, f_j (j\in J),$$
so that the index set is $\hat{J}=\{0\}\cup J$,
and $\hat{p}(0)=p(\theta),\ \hat{p}(j)=p(j)$ for $j\in J$ 
(see~\cite{Kbook} for details in the Lie algebra case).
By~\Rem{rem11},~\Thm{thmsh} applies to $\hat{\fg}$.
\end{exa}

\subsubsection{}
\begin{exa}{exagl}
If $\fg=\fgl(m,m)$, then $\fg[t,t^{-1}]\oplus \mathbb{C}K$ contains an ideal
$J=\sum_{n\not=0}\mathbb{C}I_{2m}t^n$, which intersects 
$\fh\oplus \mathbb{C}K$ trivially. Let
$\fhg=(\fg[t,t^{-1}]/J)\oplus \mathbb{C}K$. It is easy to see
that $\fhg$ extends to a contragredient Lie superalgebra as in~\Rem{rem11},
hence~\Thm{thmsh} again applies.The same is true for
$\fhg=\fg[t,t^{-1}]\oplus 
\mathbb{C}K$, where 
$\fg$=$\fsl(m,m)/ \mathbb{C}I_{2m}$. 
\end{exa}

\section{Determinant of the Shapovalov form}
\label{SectSh}
Let $\fg:=\fg(\cA,p)$ be the Lie superalgebra, attached to the data $(\cA,p)$,
with a symmetrizable Cartan matrix.
In this section we prove~\Thm{thmsh}.

\subsection{Linear factorization}
Let
$$\begin{array}{c}
\Irr:=\{\alpha\in Q^+\setminus Q_I|\ 
\alpha/n\not\in Q^+\text{ for }
n\in\mathbb{Z}_{\geq 2}\},\\
\widetilde{\Irr}:=\{(m,\alpha)\in\Irr\times \mathbb{Z}_{\geq 1}|\ 
(\alpha|\alpha)\not=0\}\cup \{(\alpha,1)|\ 
\alpha\in\Irr 
\&\ (\alpha|\alpha)=0\}.
\end{array}
$$

\subsubsection{}\label{dym}
Any simple subquotient of $M_I(\lambda)$ is of the form 
$L(\lambda-\alpha)$ for $\alpha\in Q^+\setminus Q_I$. 
From~\Prop{propCasimir} we conclude that the Casimir element acts on 
$M(\lambda)$ by the scalar $(\lambda+2\rho|\lambda)$,
and that if $L(\lambda-\alpha)$ is a subquotient of
$M_I(\lambda)$ then $(\alpha|2(\lambda+\rho)-\alpha)=0$.
Writing $\alpha=m\beta$
with $\beta\in\Irr, m\geq 1$ we obtain that
\begin{equation}\label{Vc:}\begin{array}{l}
[M_I(\lambda): L(\lambda-m\beta)]\not=0\ \Longrightarrow\ \ 
\phi_{m\beta}(\lambda)=0.
\end{array}\end{equation}
Observe that $\phi_{m\beta}=\phi_{\beta}$ if $(\beta|\beta)=0$.
Hence up to a  non-zero constant factor one has
$$\det S_{\nu}(\lambda)=\prod_{(m,\beta)\in \widetilde{\Irr}}
\phi_{m\beta}(\lambda)^{d_{m,\beta}(\nu)},$$
where $d_{m,\beta}(\nu)$ are some non-negative integers.
Note that $d_{m,\beta}(\nu)\not=0$ forces $m\beta\leq\nu$, i.e. 
$\nu-m\beta\in Q^+$.

\subsection{Jantzen filtration}
In~\ref{jancon} we recall the construction of 
the Jantzen filtration (see~\cite{Jan}).
This filtration depends
on a ``generic element'' $\rho'\in\fh^*$. 
For a semisimple Lie algebra one can take a sum of
fundamental roots:
$\rho':=\sum_{j\in J\setminus I} \omega_j$.
It is known that for semisimple Lie algebras the Jantzen filtration
does not depend on a choice of ``generic'' $\rho'$, see~\cite{BB}, 5.3.1.

\subsubsection{}\label{jancon}
Fix $\rho'\in \fh^{\perp}_I$ such that $(\rho'|\alpha)\not=0$
for any $\alpha\in Q^+\setminus Q^+_I$. 
Let $t$ be an indeterminate. 

Take $\lambda\in\fh_I^{\perp}$. Introduce the generalized Verma module
$M_I(\lambda+t\rho')$ as follows. Define the action of 
 $(\fn_++\fh+\fh_{-,I})$ on  $\mathbb{C}[t]$: 
 $(\fn_++\fh_I+\fh_{-,I})$ acts trivially and
$h\in\fh$ acts by the multiplication to 
$(\lambda+t\rho')(h)=\lambda(h)+t\rho'(h)$.
Now $M_I(\lambda+t\rho')$ is the following $\fg-\mathbb{C}[t]$ bimodule
$$M_I(\lambda+t\rho'):=
\Ind_{\fn_++\fh+\fn_{-,I}}^{\fg}\mathbb{C}[t].$$ 
The module $M_I(\lambda+t\rho')$ admits a unique invariant 
$\mathbb{C}[t]$-bilinear form
$S^{\lambda+t\rho'}: 
M_I(\lambda+t\rho')\otimes M_I(\lambda+t\rho')\to \mathbb{C}[t]$
which satisfies the properties~(\ref{defformS}); this form is symmetric.
For  $r\in\mathbb{Z}_{\geq 0}$, set
$$M_I^r(\lambda+t\rho'):=\{v\in M_I(\lambda+t\rho')|\ 
   S^{\lambda+t\rho'}(v,v')\in t^r \mathbb{C}[t]\ \ \forall v'\}.$$
This defines a decreasing filtration. The property (ii) of~(\ref{defformS})
insures that each $M_I^r(\lambda+t\rho')$ is a sub-bimodule
of $M_I(\lambda+t\rho')$. 
The weight spaces of $M_I(\lambda+t\rho')$ are free of finite rank
$\mathbb{C}[t]$-modules so we can define
$\det S^{\lambda+t\rho'}_{\nu}$ (up to a non-zero constant
factor). Clearly, 
$\det S^{\lambda+t\rho'}_{\nu}=\det S_{\nu}(\lambda+t\rho')$
and this is non-zero due to the linear factorization
of $\det S_{\nu}$ and ``genericity'' of $\rho'$. As a result, 
$\cap_{r=0}^{\infty} M_I^r(\lambda+t\rho')=0$.

Specializing this filtration at $t=0$ we obtain
the Jantzen filtration $\cF^r(M_I(\lambda))$ on $M_I(\lambda)$.
The weight spaces of $M^r_I(\lambda+t\rho')$ are free of finite rank
$\mathbb{C}[t]$-modules. Thus $\cF^r(M_I(\lambda))$ is just the image 
of $M_I^r(\lambda+t\rho')$ under the canonical map
$M_I(\lambda+t\rho')\to M_I(\lambda+t\rho')/tM_I(\lambda+t\rho')
\iso M_I(\lambda)$.  In particular the $\cF^r(M_I(\lambda))$
form a decreasing filtration by submodules of $M_I(\lambda)$
having zero intersection. One readily sees that 
$\cF^0(M_I(\lambda))=M_I(\lambda)$ and that
$\cF^1(M_I(\lambda))$ coincides with the maximal proper
submodule of $M_I(\lambda)$.

\subsubsection{}
Define the sets $\tilde{R}_{m,\beta}$ ($m\in\mathbb{Z}_{\geq 1},\ \beta'\in\Irr$)
and $C(\lambda)$ ($\lambda\in\fh_I^{\perp}$) as follows:
$$\begin{array}{l}
\tilde{R}_{m,\beta}:=\{ (m',\beta')\in\mathbb{Z}_{\geq 1}\times \Irr|\ \ \ 
\phi_{m\beta}|_{\fh_I^{\perp}}=\phi_{m'\beta'}|_{\fh_I^{\perp}}\},\\
C(\lambda):=\{(m,\alpha)\in\mathbb{Z}_{\geq 1}\times \Irr|\ \ 
\ \phi_{m\alpha}(\lambda)=0\}.
\end{array}$$

\subsubsection{}
The following ``sum formula'' is proven in~\cite{Jan}:
\begin{equation}\label{sumfor}
\sum_{i=1}^{\infty} \dim \cF^i(M_I(\lambda)_{\lambda-\nu})=\sum_{(m,\beta)\in
C(\lambda)\cap \widetilde{\Irr}} d_{m,\beta}(\nu),
\end{equation}
where $d_{m,\beta}(\nu)$ are exponents introduced in~\ref{dym}.
\begin{proof}
Note that the sum
$\mathop{\sum}\limits_{r=1}^{\infty}\dim \cF^r(M_I(\lambda)_{\lambda-\nu})$
is equal to the order of zero of $\det S_{\nu}$ at  the point 
$\lambda\in\fh^{\perp}_I$.
Let $A$ be the localization of $\mathbb{C}[t]$ by the maximal
ideal generated by $t$: $A=\mathbb{C}[t]_{(t)}$.
Let $N$ be a free $A$-module of finite rank,
endowed with a non-degenerate bilinear
form $D: N\otimes N\to A$.
Define a decreasing filtration
$$
F^j(N):=\{v\in N\mid   D(v, v')\in At^j\,\text{ for any $v'\in N$}\}.
$$
Taking $N$ to be the localized module
$M_I(\lambda)_{\lambda-\nu}\otimes_A \mathbb{C}[t]$ and $D$ to be
the bilinear form induced by $S^{\lambda+t\rho'}$, we see
that the filtration on $N$, induced by the Jantzen filtration, 
is just $F^j(N)$. Now
the sum formula follows from the following claim:
{\em the order of zero of  $\det D$ at the origin is equal to}
$$
\mathop{\sum}\limits_{j=1}^{\infty}\dim \bigl(F^j(N)/(F^j(N)\cap tN)\bigr).
$$
In order to prove the claim, note that $N$ has two systems of generators
$v_i$ and $v'_i$ (for $i=1, \ldots, r$) such that $D(v_i,
v'_j)=\delta_{ij}t^{s_i}$
(for $s_i\in\mathbb{Z}_{\geq 0}$).
The order of zero of  $\det D$ at the origin is
$\mathop{\sum}\limits_{i=1}^r s_i$ and
$$
\dim F^j(N)/(F^j(N)\cap tN)=|\{i\mid s_i\geq j\}|.
$$
The equality $\mathop{\sum}\limits_i
s_i=\mathop{\sum}\limits_{j=1}^{\infty}|\{i\mid s_i\geq j\}|$
implies the claim.\end{proof}

\subsubsection{}
Define the functions 
$d_{m,\beta},\tau_{\alpha}:Q\to\mathbb{Z}_{\geq 0}$ 
($\alpha,\beta\in Q$) by  
$$\tau_{\alpha}:\nu\mapsto K(\nu-\alpha),\ \ \ 
d_{m,\beta}|_{(Q\setminus Q^+)\cup Q_I}=0,\ 
d_{m,\beta}:\nu\mapsto d_{m,\beta}(\nu) 
\text{ for }\nu\in Q^+\setminus Q_I.$$
The following lemma is proven in~\cite{JNato}, 6.8 for simple Lie algebras;
however, the proof is valid in our general setup.
\subsubsection{}
\begin{lem}{sumJ}
For any $\lambda\in\fh_I^{\perp},\ m\geq 1,\ \beta\in\Irr$ 
there exist integers $a^{\lambda}_{m,\beta}, a_{m,\beta}$ such that
\begin{enumerate}

\item
$\sum_{i=1}^{\infty} \ch \cF^i(M_I(\lambda))=\sum_{(m,\beta)\in
C(\lambda)}a^{\lambda}_{m,\beta}\ch M(\lambda-m\beta),$
\item
$\sum_{(m',\beta')\in \tilde{R}_{m,\beta}\cap \widetilde{\Irr}}
d_{m',\beta'}=
\sum_{(m',\beta')\in \tilde{R}_{m,\beta}}a_{m',\beta'}
\tau_{m'\beta'}$.
\end{enumerate}
\end{lem}
\begin{proof}
Combining the fact that $\cF^i(M_I(\lambda))$ 
is a $\fg$-submodule of $M_I(\lambda)$ and formula~(\ref{Vc:}),
we deduce that
$\ch \cF^i(M_I(\lambda))=\sum_{(m,\beta)\in
C(\lambda)}a_{m,\beta}^{\lambda,i}\ch M(\lambda-m\beta)$
for some integers $a_{m,\beta}^{\lambda,i}$; note that the sum is infinite, 
but ``locally finite'': for each $\nu\in Q^+$  only finite many terms 
$M(\lambda-m\beta)_{\lambda-\nu}$ are non-zero.
Thus we obtain (i) for $a_{m,\beta}^{\lambda}:=
\sum_{i=1}^{\infty}a_{m,\beta}^{\lambda,i}$.
For (ii) fix a pair $(m,\beta)$. Let $\lambda\in\fh_I^{\perp}$ 
be a ``generic point'' of 
the hyperplane $\{\xi: \phi_{m\beta}(\xi)=0\}$
in the following sense: $\lambda$ does not belong to the hyperplanes
$\{\xi: \phi_{m'\beta'}(\xi)=0\}$ if $(m',\beta')\not\in
\tilde{R}_{m,\beta}$; in other words, $C(\lambda)=\tilde{R}_{m,\beta}$.
Combining (i) and  formula~(\ref{sumfor}) one obtains
$$\sum_{(m',\beta')\in C(\lambda)} a^{\lambda}_{m',\beta'}
\tau_{m'\beta'}=\sum_{(m',\beta')\in C(\lambda)\cap\widetilde{\Irr}}
d_{m',\beta'}.$$
Since $C(\lambda)=\tilde{R}_{m,\beta}$ one gets (ii)
for the integers $a_{m,\beta}:=a^{\lambda}_{m,\beta}$.
\end{proof}

\subsubsection{}
\begin{cor}{zamena}
$\displaystyle\prod_{(m',\beta')\in \tilde{R}_{m,\beta}\cap\tilde{Irr}}
\phi_{m'\beta'}^{d_{m',\beta'}(\nu)}=
\displaystyle\prod_{(m',\beta')\in \tilde{R}_{m,\beta}}
\phi_{m'\beta'}^{a_{m',\beta'}(\nu)K(\nu-m'\beta')}$.
\end{cor}
\begin{proof}
By definition $\phi_{m'\beta'}=\phi_{m\beta}$ for 
$(m',\beta')\in \tilde{R}_{m,\beta}$. In the light of~\Lem{sumJ} (ii),
both sides of formula are equal to 
$\phi_{m\beta}^{\sum_{(m',\beta')\in \tilde{R}_{m,\beta}}
a_{m',\beta'}(\nu)\tau_{{m'\beta'}(\nu)}}$.
\end{proof}

\subsection{Leading term}\label{lead}
Using the geometric series, we expand
$$R_I/R=\displaystyle\prod_{\alpha\in\Delta^+\setminus Q_I}
(1-s(\alpha)e^{-\alpha})^{-s(\alpha)\dim\fg_{\alpha}}=
\sum_{\alpha\in Q^+}K_I(\alpha)e^{-\alpha},\ \text{ where }
K_I(\alpha)\in\mathbb{Z}_{\geq 0}.$$
Set $K_I(\alpha)=0$ for $\alpha\in Q\setminus Q^+$; note that
$K_I(\alpha)=0$ for $\alpha\in Q_I,\ \alpha\not=0$.

Consider the natural grading on
the symmetric algebra 
$\cS(\fh/\fh_I)=\oplus_{r=0}^{\infty}\cS^r(\fh/\fh_I)$.
The following proposition is a particular case of~\cite{GS} Thm. 3.1.

\begin{prop}{claimlead}
Up to a non-zero  constant factor,
the leading term of $\det S_{\nu}$ is
$$\gr\det S_{\nu}=\prod_{\alpha\in {\Delta}^+\setminus Q_I} 
h_{\alpha}^{(\dim\fg_{\alpha})\sum_{r\geq 1} (-1)^{(r-1)p(\alpha)
K_I(\nu-r\alpha)}},
$$
where $h_{\alpha}\in\fh/\fh_I$ is 
such that $\mu(h_{\alpha})=(\mu|\alpha)$ for any 
$\mu\in\fh_I^{\perp}$.
\end{prop}
\begin{proof}
Denote by $\tilde{\Delta}^+_0, \tilde{\Delta}^+_1$ 
the corresponding multisets of roots, 
where the multiplicity of $\gamma$ is equal to $\dim\fg_{\gamma}$.
Set $\tilde{\Delta}^+:=\tilde{\Delta}^+_0\cup\tilde{\Delta}^+_1$.
Define similarly $\tilde{\Delta}^+_I$ (the multiset corresponding
to $\Delta^+_I$). Fix a total ordering on $\tilde{\Delta}^+$
such that $\gamma_1\geq \gamma_2$ if  $\gamma_1-\gamma_2\in Q^+$.
\subsubsection{}
A vector $\mathbf{m}=
\{ m_{\gamma}\}_{\gamma\in\tilde{\Delta}^+\setminus\tilde{\Delta}^+_I}$ 
is called a {\em partition of $\alpha\in Q^+\setminus Q_I$} if 
$\alpha=\displaystyle\sum_{\gamma\in \tilde{\Delta}^+}
m_{\gamma}\gamma:\ \ \ 
m_{\gamma}\in \mathbb{Z}_{\geq 0}\text{ for }
\gamma\in\tilde{\Delta}^+_0\setminus\tilde{\Delta}^+_I, 
 \text{ and }
m_{\gamma}\in \{0,1\}\text{ for }
\gamma\in \tilde{\Delta}_1^+\setminus\tilde{\Delta}^+_I$.
Denote by ${\cP}(\alpha)$ the set of all partitions of $\alpha$. 
One has $|{\cP}(\alpha)|=K_I(\alpha)$.

\subsubsection{}
For $\gamma\in\tilde{\Delta}^+$ denote by $\ol{\gamma}$
the corresponding element in the set $\Delta$.
Choose bases $\{f_{\gamma}\}_{\gamma\in\tilde{\Delta}^+}$
of $\fn_-$ and $\{e_{\gamma}\}_{\gamma\in\tilde{\Delta}^+}$
of $\fn_+$
such that 
$f_{\gamma}\in\fg_{-\ol{\gamma}},\ e_{\gamma}\in\fg_{\ol{\gamma}}\,$.
In the light of~\Prop{propDA} (ii), for each $\alpha\in\Delta^+$
the entries of the matrix 
$$D_{\alpha}=
\bigl([f_{\gamma_i},e_{\gamma_j}] \bigr)_{\ol{\gamma_i}=
\ol{\gamma_j}=\alpha}$$
are proportional to $h_{\alpha}$ and $\det D_{\alpha}\not=0$.
Hence we can choose the bases in such a way that all
matrices $D_{\alpha}$ are diagonal: 
$D_{\alpha}=\bigl(\delta_{ij}h_{\alpha}\bigr)_{i,j}$.

For every $\mathbf{m}\in \cP(\alpha)$, define the
monomial
$$
\mathbf{f}^{\mathbf{m}}:=
\prod_{\alpha} f_{\alpha}^{m_{\alpha}},\ \ \mathbf{e}^{\mathbf{m}}:=
\prod_{\alpha} e_{\alpha}^{m_{\alpha}},
$$
where the order of factors is given by the total ordering fixed above.
Take $\lambda\in\fh_I^{\perp}$ and let $v_{\lambda}$ be the highest
weight vector of $M_I(\lambda)$.
The set 
$\{\mathbf{f}^{\mathbf{m}}v_{\lambda}\mid \mathbf{m}\in \cP(\nu)\}$ 
forms a basis of $M_I(\lambda)_{\lambda-\nu}$. By definition
given in~\ref{HCpr},
$$\det S_{\nu}=\det
\bigl(\HC_I(\sigma(\mathbf{f}^{\mathbf{m}})
\mathbf{f}^{\mathbf{s}})\bigr)_{\mathbf{m},\mathbf{s}\in \cP(\nu)}.$$
Since both $\{\sigma(\mathbf{f}^{\mathbf{m}})\}_{\mathbf{m}\in \cP(\nu)}$
and $\{\mathbf{e}^{\mathbf{m}}\}_{\mathbf{m}\in \cP(\nu)}$
are bases of the same vector space,
one has
$$\det S_{\nu}=\det
\bigl(\HC_I(\mathbf{e}^{\mathbf{m}}
\mathbf{f}^{\mathbf{s}})\bigr)_{\mathbf{m},\mathbf{s}\in \cP(\nu)},$$
up to a non-zero constant factor.

\subsubsection{}
Set $|\mathbf{k}|=\displaystyle\sum_{\alpha\in\Delta^+} k_{\alpha}$.
For $u\in \cU(\fg)$ denote by $\gr u$ the image of $u$ in
the symmetric algebra $\Sg$.

\begin{lem}{}
For any $\mathbf{m}, \mathbf{s}\in\cP(\nu)$, we have
\begin{enumerate}
\item
$\deg\HC(\mathbf{e}^{\mathbf{m}}\mathbf{f}^{\mathbf{s}})
\leq\min(|\mathbf{m}|, |\mathbf{s}|);$
\item
if $|\mathbf{m}|=|\mathbf{s}|$, we have
$$
\deg\HC(\mathbf{e}^{\mathbf{m}}\mathbf{f}^{\mathbf{s}})
=|\mathbf{m}\mid
\Longleftrightarrow\ \mathbf{m}=\mathbf{s};
$$
\item up to a non-zero constant factor,
$$
\gr\HC(\mathbf{e}^{\mathbf{m}}\mathbf{f}^{\mathbf{m}})=
\mathop{\prod}\limits_{\gamma\in\tilde{\Delta}^+}h_{\ol{\gamma}}^{m_{\gamma}}.
$$
\end{enumerate}
\end{lem}
Proof is by induction  on $\nu\in Q^+$
with respect to the partial order (see~\ref{partord}).

\subsubsection{}
\begin{cor}{estimshap1}
Up to a non-zero constant factor,  the leading term of $\det S_{\nu}$
is equal to
$$
\mathop{\prod}\limits_{\alpha\in{\Delta}^+\setminus
{\Delta}_I}h_{\alpha}^{r_{\alpha}(\nu)},
$$
where
$$r_{\alpha}(\nu):=\mathop{\sum}\limits_{\gamma\in\tilde{\Delta}^+:
\ol{\gamma}=\alpha}
\mathop{\sum}\limits_{\mathbf{m}\in\cP(\nu)}m_{\gamma}.$$
\end{cor}

\subsubsection{}
\begin{lem}{estimshap2}
For any $\gamma\in \tilde{\Delta}^+\setminus\Delta_I$ one has
$$
\displaystyle\mathop{\sum}_{\mathbf{m}\in \cP(\nu)}m_{\gamma}=
\displaystyle\mathop{\sum}_{r=1}^{\infty}(-1)^{(r-1)p(\ol{\gamma})}
K_I(\nu-r\ol{\gamma}).
$$
\end{lem}
Proof is by induction  on $\nu\in Q^+\setminus Q_I$
with respect to the partial order (see~\ref{partord}).

\subsubsection{}
Combining~\Cor{estimshap1} and~\Lem{estimshap2}, we obtain the required
assertion.
\end{proof}

\subsection{Computation of $a_{m,\beta}$}
Since $\coprod \tilde{R}_{m,\beta}=\mathbb{Z}_{\geq 1}\times \Irr$, 
 \Lem{sumJ} (ii) gives
\begin{equation}\label{datau}
\sum_{(m,\beta)\in \widetilde{\Irr}}d_{m,\beta}=
\sum_{(m,\beta)\in \mathbb{Z}_{\geq 1}\times \Irr}a_{m,\beta}
\tau_{m\beta}.
\end{equation}
Both sides of the above formula are well-defined functions on $Q$:
for each $\nu\in Q$ only summands
indexed by the pairs $(m,\beta)$, where $\nu-m\beta\in Q^+$, are non-zero
at $\nu$, and thus only finitely many summands 
 are  non-zero for each $\nu\in Q$.

\subsubsection{}
From~\Lem{lead},
$$\sum_{(m,\beta)\in \widetilde{\Irr}}d_{m,\beta}(\nu)=
\sum_{\alpha\in {\Delta}^+\setminus Q_I}\sum_{r=1}^{\infty} 
(-1)^{(r+1)p(\alpha)}(\dim\fg_{\alpha}) K_I(\nu-r\alpha),
$$
and thus, using~(\ref{datau}) and $\tau_{m\beta}(\nu)=K(\nu-m\beta)$, we get
$$\sum_{(m,\beta)\in \mathbb{Z}_{\geq 1}\times \Irr}a_{m,\beta}\sum_{\nu}
K(\nu-m\beta)e^{-\nu}=
\sum_{\alpha\in {\Delta}^+\setminus Q_I}\sum_{r=1}^{\infty} \sum_{\nu}
(-1)^{(r+1)p(\alpha)}(\dim\fg_{\alpha}) K_I(\nu-r\alpha)e^{-\nu},$$
which can be rewritten as
$$\sum_{(m,\beta)\in \mathbb{Z}_{\geq 1}\times \Irr}a_{m,\beta}\sum_{\nu}
K(\nu)e^{-\nu-m\beta}=
\sum_{\alpha\in {\Delta}^+\setminus Q_I}\sum_{r=1}^{\infty} \sum_{\nu}
(-1)^{(r+1)p(\alpha)}(\dim\fg_{\alpha})K_I(\nu)e^{-\nu-r\alpha},
$$
that is
$$R^{-1}\sum_{(m,\beta)\in \mathbb{Z}_{\geq 1}\times \Irr}a_{m,\beta} 
e^{-m\beta}=R^{-1}R_I\sum_{\alpha\in {\Delta}^+\setminus Q_I}
\sum_{r=1}^{\infty}  (-1)^{(r+1)p(\alpha)} (\dim\fg_{\alpha}) e^{-r\alpha}.$$
Therefore the integer $a_{m,\beta}$ is
equal to the coefficient of $e^{-m\beta}$ in the expression
$$R_I \sum_{\gamma\in {\Delta}^+\setminus Q_I}
\sum_{r=1}^{\infty}  (-1)^{(r+1)p(\gamma)}(\dim\fg_{\gamma})
 e^{-r\gamma}.$$
Hence
$$a_{m,\beta}=\sum_{\gamma\in {\Delta}^+\setminus Q_I}
\sum_{r=1}^{\infty}  (-1)^{(r+1)p(\gamma)} (\dim\fg_{\gamma})
k_I(m\beta-r\gamma).$$

\subsubsection{}
Substituting the formula for $a_{m,\beta}$ into~\Cor{zamena} we get

$$\begin{array}{rl}
\det S_{\nu}&=\displaystyle\prod_{(m,\beta)\in\mathbb{Z}_{\geq 1}\times\Irr}
\phi_{m\beta}^{K(\nu-m\beta)\sum_{\gamma\in {\Delta}^+\setminus Q_I}
\sum_{r=1}^{\infty}  (-1)^{(r+1)p(\gamma)} (\dim\fg_{\gamma})
k_I(m\beta-r\gamma)}\\
 &=
\displaystyle\prod_{\alpha\in Q^+}\phi_{\alpha}^{K(\nu-\alpha)
\sum_{\gamma\in {\Delta}^+\setminus Q_I}
\sum_{r=1}^{\infty}  (-1)^{(r+1)p(\gamma)} 
(\dim\fg_{\gamma}) k_I(\alpha-r\gamma)}\\
&=\displaystyle\prod_{\alpha\in Q^+}\prod_{r=1}^{\infty}
\prod_{\gamma\in {\Delta}^+\setminus Q_I}
\phi_{\alpha}^{(-1)^{(r+1)p(\gamma)} (\dim\fg_{\gamma}) 
k_I(\alpha-r\gamma)K(\nu-\alpha)}\\
&=\displaystyle\prod_{r=1}^{\infty}\prod_{\alpha\in Q^+}
\prod_{\gamma\in {\Delta}^+\setminus Q_I}
\phi_{\alpha+r\gamma}^{(-1)^{(r+1)p(\gamma)}(\dim\fg_{\gamma})
 k_I(\alpha)K(\nu-\alpha-r\gamma)}.
\end{array}$$
Recalling that  $k_I(\alpha)=0$ for $\alpha\not\in Q^+_I$,
 this completes the proof of~\Thm{thmsh}.

\section{Vacuum determinant}
Let $\fg=\fg(\cA,p)$ be a finite dimensional contragredient Lie superalgebra
and let $\fg(\hat{\cA},\hat{p})=\mathbb{C}D\ltimes\hat{\fg}$ 
be its (untwisted) affinization described
in~\Exa{exa13}, which notation we retain, except that here we take 
$\fhh=\fh+\mathbb{C}K$.

We denote $\Delta(\hat{\cA},\hat{p})$ by $\hat{\Delta}$ and 
$\Delta_I=\Delta(\cA,p)$ by ${\Delta}$,
$Q^+(\hat{\cA},\hat{p})$ by $\hat{Q^+}$ and $Q^+_I=Q^+(\cA,p)$ 
by $Q^+$, and so on.

Denote the Weyl group of $\fg$ (resp., of $\fhg$) by $W$ (resp., by
$\hat{W}$). Introduce the twisted action $\hat{W}$ on $\fhh^*$ as
$w.\lambda:=w(\lambda+\hat{\rho})-\hat{\rho}$; notice that
$w.\lambda:=w(\lambda+\rho)-{\rho}$ if $w\in W$. Recall that~\cite{Kbook}
\begin{equation}\label{rhoha}
(\delta|\hat{\rho})=h^{\vee}_B=(\rho|\theta)+\frac{1}{2}(\theta|\theta).
\end{equation}

Recall that $\hat{J}=\{0\}\cup J$.
We apply~\Thm{thmsh} to $\fhg$ (see~\Rem{rem11}) and $I=J\subset
\hat{J}$.

\subsection{}
Introduce  $\Lambda_0\in\fhh^*$ such that 
$\Lambda_0(h)=0$ for $h\in\fh$, 
$\Lambda_0(K)=1$.
Any $\lambda\in\fhh^{\perp}_I$ takes the form
$\lambda=k\Lambda_0$ for some $k\in\mathbb{C}$. 
Thus, $\det S_{\nu}$ is a polynomial in one variable $k$.

\subsection{}\label{vac1}
The generalized Verma module $M_I(k\Lambda_0)$ is the vacuum module $V^k$.
Using~\Thm{thmsh} for $\hat{\fg}$, we obtain the formula
for the {\em vacuum determinant}:
\begin{equation}\label{cordet}
\det S_{\nu}(k)=\prod_{r=1}^{\infty}
\prod_{\gamma\in \hat{\Delta}^+\setminus \Delta}
\prod_{\alpha\in Q^+}\phi_{r\gamma+\alpha}(k)^{(\dim\fhg_{\gamma})
\hat{K}(\nu-\alpha-r\gamma)
(-1)^{(r+1)p(\gamma)} k_J(\alpha)},
\end{equation}
where
$$\sum_{\alpha\in Q^+} k_J(\alpha)e^{-\alpha}=
\prod_{\alpha\in\Delta^+_0}(1-e^{-\alpha})
\prod_{\alpha\in\Delta^+_1}(1+e^{-\alpha})^{-1}.$$

Recall that $\dim\fhg_{\gamma}=\dim\fh$ if $\gamma\in\mathbb{Z}\delta$ 
(unless $\fg=\fgl(m,m)$, when $\dim\fhg_{s\delta}=2m-1$, see~\Exa{exagl}),
and $\dim\fhg_{\gamma}=1$ if $\gamma\not\in\mathbb{Z}\delta$.

Write $\gamma\in\hat{\Delta}^+\setminus\Delta$
as $\gamma=u\delta+\gamma'$, where $u\in\mathbb{Z}_{\geq 1}$
and $\gamma'\in\Delta$. Then, by~(\ref{rhoha}):
\begin{equation}\label{phira}
\frac{1}{ru}\phi_{r\gamma+\alpha}(k)=k+h^{\vee}_B+
\frac{(\rho-\alpha|\gamma')-r(\gamma'|\gamma')/2}{u}+
\frac{(\alpha|\rho)-(\alpha|\alpha)/2}{ru}.
\end{equation}

\subsubsection{}
\begin{rem}{remdet}
It is easy to see that $\phi_{\xi}=
\phi_{w(\xi-\rho)+\rho}$ for any $w\in W$ since
$(\lambda|w\xi)=(\lambda|\xi)$ if $\lambda\in\fhh^{\perp}_I$,
for any $w\in W$.
\end{rem}

\subsubsection{}
Apart from the case $D(2,1,a)$ with irrational $a$, 
a finite-dimensional contragredient
Lie superalgebra $\fg$ admits
a symmetrizable Cartan matrix with integer entries. As a consequence, 
we can (and will) normalize
the bilinear form $(.|.)$ in such a way that 
the scalar product of any two roots is rational. 
Unless otherwise stated, we will assume that $\fg\not=D(2,1,a)$
 with irrational $a$.

\begin{cor}{rat}
If $\fg$ is not of the type $D(2,1,a)$ with irrational $a$
then the vacuum module
$V^k$ is simple for $k\not\in\mathbb{Q}$.
\end{cor}

\subsubsection{}\label{defM}
Write $\det S_{\nu}(k)=\prod_{b\in\mathbb{C}} 
(k+h^{\vee}_B-b)^{m_{b}(\nu)}$
and set
$$M_{b}:=R \sum_{\nu} m_{b}(\nu) e^{-\nu}.$$
Then
$$M_b=\sum_{(r;\gamma;\alpha)\in Y(b)} 
(-1)^{(r+1)p(\gamma)}(\dim\fhg_{\gamma})
k_I(\alpha) e^{-\alpha-r\gamma},$$
where $Y(b)$ is the set of triples $(r;\gamma;\alpha)$ with
$\phi_{r\gamma+\alpha}$ proportional to $(k+h^{\vee}_B-b)$, that is
$$Y(b):=\{(r;\gamma;\alpha)|\ r\in\mathbb{Z}_{\geq 1},\ 
\gamma\in \hat{\Delta}^+\setminus \Delta,\ \alpha\in Q^+\ \text{ such that }
\frac{\phi_{r\gamma+\alpha}}{k+h^{\vee}_B-b}\in\mathbb{C}^*\}.$$

We know that $M_0\not=0$.
By~\Cor{rat}, for $\fg\not=D(2,1,a)$ with irrational $a$, one has
$M_b=0$ if $b\not\in\mathbb{Q}$. In this case,
we present a non-zero rational number $b$ 
in the form $b=p/q$, where $p,q$ are relatively prime non-zero integers
and $q\geq 1$.

\subsection{}\label{kappa2}
Consider the restriction of the bilinear form $(.|.)$ to
the real vector space 
$\fh_{\mathbb{R}}:=\sum_{\alpha\in\Delta}\mathbb{R}\alpha$;
the dimension of a maximal isotropic subspace of $\fh_{\mathbb{R}}$
is called the {\em defect} of $\fg$.

A simple finite-dimensional
contragredient superalgebra of defect zero is either a Lie algebra
or $\osp(1,2n)$.

\subsubsection{}\label{Wsm}
Let $\kappa(.|.)$ denote the Killing form. If $\kappa$ is non-zero, set
$\Delta^{\#}=\{\alpha\in\Delta|\ \kappa(\alpha|\alpha)>0\}$.
Then $\Delta^{\#}_0$ is the root system of one of 
simple components of $\fg_0$.  If $\kappa=0$ 
then $\fg$ is of type $A(n,n),\ D(n+1,n)$ or $D(2,1,a)$.
In this case the root system is a union of two mutually orthogonal
subsystems: 
$\Delta_0=A_n\cup A_n,\ D_{n+1}\cup C_n,
D_2\cup C_1$ respectively; we let $\Delta^{\#}=\Delta^{\#}_0$ 
be the first subset. Let $W^{\#}$ be the Weyl group
corresponding to $\Delta^{\#}_0$, that is the subgroup
of $W$ generated by the $r_{\alpha}$ with $\alpha\in\Delta^{\#}_0$.

\subsubsection{}\label{setS}
A subset $S$ of $\Delta$ is called {\em maximal isotropic} if it consists
of the defect $\fg$ roots that span a maximal isotropic subspace of 
$\fh_{\mathbb{R}}$. The existence of $S$ is proven in~\cite{KWn};
it is also shown that one can choose a set of simple roots $\Pi$ 
is such a way that $S\subset \Pi$. We fix $S$ and $\Pi$ which contains
$S$. We set 
$$\mathbb{N}S:=\{
\sum_{\beta\in S} n_{\beta}\beta,\ n_{\beta}\in\mathbb{Z}_{\geq 0}\}.$$
For $\alpha\in\mathbb{N}S$ denote by $\htt\alpha$ the {\em height} of 
$\alpha$: $\htt\alpha=\sum n_{\beta}$ if 
$\alpha=\sum_{\beta\in S} n_{\beta}\beta$.

\subsubsection{}
By a {\em regular exponential function} on $\fhh$ we mean a finite linear
combination of exponentials $e^{\lambda}:\ \lambda\in\fhh^*$. A {\em rational
exponential function} is a ratio $P/Q$, where $P,Q$ are 
regular exponential functions and $Q\not=0$. The Weyl group $\hat{W}$ acts
on the field of rational exponential functions by the formulas
$w(e^{\lambda})=e^{w\lambda},\ w.(e^{\lambda})=e^{w.\lambda}$.

\subsection{}
Retain notation of~\ref{defM}.

\begin{thm}{detdef}
Assume that $W^{\#}S\subset\Delta^+$. Then
$$\begin{array}{l}
\det S_{\nu}(k)
=\displaystyle\prod_{r\geq 1}
\displaystyle\prod_{\gamma\in\hat{\Delta}^+\setminus\Delta}
\displaystyle\prod_{\alpha\in \mathbb{N}S}
\phi_{r\gamma+\alpha}(k)^{(\dim\fhg_{\gamma}) d_{r,\gamma,\alpha}(\nu)},\\
\text{ where }\ 
\phi_{r\gamma+\alpha}(k)=(\Lambda_0|\gamma)k+
(\hat{\rho}-\alpha|\gamma)-r(\gamma|\gamma)/2,\\
\text{ and } d_{r,\gamma,\alpha}(\nu) \text{ are integers, defined by }\\
\displaystyle\sum_{\nu\in \hat{Q}} d_{r,\gamma,\alpha}(\nu)e^{-\nu}=
(-1)^{\htt\alpha+(r-1)p(\gamma)}R^{-1}
\displaystyle\sum_{w\in W^{\#}}(-1)^{l(w)}e^{w.(-r\gamma-\alpha)},\
\text{ i.e.}\\

M_b=\displaystyle\sum_{(r;\gamma;\alpha)\in Y_S(b)} 
(-1)^{(r+1)p(\gamma)+\htt\alpha}(\dim\fhg_{\gamma})
\displaystyle\sum_{w\in W^{\#}}(-1)^{l(w)}e^{w.(-r\gamma-\alpha)},\\
\text{ where }\ 
Y_S(b):=\{(r;\gamma;\alpha)|\ r\in\mathbb{Z}_{\geq 1},\ 
\gamma\in \hat{\Delta}^+\setminus \Delta,\ \alpha\in \mathbb{N}S
\ \text{ such that }
\frac{\phi_{r\gamma+\alpha}}{k+h^{\vee}_B-b}\in\mathbb{C}^*\}.
\end{array}$$
Note that $r\gamma+\alpha$ uniquely determines a triple
$(r;\gamma;\alpha)$, and that $\dim\fhg_{\gamma}=1$, apart from the case when
$\phi_{r\gamma+\alpha}(k)$ is proportional to $k+h^{\vee}_B$.
\end{thm}
\begin{proof}
Since $S$ consists of simple mutually orthogonal isotropic roots 
one has $(\alpha|\alpha)=(\alpha|\hat{\rho})=0$ for any $\alpha\in\mathbb{N}S$.
This gives the formula for $\phi_{r\gamma+\alpha}$.

For $\alpha\in\mathbb{N}S$ and  $\xi:=-w(-\alpha)$ one has
$\phi_{r\gamma+\xi}=\phi_{r\gamma+\rho-w\rho+w\alpha}=
\phi_{rw^{-1}\gamma+l\beta}$ by~\Rem{remdet}. Combining 
 the formulas in~\ref{vac1} and~\Lem{lemdef}
we obtain for each $r$:
$$\begin{array}{rl}
\displaystyle\prod_{\gamma\in\hat{\Delta}^+\setminus\Delta}
\prod_{\alpha\in Q^+}
\phi_{r\gamma+\alpha}^{\hat{K}(\nu-\alpha-r\gamma)k_I(\alpha)}&=
\displaystyle\prod_{\gamma\in\hat{\Delta}^+\setminus\Delta}
\displaystyle\prod_{w\in W^{\#}}\displaystyle\prod_{\alpha\in\mathbb{N}S}
\phi_{rw^{-1}\gamma+\alpha}^{(-1)^{l(w)+\htt \alpha}
\hat{K}(\nu-r\gamma+w\rho-\rho-w\alpha)}\\
&=\displaystyle\prod_{\gamma\in\hat{\Delta}^+\setminus\Delta}
\displaystyle\prod_{w\in W^{\#}}\displaystyle\prod_{\alpha\in\mathbb{N}S}
\phi_{r\gamma+\alpha}^{(-1)^{l(w)+\htt\alpha}
\hat{K}(\nu-rw\gamma+w\rho-\rho-w\alpha)}\\
&=
\displaystyle\prod_{\gamma\in\hat{\Delta}^+\setminus\Delta}
\displaystyle\prod_{\alpha\in\mathbb{N}S}
\displaystyle\prod_{w\in W^{\#}}
\phi_{r\gamma+\alpha}^{(-1)^{l(w)+\htt\alpha}
\hat{K}(\nu+w.(-r\gamma-\alpha))}.
\end{array}$$
Now, the formula for the integers
$d_{r,\gamma,\alpha}(\nu)$ follows from~\ref{vac1}.
\end{proof}

\subsection{}
\begin{lem}{lemdef}
Assume that $W^{\#}S\subset\Delta^+$. Then
for any $\alpha\in Q$ the orbit $W^{\#}.(-\alpha)$ meets
$-\mathbb{N}S$ at most once and 
$$k_I(\alpha)=\left\{\begin{array}{ll}
0, \ \text{ if } W.(-\alpha)\cap (-\mathbb{N}S)=\emptyset,\\
(-1)^{l(w)+\htt (-w.(-\alpha))},\ \text{ if } w.(-\alpha)\in\mathbb{N}S.
\end{array}\right.$$
\end{lem}
\begin{proof}
First, let us show that $\mathbb{N}S\cap \Delta^+=S$. 
Indeed, if the defect of $\fg$ is not greater than one 
the assertion is trivial. The root systems 
of finite-dimensional contragredient Lie superalgebras are described
in~\cite{Kadv}, 2.5.4.
All exceptional superalgebras have defect one. For non-exceptional
superalgebras of non-zero defect, $\fh_{\mathbb{R}}$ has an orthogonal basis 
$\{\vareps_i|\delta_j\}_{i=1,n;j=1,m}$, where $(\vareps_i|\vareps_i)=
-(\delta_j|\delta_j)$ for any $i,j$ 
and $\Delta\subset
\{\pm t\vareps_i,\pm t\delta_j: t=1,2;\ \pm\vareps_i\pm\delta_j;
\pm\vareps_i\pm\vareps_{i'}; \pm\delta_j\pm\delta_{j'}\}$.
As a result,
$S$ is of the form $\{\pm\vareps_{i_l}\pm\delta_{j_l}\}$, 
where $i_l\not=i_{l'}, j_l\not=j_{l'}$ for  $l\not=l'$.
This implies $\mathbb{N}S\cap \Delta^+=S$.

Recall that all root spaces of $\fg$ are one-dimensional and thus
$$R:=\displaystyle\prod_{\alpha\in\Delta^+_0}
\frac{(1-e^{-\alpha})}{(1+e^{-\alpha})}=\sum_{\alpha\in Q^+} 
k_I(\alpha)e^{-\alpha}.$$
Since $\mathbb{N}S\cap \Delta^+=S$ we have
$k_I(\alpha)=(-1)^{\htt\alpha}$ for $\alpha\in \mathbb{N}S$.
Thm. 2.1 of~\cite{KWn} states that
$$
e^{\rho}R=\sum_{w\in W^{\#}} (-1)^{l(w)} w\bigl(
\frac{e^{\rho}}{\prod_{\beta\in S} (1+e^{-\beta})}
\bigr).
$$
The assumption  $W^{\#}S\subset\Delta^+$ forces 
$W^{\#}(\mathbb{N}S)\subset Q^+$ and the above formula gives
\begin{equation}\label{Thm21}
R=\sum_{w\in W^{\#}} (-1)^{l(w)}
\sum_{\alpha\in\mathbb{N}S} (-1)^{\htt \alpha}e^{w.(-\alpha)}.
\end{equation}
We see that
$k_I(\alpha)=0$ if $W^{\#}.(-\alpha)$ does not meet $\mathbb{N}S$,
and, moreover,
$w(e^{\rho}R)=(-1)^{l(w)}e^{\rho}R$, that is
$$k_I(\alpha)=(-1)^{l(w)}k_I(-w.(-\alpha))\ \text{ for any }
\alpha\in Q,w\in W^{\#}.$$

We already know that $k_I(\alpha)=(-1)^{\htt\alpha}$ for 
$\alpha\in \mathbb{N}S$.
It remains to verify that for any $\xi\in Q$ the orbit $W.(-\xi)$ meets
$-\mathbb{N}S$ at most once or, equivalently, that
for any $\alpha\in\mathbb{N}S$ one has
$W^{\#}.(-\alpha)\cap (-\mathbb{N}S)=\{-\alpha\}$. Indeed, from~(\ref{Thm21})
for any $\alpha\in\mathbb{N}S$ one has 
$$k_I(\alpha)=\sum_{w\in W^{\#}: -w.(-\alpha)\in
\mathbb{N}S}(-1)^{l(w)+\htt(-w.(-\alpha))}.$$
However if $-w.(-\alpha)\in\mathbb{N}S$ then 
$k_I(-w.(-\alpha))=(-1)^{\htt(-w.(-\alpha))}$ and 
$k_I(-w.(-\alpha))=(-1)^{l(w)}k_I(\alpha)$ so
$k_I(\alpha)=(-1)^{l(w)+\htt(-w.(-\alpha))}$.
Hence $k_I(\alpha)=\sum_{w\in W^{\#}: -w.(-\alpha)\in
\mathbb{N}S}k_I(\alpha)$ so $W^{\#}.(-\alpha)\cap (-\mathbb{N}S)=\{-\alpha\}$
as required.
\end{proof}

\section{Virasoro algebra}
In this section we prove formula~(\ref{04}) and~\Thm{thm05} (see
~\Thm{thmvacvir} and~\Prop{propmono} respectively).

\subsection{Notation}
Denote by $\Vir_{\geq k}$ (resp., $\Vir_{<k}$) the subspace spanned by
$L_j, j\geq k$ (resp., $L_j, j<k$). 
Notice that $\Vir_{\geq -1}, \Vir_{<-1}$ are subalgebras. 
A {\em Verma module} $M(h;c)$ ($h,c\in\mathbb{C}$) over $\Vir$
is induced from the one-dimensional
module $\mathbb{C}|h;c\rangle$
of $\Vir_{\geq 0}+\mathbb{C}C$, where $\Vir_{> 0}$ acts trivially,
$L_{0}$ acts by the scalar $h$ and $C$ acts by the scalar $c$.
The weight spaces of $M(h;c)$ are eigenspaces of $L_0$ with eigenvalues
$h+n, n\in\mathbb{Z}_{\geq 0}$.

A {\em vacuum module} $V^c$ is induced from the one-dimensional
module $\mathbb{C}|0;c\rangle$
of $\Vir_{\geq -1}+\mathbb{C}C$, where $\Vir_{\geq -1}$ acts trivially
and $C$ acts by the scalar $c$. Clearly, $V^c=M(0;c)/M(1;c)$.

In this section we use letters $r,s,p,q,k,m$ for non-negative integers.
For positive integers $p,q$ we denote, as before, by 
$(p,q)$ their greatest common
divisor. We denote the maximal proper submodule of $M(h;c)$ by 
$\ol{M}(h;c)$ and the simple quotient of  $M(h;c)$ by $L(h;c)$.

\subsection{Main result}
Introduce the anti-involution $\sigma$ on $\Vir$ by 
the formulas $\sigma(L_n)=L_{-n},\ \sigma(C)=C$. 
Define the triangular
decomposition $\Vir=\Vir_{<0}\oplus(\mathbb{C}L_0+\mathbb{C}C)
\oplus\Vir_{\geq 1}$, and introduce
the Harish-Chandra
projection with respect to this triangular
decomposition. Define the contravariant forms on Verma modules
and on vacuum modules as in~\ref{HCpr}. Define the Jantzen filtrations
on these modules as in~\ref{jancon} and observe
that the ``sum formula''~(\ref{sumfor})
 holds in this setup. We denote the determinant of the  contravariant
form on the eigenspace of $L_0$ with the eigenvalue
$h+N$ ($N\in\mathbb{Z}_{\geq 0}$)
in $M(h;c)$ by ${\det}_{h+N}$ (resp., on the eigenspace $V^c_N$
of $L_0$ in $V^c$ with the eigenvalue $N$
by  ${\det}'_N$). These are polynomials in $h$ and $c$ (resp., $c$).

\subsubsection{}
\begin{thm}{thmvacvir}
Let $c_{p,q}=1-\frac{6(p-q)^2}{pq}$.
\begin{enumerate}
\item
Up to a non-zero scalar factor, the vacuum determinant is as follows:
$${\det}'_N(c)=\prod_{p>q>1, (p,q)=1}
\bigl(c-c_{p,q})\bigr)^{\dim
L((p-1)(q-1);c_{p,q})_{N}},$$
where $\dim L((p-1)(q-1);c_{p,q})_{N}$ is given by the right-hand side 
of~(\ref{05}).

\item
A vacuum module $V^c$ is simple iff 
$c\not\in\{c_{p,q}\}_{p,q\in\mathbb{Z}_{\geq 2}, (p,q)=1}$.
\item
If  $V^c$ is not simple, then $\cF^1(V^c)=L((p-1)(q-1);c_{p,q}), \cF^2(V^c)=0$,
where $c=c_{p,q},\ p,q\in\mathbb{Z}_{\geq 2}, (p,q)=1$.

\item The vertex algebra $\Vir_c$ satisfies Zhu's $C_2$ condition 
iff the vacuum module $V^c$ is not simple.

\end{enumerate}
\end{thm}

We prove (iv) in~\ref{singvir} and (i) in~\ref{vacvir} below;
(ii), (iii) follow from (i) and Jantzen sum formula~(\ref{sumfor}).

\subsection{Singular vectors in $V^c$}\label{singvir}
Since $L_{-1}\vac=0$ in $V^c$, it is clear that
$C_2$ holds  iff the vectors $L_{-2}^k\vac$ ($k\geq 1$)
are linearly dependent over $C_2(\Vir_c):=\spn\{L_{-k}v|\ k>2, v\in\Vir_c\}$.
If $V^c$ is simple then $\Vir^c=\Vir_c$
and the vectors $L_{-2}^k\vac$ are linearly independent over $C_2(\Vir_c)$ and
thus $\Vir_c$ does not satisfy $C_2$ condition.
Take $c$ such that $V^c$ is not simple. In order
to check the $C_2$ condition, it is enough to verify
that a singular vector in $V^c$ is of the form
$(L_{-2}^k+a)\vac$, where $a\in\cU(\Vir_{<-1})$ lies
in the right ideal generated by $L_{-i}, i>2$. This will be shown
in~\Prop{propmono}.

\subsubsection{A total ordering on monomials}\label{monor}
For $v\in V^c$ call {\em the monomials of 
$v$} all the ordered monomials appearing in $u$, where $u\in\cU(\Vir_{<-1})$
is such that $v= u\vac$.

Introduce the following (lexicographic) total order (cf.~\cite{Kt})
on the ordered monomials of $\cU(\Vir_{<-1})$
with given $\ad L_0$-eigenvalue:
for $L_{-i_s}\ldots L_{-i_1}$ and $L_{-j_r}\ldots L_{-j_1}$
with $i_s\geq \ldots \geq i_1\geq 2,\ j_r\geq  \ldots \geq j_1\geq 2$
and $\sum i_m=\sum j_n=N$, put
$
L_{-i_s}\ldots L_{-i_1}<L_{-j_r}\ldots L_{-j_1}\ \text{ if either }
i_1<j_1, \text{ or } i_1=j_1,\ i_2=j_2\ldots, i_m=j_m, i_{m+1}<j_{m+1}.
$
For example, $L_{-4}^2>L_{-5}L_{-3}>L_{-3}L_{-3}L_{-2}>L_{-2}^4$.

\subsubsection{}
\begin{prop}{propmono}
The minimal monomial of a singular vector $v$ of $V^c$, 
not proportional to $\vac$, is $L_{-2}^m$, where m is a positive integer.
\end{prop}
\begin{proof}
Observe that for $u\in\Vir_{<-1}$ 
and $k\geq -1$, one has $L_ku\vac=[L_k,u]\vac$.
In particular, if $v=u\vac$ then
the monomials of $L_kv$ for $k\geq -1$ are the monomials
of $[L_k,u]$, which lie in $\cU(\Vir_{<-1})$.

The minimal monomial
of $[L_1,L_{-i_s}\ldots L_{-i_1}]$ is 
$L_{-i_s}\ldots L_{-i_2}L_{-i_1+1}$. In particular,
if $X,Y$ are monomials in $\Vir_{<-1}$ and $X<Y$ then
the minimal monomial of $[L_1,X]$ is less
than the minimal monomial of $[L_1,Y]$. If $i_1>2$ then
$L_{-i_s}\ldots L_{-i_2}L_{-i_1+1}$ belongs to $\Vir_{<-1}$.
As a consequence, the minimal monomial of a singular vector $v$
is of the form $L_{-i_s}\ldots L_{-i_1}$, where $i_1=2$.
Indeed, suppose that  $i_1>2$;
then the minimal monomial of $L_1v$ is 
$L_{-i_s}\ldots L_{-i_2}L_{-i_1+1}$, which belongs to $\Vir_{<-1}$,
and thus $L_1v\not=0$, so $v$ is not singular.

Now it remains to show that the minimal monomial of $v$
is not of the form $X=X'L_{-r}^sL_{-2}^m$  for some $r>2, s>0$.

Let $X=X'L_{-r}^sL_{-2}^m$ be a monomial ($r>2$ and $X'$ does not contain
$L_{-2}$ and $L_{-r}$). Then the minimal monomial of $L_{r-2}X\vac$ 
is $X'L_{-r}^{s-1}L_{-2}^{m+1}$. Suppose that
$X$ is the minimal monomial of a singular vector $v$ (we have
shown that $m\geq 1$). Since $L_{r-2}v=0$, the monomial 
$$Z:=X'L_{-r}^{s-1}L_{-2}^{m+1}$$
should appear as a monomial in $L_{r-2}Y\vac$ for some $Y>X$.
Write $Y=Y''L_{-2}^k$, where $Y''$ does not contain $L_{-2}$.
Recall that
$$Y>X\ \Longleftrightarrow\ \ \left[ \begin{array}{l}
k<m,\\
k=m\text{ and } Y''>X'L_{-r}^s.  
\end{array}\right.$$

The degree of $L_{-2}$ in any
monomial of $[L_{r-2},Y]$ is at most $k+1$.
Hence $L_{r-2}Y\vac$ does not contain $Z$ if $k<m$.

In the remaining case $Y=Y''L_{-2}^m$ for $Y'>X'L_{-r}^s$, write
$Y=Y'L_{-t}^pL_{-2}^m$, where $Y'$ does not contain $L_{-t}$. Then $t>2$ and
the condition $Y>X$ forces that either $t>r$ or $t=r,\ Y'L_{-r}^p>X'L_{-r}^s$.

If $Y=Y'L_{-t}^pL_{-2}^m$ for some $t>r$ then
the degree of $L_{-2}$ in any
monomial of $[L_{r-2},Y]$ is at most $m$
so $L_{r-2}Y\vac$ does not contain the monomial $Z$.

If $Y=Y'L_{-r}^pL_{-2}^m$, then the only monomial of
$[L_{r-2},Y]$, having a factor $L_{-2}^{m+1}$,
is $Y'L_{-r}^{p-1}L_{-2}^{m+1}$. Since $Y'L_{-r}^p>X'L_{-r}^s$,
one has $Y'L_{-r}^{p-1}>X'L_{-r}^{s-1}$ and so
$Y'L_{-r}^{p-1}L_{-2}^{m+1}>Z$. Hence $L_{r-2}Y\vac$ does not contain $Z$,
a contradiction. The assertion follows.
\end{proof}

\subsection{Proof of~\Thm{thmvacvir} (ii)}\label{vacvir}
\subsubsection{Outline of the proof}\label{outvir}
In~\Lem{cnot1} we will show that $V^c$ has a subquotient, isomorphic
to $L((p-1)(q-1);c)$ if $c=c_{p,q}$, where
$p>q\geq 2, (p,q)=1$. Using the sum formula~(\ref{sumfor})
we conclude that  ${\det}'_N(c)$ is divisible by the polynomial
$$P_N(c):=\prod_{p>q\geq 2, (p,q)=1}\bigl(c-c_{p,q})\bigr)^{\dim
L((p-1)(q-1);\, c_{p,q})_{N}}.$$
In~\ref{vir1}--\ref{apq} we will show that the degree of 
${\det}'_N(c)$ coincides with the degree of $P_N(c)$ so ${\det}'_N(c)=a P_N(c)$
for $a\in\mathbb{C}^*$.
This proves~\Thm{thmvacvir} (ii).

\subsubsection{}
\begin{lem}{cnot1}
Let $c=c_{p,q}$, where
$p>q\geq 2$ are relatively prime integers.
Then $V^c$ has a subquotient isomorphic
to $L((p-1)(q-1);c)$.
\end{lem}
\begin{proof}
Recall that $V^c=M(0;c)/M(1;c)$. We will  show that 
$M(0;c)$ has a singular vector of 
weight $(p-1)(q-1)$, whereas $M(1;c)$ does not have such a vector.

Recall (\cite{K3};\ \cite{KR}, 8.1--8.4) 
that the determinant of the contravariant (=Shapovalov) form
for a Verma module over the Virasoro algebra is, up
to a non-zero constant factor:
$${\det}_{N+h}(c,h)=\prod_{r,s\in\mathbb{Z}_{\geq 1}} 
(h-h_{r,s}(c))^{p_{cl}(N-rs)},$$
where $p_{cl}(m)$ is the classical partition function:
$$
\prod_{k=1}^{\infty}(1-x^k)^{-1}=
\sum_{m\in\mathbb{Z}} p_{cl}(m)x^m,
$$
and the functions $h_{r,s}(c)$ can be described as follows:
$$h_{r,r}(c)=\frac{(r^2-1)(c-1)}{24}, \ \ 
\ h_{r,s}(c)=\frac{(p'r-q's)^2-(p'-q')^2}{4p'q'},$$
where  $p',q'\in\mathbb{C}$ are such that 
$c=1-\frac{6(p'-q')^2}{p'q'}$.

Let $c=c_{p,q}$, where
$p>q\geq 2$ are relatively prime integers.
One has $h_{r,r}(c)\not=0$ for $r\geq 2$ and
$h_{r,s}(c)=0$ iff $pr-qs=\pm(p-q)$; $h_{r,r}(c)\not=1$ for $r\geq 1$
and $h_{r,s}(c)=1$ iff $pr-qs=\pm(p+q)$.
As a result, $h_{p-1,q-1}(p,q)=0$ and $h_{r,s}(p,q)\not=1$
if $rs<(p+1)(q-1)$. Hence $M(0;c)$ has a singular vector of 
weight $(p-1)(q-1)$, whereas
the minimal weight of a singular vector in $M(1;c)$ is
$1+(p+1)(q-1)=pq-p+q$. The claim follows. 
\end{proof}

\subsubsection{}\label{vir1}
The leading term of ${\det}'_N(c)$ is $c^{d(N)}$, where
$$d(n)=
\displaystyle\sum_{\lambda\vdash n, 1\not\in\lambda} l(\lambda).$$
Here $\lambda\vdash n$ stands for a partition of $n$ (we will
write $|\lambda|=n$),
$l(\lambda)$ is the number of parts of $\lambda$,
 and $1\in\lambda$ means that
$\lambda$ contains a part equal to $1$. One has
$$
\displaystyle\sum_{\lambda:\ 1\not\in\lambda} 
t^{l(\lambda)}x^{|\lambda|}=\prod_{m=2}^{\infty}(1-tx^m)^{-1}$$
and this allows to express the generating function $\sum_{n} d(n)x^n$
as follows:
$$\sum_{n} d(n)x^n=
\displaystyle\sum_{1\not\in\lambda} l(\lambda)x^{|\lambda|}
=\frac{\partial \prod_{m=2}^{\infty}(1-tx^m)^{-1}}
{\partial t}|_{t=1}.$$
Therefore
$$\sum_n d(n)x^n=\prod_{m=2}^{\infty}(1-x^m)^{-1}
\sum_{r=2}^{\infty} \frac{x^{r}}{1-x^{r}}=\prod_{m=2}^{\infty}(1-x^m)^{-1}
\displaystyle\sum_{\scriptstyle{r\geq 2,\ s\geq 1}}
x^{rs},$$
which can be rewritten as
$$\prod_{m=1}^{\infty}(1-x^m)\sum_n d(n)x^n=
\sum_{r\geq 2,s\geq 1} (x^{rs}-x^{rs+1}).$$

\subsubsection{}\label{vir2}
Take $c=c_{p,q}$, where
$p,q\in\mathbb{Z}$ are such that $p>q\geq 2, (p,q)=1$. One has
$$\prod_{k\geq 1}(1-x^k)\ch L((p-1)(q-1);c)=
\sum_{k\in\mathbb{Z}\setminus\{0\}} (x^{(1+kq)(1+kp)}-x^{(kq-1)(1+kp)+1}),$$
see~\cite{FF}, \cite{Ast}. In order to prove that 
the degree of 
${\det}'_N(c)$ coincides with the degree of $P_N(c)$ (see~\ref{outvir})
it remains to verify the following identity of formal power series in $x$:
\begin{equation}\label{vircon1}
\sum_{r\geq 2,s\geq 1} (x^{rs}-x^{rs+1})=\sum_{p>q\geq 2, (p,q)=1} 
\sum_{k\in\mathbb{Z}\setminus\{0\}} (x^{(1+kq)(1+kp)}-x^{(kq-1)(1+kp)+1}).
\end{equation}

\subsubsection{}\label{apq}
One has
$$\sum_{r\geq 2,s\geq 1}x^{rs}=\frac{1}{2}\sum_{k,l\geq 1} x^{(k+1)(l+1)}
+\frac{1}{2}\sum_{k,l\geq 2} x^{(k-1)(l-1)}-x/2.$$
Writing $k=jp, l=jq,$ where $j:=(k,l)$ and $(p,q)=1$, we obtain
$$\begin{array}{rl}
\displaystyle\sum_{k,l\geq 1} 
x^{(k+1)(l+1)}&=\displaystyle\sum_{j\geq 1}\sum_{p,q\geq 1,\
 (p,q)=1}
x^{(jp+1)(jq+1)}\\
&=\displaystyle\sum_{j\geq 1}\bigl(\sum_{ p>q\geq 2,\ (p,q)=1} 
2x^{(jp+1)(jq+1)}+
2\displaystyle\sum_{p\geq 2}x^{(j+1)(jp+1)}+x^{(j+1)^2}\bigr)
\end{array}$$
and similarly
$$\begin{array}{rl}
\displaystyle\sum_{k,l\geq 2} x^{(k-1)(l-1)}&=
\displaystyle\sum_{j\geq 2}\sum_{p,q\geq 1,\ (p,q)=1}
x^{(jp-1)(jq-1)}+\displaystyle\sum_{p,q\geq 2,\ (p,q)=1}x^{(p-1)(q-1)}
\\
&=2\displaystyle\sum_{j\geq 1}\sum_{(p,q)=1, p>q\geq 2} x^{(jp-1)(jq-1)}
+\displaystyle\sum_{j\geq 2}\bigl(2\sum_{p\geq 2}
x^{(j-1)(jp-1)}+x^{(j-1)^2}\bigr)\\
&=
2\displaystyle\sum_{j\geq 1}\sum_{(p,q)=1, p>q\geq 2} x^{(jp-1)(jq-1)}
+\displaystyle\sum_{j\geq 2}\bigl(2\displaystyle\sum_{p\geq 2}x^{(j-1)(jp-1)}+
x^{j^2}\bigr)+x.
\end{array}
$$
Therefore
$$\begin{array}{rl}
\displaystyle\sum_{r\geq 2,s\geq 1}x^{rs}&=
\displaystyle\sum_{j\geq 1}\sum_{(p,q)=1, p>q\geq 2} 
\bigl(x^{(jp+1)(jq+1)}+x^{(jp-1)(jq-1)}\bigr)\\
&\ \ \ \ \ \ \ \ +
\displaystyle\sum_{j\geq 2} x^{j^2}+
\displaystyle\sum_{j\geq 1}\sum_{p\geq 2}x^{(j+1)(jp+1)}
+\displaystyle\sum_{j\geq 2}\sum_{p\geq 2}x^{(j-1)(jp-1)}\\
&=
\displaystyle\sum_{j\in\mathbb{Z}\setminus\{0\}}\sum_{(p,q)=1, p>q\geq 2} 
x^{(jp+1)(jq+1)}+\displaystyle\sum_{j\in\mathbb{Z}\setminus\{0,-1\}}
\sum_{p\geq 2}x^{(j+1)(jp+1)}+\displaystyle\sum_{j\geq 2} x^{j^2}.
\end{array}$$
Moreover,
$$\begin{array}{rl}
\displaystyle\sum_{r\geq 2,s\geq 1}x^{rs+1}&=\!\!\!\displaystyle
\sum_{k\geq 2,l\geq 1} x^{(k-1)(l+1)+1}=\!\!\!
\displaystyle\sum_{j\geq 2}\sum_{p,q\geq 1,\ (p,q)=1} x^{(jp-1)(jq+1)+1}
+\!\!\!\!\!\!\!\!\displaystyle\sum_{p\geq 2,\ q\geq 1,\ (p,q)=1}\!\!\!\!\!\!
\!\!\! x^{(p-1)(q+1)+1}\\
& =
\displaystyle\sum_{j\geq 1}\sum_{p>q\geq 2, (p,q)=1} \bigl(x^{(jp-1)(jq+1)+1}
+x^{(jp+1)(jq-1)+1}\bigr)+\displaystyle\sum_{j\geq 2} x^{j^2}\\
&\ \ \ \ \ \ \ \ +
\displaystyle\sum_{j\geq 1}\sum_{p\geq 2}x^{(jp-1)(j+1)+1}+
\displaystyle\sum_{j\geq 2}\sum_{p\geq 2}
x^{(jp+1)(j-1)+1}\\
&=
\displaystyle\sum_{j\in\mathbb{Z}\setminus\{0\}}\sum_{(p,q)=1, p>q\geq 2} 
x^{(jp+1)(jq-1)+1}+\displaystyle\sum_{j\in\mathbb{Z}\setminus\{0,-1\}}
\sum_{p\geq 2}x^{(j+1)(jp-1)+1}+\displaystyle\sum_{j\geq 2} x^{j^2}.
\end{array}$$
Now~(\ref{vircon1}) follows from the following identities:
$$\displaystyle\sum_{j\in\mathbb{Z}\setminus\{0,-1\}}x^{(jp-1)(j+1)+1}=
\displaystyle\sum_{j\in\mathbb{Z}\setminus\{0,-1\}}x^{(-j)(-(j+1)p+1)}=
\displaystyle\sum_{j\in\mathbb{Z}\setminus\{0,-1\}}x^{(i+1)(ip+1)}.\ \qed$$

\section{Neveu-Schwarz algebra}
The Neveu-Schwarz superalgebra $\NS$ is a Lie superalgebra with the basis 
$\{C;L_i\}_{i\in \frac{1}{2}\mathbb{Z}}$, such that
its even part is the Virasoro algebra
(with the basis $\{C;L_i\}_{i\in\mathbb{Z}}$), the element $C$ is central, and
apart from the relations~(\ref{03})  of the Virasoro algebra, 
the following commutation relations for $m\in\mathbb{Z}$ and 
$i,j\in \frac{1}{2}\mathbb{Z}\setminus\mathbb{Z}$ hold:
$$
[L_i,L_j]=2L_{i+j}+\delta_{0,i+j}\frac{4i^2-1}{12}C,\ \
[L_m,L_j]=(\frac{m}{2}-j)L_{j+m}.
$$

\subsection{Notation}
Denote by $\NS_{>k}$ the subspace spanned by
$L_j, j>k$. 
A {\em Verma module} $M(h;c)$ ($h,c\in\mathbb{C}$) over $\NS$
is induced from the one-dimensional
module $\mathbb{C}|h;c\rangle$
of $\NS_{\geq 0}+\mathbb{C}C$, where $\NS_{> 0}$ acts trivially,
$L_{0}$ acts by the scalar $h$ and $C$ acts by the scalar $c$.
The weight spaces of $M(h;c)$ are eigenspaces of $L_0$ with eigenvalues
$h+n, n\in \frac{1}{2}\mathbb{Z}_{\geq 0}$.

Notice that $\NS_{\geq -1}$ is a subalgebra. 
A {\em vacuum module} $V^c$ over $\NS$ is induced from the one-dimensional
module $\mathbb{C}|0;c\rangle$
of $\NS_{\geq -1}+\mathbb{C}C$, where $\NS_{\geq -1}$ acts trivially
and $C$ acts by the scalar $c$. Clearly, $V^c=M(0;c)/M(1/2;c)$. 
Recall that it carries a canonical structure of a vertex algebra,
denoted by $\NS^c$. Its unique simple quotient is denoted by $\NS_c$.

Let 
\begin{equation}\label{Y}
Y:=\{(p,q)\in\mathbb{Z}_{\geq 1}\times \mathbb{Z}_{\geq 1}| 
\ p\equiv q\mod 2\, , (\frac{p-q}{2},q)=1\}.
\end{equation}

Set 
$$\psi(x,t):=
\prod_{n=0}^{\infty} (1+tx^{n+1/2})^{-1}\prod_{n=1}^{\infty}(1-tx^n).$$
The function $\psi(x,1)$ is the super analogue of the Virasoro denominator
$\prod_{n=1}^{\infty}(1-x^n)$, namely
one has: $\psi(x,1)^{-1}=\sum_{N\in \frac{1}{2}\mathbb{Z}} \dim M(h;c)_{h+N}x^N$.



\subsection{Main result}
Introduce the anti-involution $\sigma$ on $\NS$ by 
the formulas $\sigma(L_n)=L_{-n},\ \sigma(C)=C$. 
Define the triangular
decomposition $\NS=\NS_{<0}\oplus(\mathbb{C}L_0+\mathbb{C}C)
\oplus\NS_{>0}$, and introduce
the Harish-Chandra
projection with respect this triangular
decomposition. Define the contravariant forms and the Jantzen filtarions
on Verma modules
and on vacuum modules as in~\ref{HCpr},~\ref{jancon}; observe
that the ``sum formula''~(\ref{sumfor})
 holds in this setup. We denote the determinant of the  contravariant
form on the eigenspace of $L_0$ with the eigenvalue
$h+N$ ($N\in1/2\mathbb{Z}_{\geq 0}$)
in $M(h;c)$ by ${\det}_{h+N}$ (resp., on the eigenspace $V^c_N$
of $L_0$ in $V^c$ with the eigenvalue $N$
by  ${\det}'_N$). These are polynomials in $h$ and $c$ (resp., $c$).

\subsubsection{}
\begin{thm}{thmvacns}
Let $c^S_{p,q}=\frac{3}{2}(1-\frac{2(p-q)^2}{pq})$, and recall notation 
~(\ref{Y}).

\begin{enumerate}
\item
Up to a non-zero scalar factor, the vacuum determinant for $\NS$ is as follows:
\begin{equation}\label{detvns}
{\det}'_N(c)=\displaystyle\prod_{p>q\geq 2, (p,q)\in Y}
\bigl(c-c^S_{p,q})\bigr)^{\dim L((p-1)(q-1)/2; c^S_{p,q})_{N}}.
\end{equation}

\item
A vacuum $\NS$-module $V^c$ is simple iff $c\not=c^S_{p,q}$, where 
$p>q\geq 2, (p,q)\in Y$. If $V^c$ is not simple then its unique proper 
submodule is $L((p-1)(q-1)/2;c^S_{p,q})$
and
$$\ch L((p-1)(q-1)/2;c^S_{p,q})=\psi(x,1)^{-1}
\sum_{k\in\mathbb{Z}\setminus\{0\}}(x^{(kp+1)(kq+1)/2}-
x^{((kp+1)(kq-1)+1)/2}).$$

\item
If  the $\NS$-module $V^c$ is not simple, i.e. $c=c^S_{p,q}$,
where $p>q\geq 2, (p,q)\in Y$,
 then 
$$\cF^1(V^c)=L((p-1)(q-1)/2;c^S_{p,q}),\ \cF^2(V^c)=0.$$

\item The vertex algebra $\NS_c$ satisfies Zhu's $C_2$ condition 
iff the vacuum module $V^c$ is not simple.
\end{enumerate}
\end{thm}

\subsection{Superpartitions}
Let us call $\lambda=(\lambda_1,\lambda_2,\ldots,\lambda_m)$
a {\em superpartition} of $N$ if $\sum_{i=1}^m\lambda_i=N$,
$\lambda_1\leq\lambda_2\leq\ldots\leq\lambda_m$,
$2\lambda_i\in\mathbb{Z}_{\geq 1}$
for any $i$, and $\lambda_i\not=\lambda_{i\pm 1}$ if
 $\lambda_i$ is not integer (i.e., any half-integer appears
at most once in the multiset $\{\lambda_i\}_{i=1}^m$).
Write $\lambda\vdash N$ if $\lambda$ is a superpartition of $N$;
set $|\lambda|=N$ and $l(\lambda)=m$ if 
$\lambda=(\lambda_1,\lambda_2,\ldots,\lambda_m)$.
For $b\in \frac{1}{2}\mathbb{Z}$ write $b\in\lambda$ if $\lambda_i=b$
for some $i$.

Note that $\psi(x,t)^{-1}$ is 
the generating function for superpartitions:
$\psi(x,t)^{-1}=\sum_{\lambda} t^{l(\lambda)}x^{|\lambda|}$.

\subsection{Proof of~\Thm{thmvacns} (i),(ii),(iii)} It is straightforward
to deduce (ii) from~\cite{Ast}, 8.2 (which relies on the Kac determinant 
formula for $\NS$ \cite{K3}). It follows that
if $c=c^S_{p,q}$, where $p>q\geq 2, (p,q)\in Y$, then 
$\cF^1(V^c)=L((p-1)(q-1)/2;c^S_{p,q})$. From the sum formula~(\ref{sumfor}),
it follows that
${\det}'_N(c)$ is divisible by the right-hand side of~(\ref{detvns})
and, moreover, that (i) implies (iii). We prove~(\ref{detvns})
by showing that the degree of ${\det}'_N(c)$ is equal to the degree
of the right-hand side of~(\ref{detvns}). Let $d(N)$
be the  degree of ${\det}'_N(c)$. In terms of generating functions,
we need to show that
$$\sum_{n\in \frac{1}{2}\mathbb{Z}} d(n)x^n=\sum_{p>q\geq 2, (p,q)\in Y} 
\dim L((p-1)(q-1)/2;c^S_{p,q})_{n}x^n,$$
which can be rewritten as
$$\psi(x,1)\sum_{n\in \frac{1}{2}\mathbb{Z}} d(n)x^n=\sum_{p>q\geq 2, 
(p,q)\in Y} 
\sum_{k\in\mathbb{Z}\setminus\{0\}}
(x^{(kp+1)(kq+1)/2}-x^{((kp+1)(kq-1)+1)/2}).$$

\subsubsection{}\label{ns1}
One has
$$d(n)=
\displaystyle\sum_{\lambda\vdash n, 1\not\in\lambda\, ,\frac{1}{2}
\not\in\lambda}l(\lambda).$$
Observe that
$$\psi(x,t)^{-1}\frac{(1-tx)}{1+tx^{1/2}}=
\displaystyle\sum_{\lambda\vdash n, 1\not\in\lambda\,,\frac{1}{2}
\not\in\lambda}t^{l(\lambda)}x^{|\lambda|},$$
and this allows to express the generating function
$\sum_{n\in \frac{1}{2}\mathbb{Z}} d(n)x^n$ as follows:
$$\sum_{n\in \frac{1}{2}\mathbb{Z}} d(n)x^n=
\displaystyle\sum_{\lambda:\ 1\not\in\lambda,\frac{1}{2}\not\in\lambda} 
l(\lambda)x^{|\lambda|}=\frac{\partial}{\partial t}
\frac{(1-tx)}{(1+tx^{1/2})\psi(x,t)}|_{t=1}.$$

Since $\frac{\partial}{\partial t} \frac{(1-tx)}{(1+tx^{1/2})\psi(x,t)}
=\frac{(1-tx)}{(1+tx^{1/2})\psi(x,t)}
\bigl(\sum_{n=1}^{\infty} \frac{x^{n+1/2}}{1+tx^{n+1/2}}+
\sum_{n=2}^{\infty} \frac{x^{n}}{1-tx^{n}}
 \bigr)$, we obtain
$$\psi(x,1)\sum_{n\in \frac{1}{2}\mathbb{Z}} d(n)x^n=(1-x^{1/2})
\bigl(\sum_{n=1}^{\infty} \frac{x^{n+1/2}}{1+x^{n+1/2}}+
\sum_{n=2}^{\infty} \frac{x^{n}}{1-x^{n}}
 \bigr)=(1-x^{1/2})
\displaystyle\sum_{\scriptstyle{r\geq 2,\ s\geq 1\atop r\equiv s\mod 2}}
x^{\frac{rs}{2}}.$$

Put $y:=x^{1/2}$.
Let $a$ be a rational number greater than $1$ which is not an odd integer.
Then $a$ can be uniquely written as
$a=p/q$, where $p>q\geq 2$ are integers of the same parity
and $q$ is the smallest one with this property i.e., $(\frac{p-q}{2},q)=1$.

\subsubsection{}\label{apqn}
One has
$$\displaystyle\sum_{\scriptstyle{r\geq 2,\ s\geq 1\atop r\equiv s\mod 2}}
y^{rs}
=\frac{1}{2}\sum_{\scriptstyle{k,l\geq 1\atop k\equiv l\mod 2}}
y^{(k+1)(l+1)}
+\frac{1}{2}\sum_{\scriptstyle{k,l\geq 2\atop k\equiv l\mod 2}}
y^{(k-1)(l-1)}-y/2.$$
Here and further, $r,s,k,l,p,q$ are integers.

For $k\equiv l\mod 2$ we can write
$k=jp, l=jq,$ where $j:=(\frac{k-l}{2},l)$ and $(p,q)\in Y$. We get
$$\begin{array}{rl}
\displaystyle\sum_{\scriptstyle{k,l\geq 1\atop k\equiv l\mod 2}}
y^{(k+1)(l+1)}&=\displaystyle\sum_{j\geq 1}\sum_{p,q\geq 1,\
 (p,q)\in Y}
y^{(jp+1)(jq+1)}\\
&=\displaystyle\sum_{j\geq 1}\bigl(\sum_{ p>q\geq 2,\ (p,q)\in Y} 
2y^{(jp+1)(jq+1)}+
2\displaystyle\sum_{p\geq 2, (p,1)\in Y}  y^{(j+1)(pj+1)}+y^{(j+1)^2}\bigr)
\end{array}$$
and similarly
$$\begin{array}{rl}
\displaystyle\sum_{\scriptstyle{k,l\geq 2\atop k\equiv l\mod 2}}
y^{(k-1)(l-1)}&=
\displaystyle\sum_{j\geq 2}\sum_{p,q\geq 1,\ (p,q)\in Y}
y^{(jp-1)(jq-1)}+\displaystyle\sum_{p,q\geq 2,\ (p,q)\in Y}
y^{(p-1)(q-1)}\\
&=2\displaystyle\sum_{j\geq 1}\sum_{p>q\geq 2,\ (p,q)\in Y} y^{(jp-1)(jq-1)}
+\displaystyle\sum_{j\geq 2}\bigl(2\displaystyle\sum_{p\geq 2, (p,1)\in Y} 
y^{(j-1)(pj-1)}+y^{(j-1)^2}\bigr)\\
&=
2\displaystyle\sum_{j\geq 1}\sum_{p>q\geq 2,\ (p,q)\in Y} y^{(jp-1)(jq-1)}
+\displaystyle\sum_{j\geq 2}\bigl(2\displaystyle\sum_{p\geq 2, (p,1)\in Y} 
y^{(j-1)(pj-1)}+y^{j^2}\bigr)+y.
\end{array}
$$

Therefore
$$\begin{array}{rl}
\displaystyle\sum_{\scriptstyle{r\geq 2,s\geq 1\atop r\equiv s\mod 2}}
y^{rs}&=\displaystyle\sum_{j\geq 1}\sum_{p>q\geq 2, (p,q)\in Y} 
\bigl(y^{(jp+1)(jq+1)}+y^{(jp-1)(jq-1)}\bigr)\\
&\ \ \ \ \ \ \ \ +
\displaystyle\sum_{j\geq 2} y^{j^2}+
\displaystyle\sum_{j\geq 1}\sum_{m=1}^{\infty}y^{(j+1)((2m+1)j+1)}
+\displaystyle\sum_{j\geq 2}\sum_{m=1}^{\infty}y^{(j-1)((2m+1)j-1)}\\
&=\displaystyle\sum_{j\in\mathbb{Z}\setminus\{0\}}
\sum_{p>q\geq 2, (p,q)\in Y} y^{(jp+1)(jq+1)}+
\displaystyle\sum_{j\in\mathbb{Z}\setminus\{0,-1\}}
\sum_{m=1}^{\infty}y^{(j+1)((2m+1)j+1)}\bigr)+
\displaystyle\sum_{j\geq 2} y^{j^2}.\end{array}$$

Moreover,
$$\begin{array}{rl}
\displaystyle\sum_{\scriptstyle{r\geq 2,s\geq 1\atop r\equiv s\mod 2}}
y^{rs+1}&=\!\!\!\!\displaystyle
\sum_{\scriptstyle{k\geq 2,l\geq 1\atop k\equiv l\mod 2}}
y^{(k-1)(l+1)+1}=\!\!\!\!
\displaystyle\sum_{j\geq 2}\sum_{p,q\geq 1,\ (p,q)\in Y} y^{(jp-1)(jq+1)+1}
+\!\!\!\!\!\!\!\!\!\!
\displaystyle\sum_{p\geq 2,\ q\geq 1,\ (p,q)\in Y} \!\!\!\!\!\!\!\!\!\!\!\!
y^{(p-1)(q+1)+1}\\
& =
\displaystyle\sum_{j\geq 1}\sum_{p>q\geq 2, (p,q)=1} \bigl(y^{(jp-1)(jq+1)+1}
+y^{(jp+1)(jq-1)+1}\bigr)+\displaystyle\sum_{j\geq 2} y^{j^2}\\
&\ \ \ \ \ \ \ \ +
\sum_{p\geq 2,\ (p,1)\in Y}\bigl(\displaystyle\sum_{j\geq 1}y^{(jp-1)(j+1)+1}+
\displaystyle\sum_{j\geq 2}y^{(jp+1)(j-1)+1}\bigr)\\
&=\!\!\!\!\displaystyle\sum_{j\in\mathbb{Z}\setminus\{0\}}
\sum_{p>q\geq 2, (p,q)\in Y}\!\!\!\! y^{(jp+1)(jq-1)+1}+\!\!\!\!
\displaystyle\sum_{j\in\mathbb{Z}\setminus\{0,-1\}}\sum_{p\geq 2,\ (p,1)\in Y}
\!\!\!\! y^{(jp+1)(j-1)+1}
+\displaystyle\sum_{j\geq 2} y^{j^2}.
\end{array}$$
One has
$$\displaystyle\sum_{j\in\mathbb{Z}\setminus\{0,-1\}}y^{(jp-1)(j+1)+1}=
\displaystyle\sum_{j\in\mathbb{Z}\setminus\{0,-1\}}y^{(-j)(-(j+1)p+1)}=
\displaystyle\sum_{j\in\mathbb{Z}\setminus\{0,-1\}}y^{(i+1)(ip+1)}.$$

Hence we obtain the required equality:
$$
\displaystyle\sum_{\scriptstyle{r\geq 2,s\geq 1\atop r\equiv s\mod 2}}
 (y^{rs}-y^{rs+1})=
\displaystyle\sum_{j\geq 1}\sum_{p>q\geq 2,\ (p,q)\in Y}
y^{(jp+1)(jq+1)}- y^{(jp-1)(jq+1)+1}. \qed
$$

\subsection{Proof of~\Thm{thmvacns} (iv)} Let 
$C_2(\NS_c):=
\spn\{L_{-k}v|\ k>2, v\in\NS_c\}$. Recall that the $C_2$ condition for 
$\NS_c$ means
that $C_2(\NS_c)$ has finite codimension in $V_c$.
Since $L_{-1}\vac=L_{-1/2}\vac=0$ in $V^c$, it is clear that
the $C_2$ condition holds  iff the vectors $L_{-2}^k\vac$ ($k\geq 1$)
are linearly dependent over $C_2(\NS_c)$.

If $V^c$ is simple then $\NS^c=\NS_c$
and the vectors $L_{-2}^k\vac$ are linearly independent over $C_2(\NS_c)$, and
thus $\NS_c$ does not satisfy the $C_2$ condition.

Take $c$ such that $V^c$ is not simple. Then $V^c$ has a unique proper
submodule $\ol{V}^c$. In order
to check the $C_2$ condition, it is enough to verify
that  $\ol{V}^c$ contains a vector of the form
$(L_{-2}^k+a)\vac$, 
where $a\in\cU(\NS_{<-1})$ lies
in the right ideal generated by $L_{-i}, i>2$.
Let $v$ be a singular vector of $\ol{V}^c$ (it is unique up to a scalar).
In~\Cor{corns} we will show that either $v$ or
$L_{-1/2}v$ is of the form $(L_{-2}^k+a)\vac$.
This will prove (iv).

\subsubsection{}
Consider the PBW basis of $\cU(\NS_{<-1})$ which consists
of the monomials of the form
$L_{-i_s}^{m_s}\ldots L_{-i_1}^{m_1}$, where 
$i_j\in \frac{1}{2}\mathbb{Z},\ 1<i_1<i_2\ldots<i_s,\ m_j\geq 1$ and
$m_j=1$ if $i_j\not\in \mathbb{Z}$.

Define the (lexicographic) total order 
on the PBW basis of 
$\cU(\NS_{<-1})$ 
with given $\ad L_0$-eigenvalue
in the same way as in~\ref{monor},
and retain conventions of~\ref{monor}.

\subsubsection{}
\begin{lem}{lemmonons}
Let $v \vac \in V^c$, where $v \in\cU(\NS_{<-1})$, be a singular vector,
not proportional to $\vac$. Then $v$ contains either a monomial 
$L_{-2}^k$ ($k>0$)
or a monomilal $L_{-2}^mL_{-3/2}$ ($m\geq 0$) with a non-zero coefficient.
\end{lem}
\begin{proof}
For $u\in \cU(\NS_{<-1})$ denote by $\supp u$ the set of monomials,
which $u$ contains. In this proof the letters $U,X,Y,Z$
stand for monomials in $\cU(\NS_{<-1})$.

Let $X,Y$ be monomials in $\cU(\NS_{<-1})$ and let $Y$ does not contain 
$L_{-3/2}$ and $L_{-2}$. Let 
$L_r\in\NS_{>0}$. Then one has
$$\begin{array}{lll}
(i) & YL_{-3/2}\in\supp [L_r,X] &\Longrightarrow\ \ \left[ \begin{array}{l}
X=X'L_{-3/2},\ Y\in\supp[L_r,X']\\
X=YL_{-(r+3/2)};\end{array}\right.\\
(ii)&  YL_{-2}^sL_{-3/2}\in\supp [L_r,X], s>0 &\Longrightarrow \ 
\ \left[ \begin{array}{l}
X=X'L_{-2}^sL_{-3/2},\ Y\in\supp[L_r,X']\\
X=YL_{-(r+2)}L_{-2}^{s-1}L_{-3/2}\\
X=YL_{-(r+3/2)}L_{-2}^s;\end{array}\right.\\
(iii) & YL_{-2}^s\in\supp [L_r,X], s>0 &\Longrightarrow \ 
\ \left[ \begin{array}{l}
X=X'L_{-2}^s,\ Y\in\supp[L_r,X']\\
X=YL_{-(r+2)}L_{-2}^{s-1}\\
X=YL_{-(r+1/2)}L_{-2}^{s-1}L_{-3/2}.\end{array}\right.
\end{array}$$

Let $M$ be the minimal element in $\supp v$.
Arguing as in~\Prop{propmono}, we see that $M$ contains either
$L_{-3/2}$ or $L_{-2}$. 

Assume that $M$ does not contain $L_{-3/2}$. Write $M=YL_{-2}^s$,
where $Y$ does not contain  $L_{-3/2},L_{-2}$; by the above $s>0$.
Then $YL_{-2}^{s-1}L_{-3/2}\in\supp [L_{1/2},M]$. Since
$v$ is singular, $L_{1/2}v=0$ and thus the monomial
$YL_{-2}^{s-1}L_{-3/2}\in\supp [L_{1/2},X]$
for some $X\in\supp v, X\not=M$. Since $M$ is minimal in $\supp v$,
$X$ does not contain $L_{-3/2}$. From (i), (ii) above we conclude that
$X=M$, a contradiction.

Hence $M$ contains $L_{-3/2}$. If
$M=L_{-2}^nL_{-3/2}$ the assertion of the lemma holds
so we assume that 
$$M=X'L_{-r}^sL_{-2}^nL_{-3/2},$$
where  $r\in 1/2\mathbb{Z},\ r>2,\ s\geq 1,\ n\geq 0$ 
and $X'$  does not contain $L_{-r}$. 

Note that the minimal monomial of $[L_{r-2},M]$, which 
belongs to $\cU(\NS_{<-1})$ is $Z:=X'L_{-r}^{s-1}L_{-2}^{n+1}L_{-3/2}$.
Hence $Z$ should appear as $[L_{r-2},U]$ for some
monomial $U\in\supp v, U>M$. From (ii) above we concude that
 $U=X'L_{-r}^{s-1} L_{-(r-1/2)} L_{-2}^{n+1}$.

Let $\supp_0 v$ consist of the monomials in $\supp v$, which
contain $L_{-2}$ and do not contain $L_{-3/2}$. 
By the above, $U\in\supp_0 v$
and so $\supp_0 v$ is not empty.
Let $M_0$ be the minimal element
in $\supp_0 v$. By the above, $M_0<U$, i.e.
$$M_0<X'L_{-r}^{s-1} L_{-(r-1/2)} L_{-2}^{n+1}.$$
It remains to show that $M_0=L_{-2}^k$ for some $k>0$.

Suppose that $M_0\not=L_{-2}^k$ and write $M_0=Y L_{-j}^p L_{-2}^k$, where  
$j\in 1/2\mathbb{Z},\ j>2,\ p,k\geq 1$ 
and $Y'$  does not contain $L_{-j}$. Observe that $k>n$.
The minimal monomial of $[L_{j-2},M_0]$, 
which belongs to $\cU(\NS_{<-1})$ is $Z:=YL_{-j}^{p-1}L_{-2}^{k+1}$.
Therefore $Z$ should appear as $[L_{j-2},Y]$ for some
monomial $U\in\supp v, U\not=M_0$. By (iii) above,
$U$ is either of the form $U_1=X'L_{-2}^{k+1}$ or 
$U_2=X'L_{-2}^{k}L_{-3/2}$. Since $k>n$ one has $U_2<M$ and thus
$U_2\not\in\supp v$. Moreover, $U_1<M_0$ and thus
$U_1\not\in\supp v$ as well.  Hence $M_0=L_{-2}^k$ as required.
\end{proof}

\subsubsection{}
\begin{cor}{corns1}
Let $c=c^S_{p,q}$, $p>q\geq 2,\ (p,q)\in Y$.
The minimal monomial of a singular vector $v$ of $V^c$, which is
not proportional to $\vac$, is $L_{-2}^mL_{-3/2}$ if $p,q$ are even,
and is $L_{-5/2}L_{-2}^mL_{-3/2}$ if $p,q$ are odd ($m\geq 0$).
\end{cor}
\begin{proof}
Note that if $v$ contains $L_{-2}^mL_{-3/2}$ or $L_{-5/2}L_{-2}^mL_{-3/2}$,
then the corresponding monomial is the minimal monomial in $v$.
Moreover, for $c=c^S_{p,q}$ the weight of $v$ is $(p-1)(q-1)/2$
and thus if $v$ contains $L_{-2}^mL_{-3/2}$, then $p,q$ are even
and if  $v$ contains $L_{-5/2}L_{-2}^mL_{-3/2}$, then $p,q$ are odd.

Suppose that $v$ does not contain $L_{-2}^mL_{-3/2}$.
Then, by~\Lem{lemmonons},  $v$ contains $L_{-2}^k$. 
Since $[L_{1/2},L_{-2}^k]$ contains $L_{-2}^{k-1}L_{-3/2}$, 
we conclude, using (ii)
above, that $k>1$ and that  $v$ contains $L_{-5/2}L_{-2}^mL_{-3/2}$.
\end{proof}

\subsubsection{}
\begin{cor}{corns}
Let $J$ be the right ideal in $\cU(\NS_{<-1})$
generated by $L_{-i}, i>2$.
Let $c=c^S_{p,q}$, $p>q\geq 2,\ (p,q)\in Y$ and $v$ be the singular vector
of the proper submodule of $V^c$.
 If $p,q$ are odd, then $v=(L_{-2}^{m+1}+a)\vac$ for some $a\in J, m\geq 0$.
If $p,q$ are even, then $L_{-1/2}v=(L_{-2}^{m+1}+a)\vac$ 
for some $a\in J, m\geq 0$.
\end{cor}
\begin{proof}
The monomials which do not lie in 
the right ideal generated by $L_{-i}, i>2$
are of the form $L_{-2}^mL_{-3/2},\ L_{-2}^m$ for $m\geq 0$.
Therefore if $v'\in V^c$ has integer weight then
either $v'\in J\vac$ or $v'=(L_{-2}^{m+1}+a)\vac$ for $a\in J, m\geq 0$;
in other words, if $v'\in V^c$ contains the monomial $L_{-2}^{m+1}$
then $v'=(L_{-2}^{m+1}+a)\vac$ for $a\in J, m\geq 0$.
Now the assertion for $p,q$ odd immediately follows from~\Lem{lemmonons}.
If $p,q$ are even, then $v$ contains $L_{-2}^mL_{-3/2}$
and $L_{-1/2}v$ contains $L_{-2}^{m+1}$, and the assertion follows.
\end{proof}

\section{Lie algebra case}
In this section we will prove~\Thm{thm01}.

Let $\fg$ be a simple finite-dimensional Lie algebra. 
In this section we will use the following (non-standard)
normalization 
of the form $B=(.|.)$ on $\fh^*$: $(\alpha|\alpha)=2$
if $\alpha$ is a {\em short} root. This normalization is convenient
since $(\beta|\beta)/2,\ (\rho|\beta)$ are positive integers for
any root $\beta$. In this normalization~\Thm{thm01}
takes the form:

$$V^k\ \text{  is not irreducible }\Longleftrightarrow 
(k+h^{\vee}_B)\in\{0\}\cup\{\frac{p}{q}|\ p\in\mathbb{Z}_{\geq 2},
 q\in\mathbb{Z}_{\geq 1}, (p,q)=1\}.$$

In the notation of~\ref{defM} this can be written as

$$M_{p/q}\not=0\ \Longleftrightarrow\ p\geq 2\ \text{ or }p=0.$$

We will check the last equivalence in~\ref{st1}---\ref{p>1} below.

\subsection{}\label{st1}
Retain notation of~\ref{kappa2}.
Since $\fg_1=0$ the set $S$ (introduced in~\ref{setS})
is empty and the group $W^{\#}$ (introduced in~\ref{Wsm})
coincides with $W$. \Thm{detdef} gives
$$
\begin{array}{l}
\det S_{\nu}(k)=\displaystyle
\prod_{r\geq 1}\prod_{\gamma\in\hat{\Delta}^+\setminus\Delta}
\phi_{r\gamma}(k)^{(\dim\hat{\fg}_{\gamma})d_{r,\gamma}(\nu)},\\
\phi_{r\gamma}(k)=
(\Lambda_0|\gamma)k+(\hat{\rho}|\gamma)-r(\gamma|\gamma)/2,\ 
\ \sum_{\nu}d_{r,\gamma}(\nu)e^{-\nu}=R^{-1}
\sum_{w\in W} (-1)^{l(w)} e^{w.(-r\gamma)}.
\end{array}$$

Using notation of~\ref{defM} we obtain 
$$\begin{array}{l}
M_{b}=\sum_{(r,\gamma): \phi_{r\gamma}=k+(\hat{\rho}|\delta)-b}
(\dim\fhg_{\gamma}) E(-r\gamma), \text{ where }\\
E(\lambda):=\sum_{w\in W} (-1)^{l(w)} e^{w.\lambda}.\end{array}$$
Note that all $\dim\hat{\fg}_{\gamma}=1$ in $M_b$ if $b\not=0$.

\subsection{}\label{start1}
Recall that $(\hat{\rho}|\alpha)=(\rho|\alpha)$ for $\alpha\in\Delta$.
For $r,s\geq 1$ and $\alpha\in{\Delta}^+$ one has
$$\begin{array}{l}
\phi_{r(m\delta)}(k)=k+(\hat{\rho}|\delta),\\
\phi_{r(m\delta-\alpha)}(k)=
(k+(\hat{\rho}|\delta))m-(\frac{r(\alpha|\alpha)}{2}+(\rho|\alpha)),\\
\phi_{s(m\delta+\alpha)}(k)=
(k+(\hat{\rho}|\delta))m-(\frac{s(\alpha|\alpha)}{2}-(\rho|\alpha)).
\end{array}$$

Therefore for $p\not=0$ the factor $k+(\hat{\rho}|\delta)-p/q$ appears as 

(i) $\phi_{r(m\delta-\alpha)}(k)$ for
$r=\frac{2(pl-(\rho|\alpha))}{(\alpha|\alpha)}, m=ql$, 
where $l$ is such that $r\in\mathbb{Z}_{\geq 1}$,
 
(ii) $\phi_{s(m\delta+\alpha)}(k)$ for
$s=\frac{2(pl+(\rho|\alpha))}{(\alpha|\alpha)}, m=ql$, where 
$l$ is such that $s\in\mathbb{Z}_{\geq 1}$.

Taking into account that
$$E(-sm\delta-s\alpha)=
- E(-sm\delta+
(s-\frac{2(\rho|\alpha)}{(\alpha|\alpha)})\alpha),$$
we obtain
\begin{equation}
\label{Mpq}
\begin{array}{rl}
M_{p/q}&=\sum_{\alpha\in{\Delta}^+}\sum_{l: \frac{2pl-2(\rho|\alpha)}
{(\alpha|\alpha)}
\in\mathbb{Z}_{\geq 1}}
E(-\frac{2pl-2(\rho|\alpha)}
{(\alpha|\alpha)}(ql-\alpha))\\
&-\sum_{\alpha\in{\Delta}^+}\sum_{l: l\geq 1,\ \frac{2pl+2(\rho|\alpha)}
{(\alpha|\alpha)}\in\mathbb{Z}_{\geq 1}}
E(-ql\frac{2pl+2(\rho|\alpha)}
{(\alpha|\alpha)} \delta+
\frac{2pl}{(\alpha|\alpha)}\alpha).\end{array}
\end{equation}

Observe that $2(\rho|\alpha)/(\alpha|\alpha)\in\mathbb{Z}_{\geq 1}$ 
for $\alpha\in{\Delta}^+$.

\subsection{}\label{p=1}
For $p=1$ we get

$$\begin{array}{rl}
M_{1/q}& =\sum_{\alpha\in{\Delta}^+}\sum_{l: \frac{2(l-(\rho|\alpha))}
{(\alpha|\alpha)}
\in\mathbb{Z}_{\geq 1}}
E(-ql\frac{2(l-(\rho|\alpha))}
{(\alpha|\alpha)}
\delta+\frac{2(l-(\rho|\alpha))}
{(\alpha|\alpha)}\alpha)\\
&-\sum_{\alpha\in{\Delta}^+}\sum_{n: \frac{2n}
{(\alpha|\alpha)}\in\mathbb{Z}_{\geq 1}}
E(-qn\frac{2(n+(\rho|\alpha))}
{(\alpha|\alpha)} \delta+
\frac{2n}{(\alpha|\alpha)}\alpha)=0\end{array}$$
via the substitution $n:=l-(\rho|\alpha)$.

\subsection{}\label{p<0}
Let us show that $M_{p/q}=0$ for $p<0$. 

The above formulas show that for $\alpha\in\Delta^+$ one has
$\phi_{r(m\delta+\alpha)}(k)=k+(\hat{\rho}|\delta)-a$ for $a\geq 1$; 
$\phi_{s(m\delta-\alpha)}(k)= k+(\hat{\rho}|\delta)-a$ for $a<0$
iff $(\rho-s\alpha|\rho-s\alpha)<(\rho|\rho)$. By~\Lem{lemstab} from the
Appendix, $E(-sm\delta-s\alpha)=0$. Hence 
$M_{p/q}=0$.

\subsection{}\label{p>1}
Finally, let us show that $M_{p/q}\not=0$
if $p>1$. Let $\alpha$ be a simple root satisfying $(\alpha|\alpha)=2$.
Take $l>>0$ and introduce 
$r:=\frac{2pl-2(\rho|\alpha)}{(\alpha|\alpha)}=pl-1$.
Then $r>>0$ and, in the light of~\Lem{lemw} from the Appendix, 
the only term in the expression~(\ref{Mpq})
which can be canceled with $E(-rm\delta+r\alpha)$ is the term
$E(-ql'\frac{2pl'+2(\rho|\alpha)}
{(\alpha|\alpha)} \delta+
\frac{2pl'}{(\alpha|\alpha)}\alpha)=E(-ql'(pl'+1)+pl'\alpha)$, where
$$-rm\delta+r\alpha=-ql'(pl'+1)+pl'\alpha.$$
The last formula gives $pl-1=pl'$, which is impossible since $p\geq 2$.
\qed

\section{The case $\mathfrak{osp}(1,2n)$}
In this section we will prove~\Thm{thm02}.

Let $\fg=\osp(1,2n)$. In this section we will normalize
the form $(.|.)$ on $\fh^*$ by the condition: $(\alpha|\alpha)=2$
for $\alpha\in\Delta_1$. In this normalization~\Thm{thm02} takes the form
\begin{equation}\label{Most}
M_{p/q}\not=0\ \Longleftrightarrow\ p\geq 0, p\not=2.
\end{equation}
where $M_{p/q}$ is introduced in~\ref{defM},
$p,q$ are relatively prime integers and $q>0$.
In this section we will check the above equivalence.

\subsection{}\label{st3}
Set
$$\ol{\hat{\Delta}}^+_0:=\{\alpha\in\hat{\Delta}^+_0|\ 
\alpha/2\not\in\hat{\Delta}_1^+\},\ \ 
\ol{\Delta}^+_0:=\{\alpha\in {\Delta}^+_0|\ 
\alpha/2\not\in {\Delta}_1^+\}.$$

\begin{prop}{proosp}
\begin{equation}\label{detosp}\begin{array}{l}
\det S_{\nu}(k)=\prod_{(r,\gamma)\in\Omega}
\phi_{r\gamma}(k)^{(\dim\hat{\fg}_{\gamma})
 d_{r,\gamma}(\nu)},\ \text{ where }\\
\Omega:=\{(r,\gamma)|\ r\in\mathbb{Z}_{\geq 1}, 
\gamma\in\ol{\hat{\Delta}}^+_0\setminus\Delta\}\cup
\{(2j-1,\gamma)|\ j\in\mathbb{Z}_{\geq 1},\ \gamma\in\hat{\Delta}^+_1\},\\
\phi_{r\gamma}(k)=
(\Lambda_0|\gamma)k+(\hat{\rho}|\gamma)-r(\gamma|\gamma)/2,\\
\sum_{\nu}d_{r,\gamma}(\nu)e^{-\nu}=(-1)^{(r-1)p(\gamma)}R^{-1}
\sum_{w\in W} (-1)^{l(w)} e^{w.(-r\gamma)}.\end{array}\end{equation}
\end{prop}
\begin{proof}
For $\fg=\osp(1,2l)$ the set 
the set $S$ (introduced in~\ref{setS})
is empty and the group $W^{\#}$ (introduced in~\ref{Wsm})
coincides with $W$.  \Thm{detdef} gives
$$\det S_{\nu}(k)=\prod_{\gamma\in\hat{\Delta}^+\setminus\Delta, r\geq 1}
\phi_{r\gamma}(k)^{(\dim\hat{\fg}_{\gamma}) d_{r,\gamma}(\nu)},$$
where
$$\sum_{\nu}d_{r,\gamma}(\nu)e^{-\nu}=(-1)^{(r-1)p(\gamma)}R^{-1}
\sum_{w\in W} (-1)^{l(w)} e^{w.(-r\gamma)}.$$
Now the assertion follows from the following
observations: $\phi_{(2r)\gamma}=\phi_{r(2\gamma)}$, and
$d_{2r,\gamma}=-d_{r,2\gamma}$ if $\gamma$ is odd.
\end{proof}

\subsection{}
Using notation of~\ref{defM} we have for $p\not=0$ 
\begin{equation}\label{Mosp}
\begin{array}{l}
M_{p/q}=\sum_{(r,\gamma)\in\Omega: 
\phi_{r\gamma}= k+(\hat{\rho}|\delta)-p/q}
E(-r\gamma),\ \text{where}\\
E(-r\gamma):=(-1)^{(r-1)p(\gamma)}\sum_{w\in W} (-1)^{l(w)} e^{w.(-r\gamma)}.
\end{array}
\end{equation}
Note that $E(-r\gamma)$ is well-defined, since, if 
$r\gamma\not\in\mathbb{Z}\delta$, then the product
$r\gamma$ uniquely determines the pair $(r,\gamma)\in\Omega$;
if $r\gamma\in\mathbb{Z}\delta$, then $\gamma\in\mathbb{Z}\delta$
so $p(\gamma)=0$ and the right-hand side of the last formula
does not depend on $r$.

Notice that, if $(r,\gamma),(r',\gamma')\in\Omega$ are such that
$r\gamma=r'\gamma'$, then $(r,\gamma)=(r',\gamma')$.

\subsection{}\label{phiosp}
The formulas for $\phi_{r(l\delta)},\ \phi_{r(l\delta\pm\alpha)}$ 
have the same form as for Lie algebra case, 
see~\ref{start1}.

Note that a root $\gamma\in\overline{\hat{\Delta}}_0$ has the form
$\gamma=l\delta\pm\alpha$ if $\alpha\in\overline{\Delta}_0^+$, 
or $\gamma=l\delta\pm 2\beta$ if $l$ is odd
and $\beta\in\Delta_1^+$. Taking into account that 
$(\beta|\beta)=2$ for $\beta\in\Delta_1^+$ and $(\alpha|\alpha)=4$ 
for $\alpha\in\overline{\Delta}_0^+$,  we see
that the factor $k+(\hat{\rho}|\delta)-p/q$ for $p\not=0$ appears as 

(i) $\phi_{r(l\delta\pm\alpha)}$ for 
$\alpha\in\overline{\Delta}_0^+$ if 
$r:=\frac{pm\pm(\rho|\alpha)}{2}\geq 1$,

(ii) $\phi_{r(l\delta\pm\beta)}$  for $\beta\in\Delta_1^+$ if 
$r:=pm\pm (\rho|\alpha)\geq 1$ and $r$ is odd,

(iii) $\phi_{r(l\delta\pm 2\beta)}$ for $\beta\in\Delta_1^+$ if
$l:=qm$ is odd and $r:=\frac{pm\pm 2(\rho|\beta)}{4}\in \mathbb{Z}_{\geq 1}$.

In all cases $l=qm$. 

\subsubsection{}\begin{rem}{reme}
Observe that $(\rho|\alpha)$ is even for 
$\alpha\in\overline{\Delta}_0^+$ and 
$(\rho|\beta)$ is odd for $\beta\in\Delta_1^+$. As a result,
 $pm$ is even in the cases (i), (ii); in the case (iii) both $q, m$ are odd 
and $pm\equiv 2 \mod 4$, so $p$ is even.
\end{rem}

\subsection{}
Let us show that $M_{p/q}\not=0$ for coprime integers $p,q$
iff $p\geq 0, p\not=2$.

We will use the letters $l,l',m,q,r,s$ for positive integers.

\subsubsection{}
Identify  $\overline{\Delta_0}\cup\Delta_1$ with
the root system of $B_n$; notice that $W$ identifies with the Weyl group of
$B_n$. Now repeating the arguments of~\ref{p<0}
we obtain  $M_{p/q}=0$ for $p<0$.

\subsubsection{}
Let us show that $M_{p/q}\not=0$ if $p\geq 1, p\not=2$.

Let $\beta$ be a simple odd root; then $(\beta,\rho)=1$.
In the light of~\Lem{lemw} from the Appendix, for $r>> 1$
the only term in the expression~(\ref{Mosp}),
which can be canceled with $E(-r(l\delta+\beta))$, is
$E(-s(l'\delta-\alpha))$ satisfying
$-sl'\delta+s\alpha'=s_{\beta}(-rl\delta-r\beta)$,
that is
$$s\alpha'=(r-1)\beta\ \& \ sl'=rl.$$

If $\alpha'=\beta$, then $r=s-1$, which is impossible
since both $r$ and $s$ should be odd (\ref{phiosp}, (ii)). 
If $\alpha'=2\beta$, then $s=\frac{r-1}{2}$. From~\ref{phiosp}, (ii), (iii)
we conclude that $r-1$ and  $4s+2=2r$ are divisible by $p$.
Moreover, by~\Rem{reme}, $p$ is even.
Hence $p=2$ as required.

\subsubsection{}
Finally, let us show that $M_{\frac{2}{q}}=0$ if $q$ is odd.
In this case,
for $\beta\in\Delta_1$ the term $E(-r(l\delta+\beta))$ 
appears if $l=qm, r=2m+(\rho,\beta)$ for some $m\geq 1$; thus
$s_{\beta}.(-r(l\delta+\beta))=-rqm\delta+2m\beta$ and so
$E(-r(l\delta+\beta))$ cancels 
with $E(-m(rq\delta-2\beta))$. 
For
$\alpha\in\ol{\Delta}_0^+$ the term $E(-r(l\delta+\alpha))$ 
appears if $l=qm, r=m+(\rho|\alpha)/2$ for some $m\geq 1$; thus
$s_{\alpha}(-r(l\delta+\alpha))=-rqm\delta+m\alpha$ and so
$E(-r(l\delta+\alpha))$ cancels
with $E(-m(rq\delta-\alpha))$.
\qed

\section{Lie superalgebras of non-zero defect}
In this section we prove~\Thm{thm04}:

\begin{thm}{}
Let $\fg$ be $\fgl(2,2)$ or a simple Lie superalgebra of  defect one i.e.,
$\fg=\fsl(1,n), \osp(2,n),  \osp(3,n)$, $\osp(n,2)$ with $n>2$,  
$D(2,1,a), F(4), G(3)$.
Then
$$V^k\text{ is not simple }\ \Longleftrightarrow\ 
\exists\alpha\in\Delta^+_0\ 
\text{ s.t. }
\frac{k+(\hat{\rho}|\delta)}{(\alpha|\alpha)}\in\mathbb{Q}_{\geq 0}\, .$$
\end{thm}

In other words, in the standard normalization of the invariant
bilinear form, for $\fg=\fsl(1,n), \osp(2,2n)$ the vacuum module
$V^k$ is not simple iff $k+h^{\vee}$ 
is a non-negative rational number; for all
other Lie superalgebras of defect one, except for
$D(2,1,a), a\not\in\mathbb{Q}$, and for $\fgl(2,2)$ the vacuum module
$V^k$ is not simple iff $k+h^{\vee}\in\mathbb{Q}$.
For $D(2,1,a), a\not\in\mathbb{Q}$ the vacuum module
$V^k$ is not simple iff $k\in\mathbb{Q}_{\geq 0}\cup 
\mathbb{Q}_{>0}a\cup\mathbb{Q}_{>0}(-1-a)$,
in the standard normalization of $(.|.)$.

Retain notation of~\ref{kappa2}.
In this section $p,q,r\in\mathbb{Z}_{\geq 1},\ s\in\mathbb{Z}_{\geq 0},
\gamma\in\hat{\Delta}^+\setminus\Delta$.

\subsection{Case $\fg=\mathfrak{gl}(2,2)$}
Let us show that $V^k$ is not simple iff $k\in\mathbb{Q}$, in the standard 
normalization of $(.|.)$.

\subsubsection{}
Choose a set simple roots $\Pi=\{\beta_1,\alpha,\beta_2\}$
which contains two odd roots
$\beta_1:=\vareps_3-\vareps_1,\ \beta_2:=\vareps_2-\vareps_4$
and the even root $\alpha:=\vareps_1-\vareps_2$.
The form is given by $(\vareps_i|\vareps_j)=0$ for $i\not=j$,
$(\vareps_i|\vareps_i)=-(\vareps_j|\vareps_j)=1$ for $i=1,2,\ j=3,4$.
Then  
$$\Delta^+_0=\{\alpha,\alpha+\beta_1+\beta_2\},\ 
\Delta^+_1=\{\beta_1,\beta_2,\alpha+\beta_1,\alpha+\beta_2\},
\ \rho=-\frac{\beta_1+\beta_2}{2}.$$
One has $S=\{\beta_1,\beta_2\},\ W^{\#}=\{\id,s_{\alpha}\}$ and 
$W^{\#}S\subset\Delta^+$. \Thm{detdef} gives
$$\det S_{\nu}(k)=\prod_{r=1}^{\infty}
\prod_{\gamma\in \hat{\Delta}^+\setminus \Delta}\prod_{j_1,j_2\geq 0}
\phi_{r\gamma+j_1\beta_1+j_2\beta_2}(k)^{(\dim\fhg_{\gamma})
d_{r\gamma;j_1;j_2}(\nu)},$$ where
$d_{r\gamma;j_1;j_2}=(-1)^{(r+1)p(\gamma)+j_1+j_2}
R^{-1}e^{-\rho}(1-s_{\alpha})\bigl(e^{\rho-j_1\beta_1-j_2\beta_2-r\gamma}
\bigr)$.

Note that $(\beta_1,\gamma')=(\beta_2,\gamma')$ for any $\gamma'\in\Delta$.
Therefore $\phi_{r\gamma+j_1\beta_1+j_2\beta_2}$ depends on
$r\gamma$ and the sum $j_1+j_2$ and thus
$$\begin{array}{l}
\det S_{\nu}=\prod_{r=1}^{\infty}
\prod_{\gamma\in \hat{\Delta}^+\setminus \Delta}\prod_{m\geq 0}
\phi_{r\gamma+m\beta_1}^{\dim\fhg_{\gamma}d_{r\gamma;m}(\nu)},
\end{array}$$
where the new exponents $d_{r\gamma;m}$ are given by 
\begin{equation}\label{Red}
Rd_{r\gamma;m}=(-1)^{(r+1)p(\gamma)+m}e^{-\rho}(1-s_{\alpha})\bigl(
e^{\rho-r\gamma}J(m)\bigr),
\end{equation}
where 
$$J(m):=\sum_{j=0}^m e^{-j\beta_1-(m-j)\beta_2}.$$

\subsubsection{}
Set $\alpha':=\alpha+\beta_1+\beta_2$. 
Write $\gamma=l\delta+\gamma'$, where $\gamma'\in\Delta$. The factor
$\phi_{r\gamma+m\beta_1}$ is proportional to $k-b(r;\gamma;m)$, where
$$b(r;\gamma;m):=
\frac{r(\gamma'|\gamma')/2-(\rho-m\beta_1|\gamma')}{l}=
\frac{r(\gamma'|\gamma')/2+(m+1)(\beta_1|\gamma')}{l}.$$
We have the following table
$$\begin{array}{c|c}
\gamma' & b(r;\gamma;m)\\ \hline
0;\pm\beta_i & 0\\
\pm(\alpha+\beta_i) & \mp\frac{m+1}{l}\\
\pm\alpha & \frac{\mp(m+1)+r}{l}\\
\pm\alpha' & \frac{\mp(m+1)-r}{l}.
\end{array}$$

\subsubsection{}\label{calE}
In the space $\mathcal{E}$ of regular exponential functions 
let  $\mathcal{E}_{y;\beta}$ (resp.,
$\mathcal{E}_{y;\delta}$)  be the subspace generated by 
$e^{\lambda},\ \lambda=x_{\delta}\delta+x_{\alpha}\alpha+
x_1\beta_1+x_2\beta_2\in\fh,$, where $x_1+x_2=y$ (resp., $x_{\delta}=y$).
Clearly, $\mathcal{E}_{y;\beta}, \mathcal{E}_{y;\delta}$ are invariant
under the linear operator $Q\mapsto e^{-\rho}s_{\alpha}(e^{\rho}Q)$.
We denote by $P_{i;\beta}$ (resp., $P_{+;\beta}$) the projection 
$\mathcal{E}\to\mathcal{E}_{i;\beta}$ (resp., 
$\mathcal{E}\to\sum_{x>0}\mathcal{E}_{x;\beta}$)
with the kernel
$\sum_{x\not=i}\mathcal{E}_{x;\beta}$ (resp., 
$\mathcal{E}\to\sum_{x\geq 0}\mathcal{E}_{-x;\beta}$)
and
by $P_{l;\delta}$ the projection 
$\mathcal{E}\to\mathcal{E}_{l;\delta}$ 
with the kernel
$\sum_{y\not=l}\mathcal{E}_{y;\delta}$.
Recall that $J(m)\in\mathcal{E}_{-m;\beta}$.

\subsubsection{Case $k=-p/q$}
Let us show that $M_{-p/q}\not=0$ if $p,q$ are positive integers.
Retain notation of~\ref{defM}, \ref{calE}. 
Notice that $b(r;\gamma;m)=-p/q$ forces that 
$\gamma'\in\Delta^+\cup\{-\alpha'\}$. The formula~(\ref{Red}) shows that 
$Rd_{r\gamma;m}\in\mathcal{E}_{i;\beta}$ for some $i\leq 0$
if $\gamma'\in\Delta^+$, and
for  $\gamma'=-\alpha'$ we have $Rd_{r\gamma;m}\in
\mathcal{E}_{2r-m;\beta}\cap\mathcal{E}_{rl;\delta}$, where $2r-m>0$,
because $\frac{r-m-1}{l}=\frac{p}{q}>0$. Therefore
$$\begin{array}{l}
P_{i;\delta}\circ P_{+;\beta}(M_{-p/q})=
\displaystyle\sum_{(m,r,l)\in X_i} x_{m;r;l},\\
X_i:=\{(m,r,l)|\ r,l\geq 1,\ m\geq 0,\ 
\frac{r-m-1}{l}=\frac{p}{q},\ rl=i\},\\
x_{m;r;l}:=Rd_{r(l\delta-\alpha');m}=
(-1)^me^{-r(l\delta-\alpha')}e^{-\rho}(1-s_{\alpha})
\bigl(e^{\rho}J(m)\bigr)
\end{array}$$
where the last equality uses $(\alpha'|\alpha)=0$.
Observe that $x_{m;r;l}\not=0$ since
$s_{\alpha}\bigl(e^{\rho}J(m)\bigr)=e^{\rho-(m+1)\alpha}J(m)$. 
One readily sees that $X_{q(p+1)}$ contains a unique triple $(0;p+1;q)$ 
and thus $P_{q(p+1);\delta}\circ P_{+;\beta}(M_{-p/q})=x_{0;p+1;q}$.
Hence $M_{-p/q}\not=0$ as required.

\subsubsection{Case $k=p/q$}
Let us show that $M_{p/q}\not=0$ if $p,q$ are positive integers.
Retain notation of~\ref{defM}, \ref{calE}. 
We will use the following formula:
\begin{equation}\label{Jim}
J(m)(e^{-k\beta_1}+e^{-k\beta_2})=J(m+k)+
e^{-k(\beta_1+\beta_2)}J(m-k)\ \text{ for }m\geq k\geq 0.
\end{equation}

Note that
$$b(r;l\delta-\alpha-\beta_i;m+r)=b(r;l\delta-\alpha';m+2r)=
b(r;l\delta-\alpha;m).$$
Combining~(\ref{Red}) and~(\ref{Jim}) we obtain
$$d_{r(l\delta-\alpha-\beta_1);m+r}+d_{r(l\delta-\alpha-\beta_2);m+r}
+d_{r(l\delta-\alpha);m}+d_{r(l\delta-\alpha');m+2r}=0.$$

Then
$$\begin{array}{rl}
M_{p/q}&=\sum_{j=1}^2\sum_{m,r,l: \frac{m+1}{l}=\frac{p}{q},\ m<r} 
Rd_{r(l\delta-\alpha-\beta_j);m}+\\
& \sum_{m,r,l: \frac{m+1-r}{l}=\frac{p}{q},\ m<2r} 
Rd_{r(l\delta-\alpha');m}
\sum_{m,r,l: \frac{r-(m+1)}{l}=\frac{p}{q}}
Rd_{r(l\delta+\alpha);m}.\end{array}$$
One has $Rd_{r(l\delta-\alpha-\beta_j);m}\in\mathcal{E}_{r-m;\beta}$
and $Rd_{r(l\delta-\alpha');m}\in\mathcal{E}_{2r-m;\beta}$,
whereas $Rd_{r(l\delta+\alpha);m}\in\mathcal{E}_{-m;\beta}$.
Therefore 
$$P_{0;\beta}(M_{-p/q})=
\sum_{r,l: \frac{r-1}{l}=\frac{p}{q}}Rd_{r(l\delta+\alpha);0}.$$ 
Hence  $P_{q(p+1);\delta}\circ 
P_{0;\beta}(M_{-p/q})=Rd_{(p+1)(q\delta+\alpha);0}\not=0$, and this gives
$M_{p/q}\not=0$.

\subsection{Superalgebras of  defect one}
Let $\fg$ be a basic classical
Lie superalgebra of  defect one:
$\fg=\fsl(1,n)$, $\osp(3,2n)$, $\osp(N,2)$, $\osp(2,2m)$, $F(4)$,
$G(3)$, $D(2,1,a)$.
The root systems of these Lie superalgebras are 
described in Sect.~\ref{rootsy};
in particular, the group $W^{\#}$ is explicitly written there. 
We retain notation of Sect.~\ref{rootsy} and
for each algebra fix a system of simple roots $\Pi$ described there.
One has $S=\{\beta\}\subset \Pi$, where $\beta$ is an isotropic
root given there.

\subsubsection{}
Write $\gamma=l\delta+\gamma'$, where $\gamma'\in\Delta\cup\{0\}$.
The factor
$\phi_{r\gamma+s\beta}(k)$ is proportional to $k+h^{\vee}-b(r;\gamma;s)$, where
\begin{equation}\label{formp<0}
b(r;\gamma;s):=\frac{r(\gamma'|\gamma')/2-(\rho-s\beta|\gamma')}{l}.
\end{equation}
For $b\not=0$ \Thm{detdef} gives
\begin{equation}\label{Madef1}\begin{array}{l}
M_{b}=\sum_{(r;s;\gamma): b(r;\gamma;s)=b}
 E(r;\gamma;s),\ \text{where}\\
E(r;\gamma;s):=(-1)^{s+(r-1)p(\gamma)}\sum_{w\in W^{\#}} (-1)^{l(w)} 
e^{w.(-r\gamma-s\beta)}.
\end{array}
\end{equation}

Using the $W$-invariance of $\delta$ and $(\hat{\rho}-\rho)$, we get
$$E(r;\gamma;s)=0\ \Longleftrightarrow\ 
\Stab_{W^{\#}}(\rho-r\gamma'-s\beta)\not=\id.$$

\subsubsection{}\label{puf}
If the term $E(r;\gamma;s)$ is non-zero, it is a sum of the form
$\sum_{i=1}^{|W^{\#}|} e^{\lambda_i}$, where all summands are distinct
and there exists a unique index $i$ such that $\lambda_i$ is 
dominant with respect to $\Pi^{\#}$ 
(i.e., $(\lambda_i|\alpha)\geq 0$ for any $\alpha\in \Pi^{\#}$).
As a result, 
$\sum_{i\in I}E(r_i;\gamma_i;s_i)=0$ iff the index set $I$ 
admits an involution
$\sigma: I\to I$ such that  $E(r_i;\gamma_i;s_i)+
E(r_{\sigma(i)};\gamma_{\sigma(i)};s_{\sigma(i)})=0$.

We will prove that $M_b\not=0$ by exhibiting the triple
$(r;s;\gamma)$ such that 
\begin{equation}\label{Manot=0}
\begin{array}{ll}
(i) & b(r;\gamma;s)=b\ \ \&\ \ E(r;\gamma;s)\not=0,\\
(ii) & b(r';\mu;s')=b,\ \mu\in\hat{\Delta}\setminus\Delta
\ \Longrightarrow\  E(r;\gamma;s)+E(r';\mu;s')\not=0.
\end{array}\end{equation}

\subsubsection{Cancelation}\label{canc}
Suppose that $\gamma,\gamma_1\in\hat{\Delta}$ are such that
$\gamma=\gamma_1+\beta$ and $\dim\fhg_{\gamma}=\dim\fhg_{\gamma_1}=1$.
Then $\phi_{r\gamma+s\beta}(k)=
\phi_{r\gamma_1+ (s+r)\beta}(k)$ 
and $d_{r;s;\gamma}=-d_{r;s+r;\gamma_1}$ since 
$(-1)^{(r-1)p(\gamma)+s}=-(-1)^{(r-1)p(\gamma_1)+r+s}$.
As a result, $\phi_{r\gamma+s\beta}^{d_{r;\gamma;s}}$ cancels with
$\phi_{r(\gamma_1)+ (s+r)\beta}^{d_{r;\gamma_1;s+r}}$:
$$\displaystyle\prod_{r\geq 1,s\geq 0}
\phi_{r\gamma+s\beta}^{d_{r;\gamma;s}(\nu)}
\phi_{r\gamma_1+s\beta}^{d_{r;\gamma_1;s}(\nu)}=
\displaystyle\prod_{r>s\geq 0}
\phi_{r\gamma_1+s\beta}^{d_{r;\gamma_1;s}(\nu)}.$$

\subsection{Case $D(2,1,a)$}
Retain notation of~\ref{D21alpha} and note that $(\hat{\rho}|\delta)=0$.
We will show that $V^k$ is not simple iff
$k\in\mathbb{Q}_{\geq 0}\cup \mathbb{Q}_{>0}a\cup
\mathbb{Q}_{<0}(1+a)$.
If $a$ is rational then $V^k$ is not simple iff $k\in\mathbb{Q}$.

\subsubsection{}
Take $\gamma\in\hat{\Delta}^+$. Note that $b(r;\gamma;s)=0$
if $\gamma'=0,\pm\beta$. Take $\gamma$ such that $\gamma'\not=0,\pm\beta$.
In the light of~\ref{canc} if $\gamma-\beta$ is 
a root then $\phi_{r\gamma+s\beta}^{r_{r;\gamma;s}}$ cancels
$\phi_{r(\gamma-\beta)+ (s+r)\beta}^{d_{r;\gamma-\beta;s+r}}$.
Observe that exactly one of the elements $\gamma-\beta, 
\gamma+\beta$ is a root. \Thm{detdef} gives
$$\det S_{\nu}(k)
=k^{d(\nu)}
\displaystyle\prod_{l\geq 1}
\displaystyle\prod_{r>s\geq 0}
\displaystyle\prod_{\scriptstyle{\gamma'\in\Delta, 
\gamma-\beta\not\in\Delta \atop \gamma'\not=\pm\beta}}
\phi_{r(l\delta+\gamma')+s\beta}(k)^{d_{r;s;l\delta+\gamma'}(\nu)}$$
for some $d(\nu)\in\mathbb{Z}_{\geq 0}$.

Set $P':=\{m\delta+\sum_{j=0}^2 m_j\vareps_j: m_1,m_2\geq 0\}$. Clearly,
for any $\mu\in \hat{Q}$ the orbit $W^{\#}\mu\cap P'$
contains a unique element $\sum_{j=0}^2 m_j\vareps_j$
and $\Stab_{W^{\#}}\mu\not=\id$ iff $m_1m_2=0$.

\subsubsection{}
Write $\gamma=l\delta+\gamma'$.
For $\gamma'=-\beta$ one has $\phi_{r\gamma+s\beta}=
k+(\hat{\rho}|\delta)$. 
For the remaining values of $\gamma'$ (i.e., $\gamma'\not=-\beta$
and $\gamma-\beta\not\in\Delta$) we have
$$\begin{array}{c|c|c}
\gamma' & b(r;\gamma;s) & W^{\#}(\rho-r\gamma-s\beta)\cap P'\\ \hline
-2\vareps_0 & (1+a)\frac{-r+s+1}{l}& 
-rl\delta+(2r-s-1)\vareps_0+(s+1)\vareps_1+(s+1)\vareps_2\\
-(\vareps_0+\vareps_1-\vareps_2) & 
a\frac{s+1}{l} & 
-rl\delta+(r-s-1)\vareps_0+(r+s+1)\vareps_1+(r-s-1)\vareps_2\\
-(\vareps_0-\vareps_1+\vareps_2) & 
\frac{s+1}{l}&
-rl\delta+(r-s-1)\vareps_0+(r-s-1)\vareps_1+(r+s+1)\vareps_2
\\
2\vareps_1 & a\frac{r-s-1}{l} &
-rl\delta-(s+1)\vareps_0+(2r-s-1)\vareps_1+(s+1)\vareps_2\\
2\vareps_2 &\frac{r-s-1}{l}&
-rl\delta-(s+1)\vareps_0+(s+1)\vareps_1+(2r-s-1)\vareps_2\\
\vareps_0+\vareps_1+\vareps_2 & (1+a)\frac{-s-1}{l}&
-rl\delta-(r+s+1)\vareps_0+(r-s-1)\vareps_1+(r-s-1)\vareps_2
\end{array}$$

\subsubsection{}
Set $X:=\mathbb{Q}_{\geq 0}\cup \mathbb{Q}_{>0}a\cup
\mathbb{Q}_{>0}(-1-a)$ and let us show that $M_b\not=0$
iff $b\in X$.

From the above table we see that for $r>s\geq 0$ the term
$b(r;\gamma;s)\in X$. Hence $M_b=0$ if
$b\not\in X$.
Moreover,  we see that for $r>s$ the vector
$\rho-r\gamma-s\beta$ 
has a non-trivial stabilizer in $W^{\#}$ iff $r=s+1$ and 
$\gamma'\in\{-(\vareps_0-\vareps_1+\vareps_2),
-(\vareps_0+\vareps_1-\vareps_2), \vareps_0+\vareps_1+\vareps_2\}$.
It is easy to see that the entries of last column are pairwise distinct i.e.
$W^{\#}(\rho-r\gamma-s\beta)\cap P'=
W^{\#}(\rho-r_1\gamma_1-s_1\beta)\cap P'$ forces 
$(r;\gamma;s)=(r_1;\gamma_1;s_1)$. In the light of~\ref{puf}
we obtain $M_b\not=0$ for
$b\in X$, as required.

\subsection{}
{\em In the remaining part of the section  $\fg$ 
has defect one and $\fg\not=D(2,1,a)$.}
\subsubsection{Notation}
In all cases, $\fh^*$ has a basis $\vareps_0; \vareps_1,\ldots,
\vareps_n$ and $W^{\#}$ stabilizes $\vareps_0$ and leaves invariant
the space $\fh^{\#}:=\sum_{i\geq 1} \mathbb{C}\vareps_i$.
For $\mu\in\fhh$ we denote by $\mu^{\delta}, \mu^{(i)}$ the corresponding
coordinates of $\mu$ and by $\mu^{\#}$ the projection of
$\mu$ on $\fh^{\#}$:
$$\mu=:\mu^{\delta}\delta+\sum_{i=0}^n \mu^{(i)}\vareps_i,\ \ 
\mu^{\#}:=\sum_{i=1}^n \mu^{(i)}\vareps_i.$$
\subsubsection{}\label{gamdel}
Recall that $W^{\#}$ stabilizes $\delta$ and $\vareps_0$. As a result,
for $\gamma,\mu\in\Delta$ we have
$$
E(r;\gamma;s)+E(r';\mu;s')=0\ \Longrightarrow \ 
r\gamma^{\delta}=r'\mu^{\delta}\ \&\ 
\left\{\begin{array}{ll} r\gamma^{(0)}+s=r'\mu^{(0)}+s' 
&\text{ for } \fg\not=F(4),\\
r\gamma^{(0)}+s/2=r'\mu^{(0)}+s'/2 &\text{ for } \fg=F(4).
\end{array}\right.
$$

\subsection{Proof that $M_{p/q}\not=0$ for $\fg=\osp(3,2)$}
We will deduce that $M_{p/q}\not=0$ from~(\ref{Manot=0}).
Observe that $b(2p+1;q\delta+\vareps_1;0)=p/q$. One has
$$\rho-r\gamma-s\beta=-r\gamma^{(\delta)}\delta-
(s+1/2+r\gamma^{(0)})\vareps_0+(s+1/2-r\gamma^{(1)})\vareps_1.$$
Therefore $E(r;\gamma;s)\not=0$ for any $\gamma\in\Delta$. 
Let us show that $E(r;\gamma;s)+E(2p+1;q\delta+\vareps_1;0)\not=0$ 
for any triple $(r,s,\gamma)$ such that $b(r;\gamma;s)\not=0$.
 Assume that
$E(r;\gamma;s)+E(2p+1;q\delta+\vareps_1;0)=0$. Then 
$(\rho-r\gamma-s\beta)\in W^{\#}(\rho-(2p+1)(q\delta+\vareps_1))$, that is
$r\gamma^{(\delta)}=(2p+1)q,\ 
r\gamma^{(0)}+s=0,\ 1/2-r\gamma^{(1)}=\pm (2p+1/2)$.
The second formula gives $\gamma^{(0)}\leq 0$. Since 
$b(r;\gamma;s)=0$ for $\gamma'\in\{0,\pm\beta\}$, we have the following
cases:
$\gamma'\in\{\pm\vareps_1,\pm 2\vareps_1\}, s=0$ or
$\gamma'\in\{-\vareps_0,-\vareps_0-\vareps_1\}, s=r$.
By~\ref{canc} the terms corresponding to $\gamma'=-\vareps_1,-2\vareps_1,\ s=0$
cancel with the terms corresponding to $\gamma'=-\vareps_0,
-\vareps_0-\vareps_1,\ r=s$. It remains to
show that $E(r;\gamma;s)+E(2p+1;l\delta+\vareps_1;0)\not=0$
for $\gamma'\in\{\vareps_1,2\vareps_1\}, s=0$.
If $\gamma'=2\vareps_1$ we get $1/2-2r=\pm (2p+1/2)$ which is impossible.
Finally, for $\gamma'=\vareps_1$ the formulas $1/2-r=\pm (2p+1/2), rl=(2p+1)q$ 
give $r=2p+1, q=l$ and thus
$E(r;\gamma;s)+E(2p+1;l\delta+\vareps_1;0)=2E(2p+1;l\delta+\vareps_1;0)\not=0$.
Now the inequality $M_{p/q}\not=0$ follows from~(\ref{Manot=0}).

\subsection{Proof that $M_{p/q}\not=0$ for $\fg\not=\osp(3,2)$}\label{def1p>0}
Fix $p,q\in\mathbb{Z}_{\geq 1}$.

Recall that there exists an isotropic root $\alpha\in\Delta^+$
satisfying $(\alpha|\beta)=-1, (\alpha|\rho)=1, \alpha^{(0)}=\beta^{(0)}$.
 One has
$b(m;q\delta-\alpha;p-1)=p/q$ for any $m\geq 1$. 
It is easy to see that $E(m;q\delta-\alpha;p-1)\not=0$
for $m>>0$. 
It remains to verify the condition (ii) of~(\ref{Manot=0}) for some $m>>0$,
i.e. to show that
\begin{equation}\label{aE} 
\exists m>>0\ \text{ s.t. }b(r;\gamma;s)=p/q\ \Longrightarrow\ 
E(m;q\delta-\alpha;p-1)+E(r;\gamma;s)\not=0.
\end{equation}
We claim that this holds if  $m>>0$ is a prime number. 

Indeed, assume that $b(r;\gamma;s)=p/q$ and
$E(m;q\delta-\alpha;p-1)+E(r;\gamma;s)=0$.
Write $\gamma=l\delta+\gamma'$. By~\ref{gamdel} one has $rl=mq$.

The assumption gives
$\frac{r(\gamma'|\gamma')/2-(\rho-s\beta|\gamma')}{l}=\frac{p}{q}$.
Notice that the numerator of the left-hand side is an integer and thus $l$
is divisible by $q$
except the case $\fg=\osp(3,2n),\osp(2n+1,2),\ \gamma'=\pm\vareps_i$.
Using $rl=mq$ we get
$(r;l)\in\{(m,q), (1,mq)\}$ if $\fg\not=\osp(3,2n),\osp(2n+1,2)$ or
$\gamma'\not=\pm\vareps_i$.

If $\fg=\osp(3,2n),\osp(2n+1,2),\ \gamma'=\pm\vareps_i$,
then $r(\gamma'|\gamma')-2(\rho-s\beta|\gamma')$ is an integer, so
$2l$ is divisible by $q$. Therefore 
$rl=mq$ gives that $(r;l)\in\{(m,q), (1,mq),
(2m,q/2), (2,mq/2)\}$. 

Now~\ref{gamdel} gives 
$$s=p-1-m-\frac{r\gamma^{(0)}}{\beta^{(0)}}.$$
Since $m>>0$ we obtain $r\gamma^{(0)}<<0$ so $\gamma^{(0)}<0$ and $r>>0$.
Examining root systems, we see that $\gamma^{(0)}<0$
implies that either $\gamma^{(0)}=-\beta^{(0)}$
or $\gamma'=-2\beta^{(0)}\vareps_0$. Finally, we obtain the following cases:
\begin{enumerate}
\item
$\gamma^{(0)}=-\beta^{(0)}$ and $(r,l,s)=(m,q,p-1)$;
\item
$\gamma'=-2\beta^{(0)}\vareps_0$ and $(r,l,s)=(m,q,m+p-1)$;
\item
$\fg=\osp(3,2n),\osp(2n+1,2), 
\gamma'=-\vareps_0$ and  $(r,l,s)=(2m,q/2,m+p-1)$.
\end{enumerate}
We will show that the cases (ii), (iii)
do not hold and (i) implies 
$(r;s;\gamma)=(m;p-1;l\delta-\alpha)$, that is
 $E(m;q\delta-\alpha;p-1)+E(r;\gamma;s)=2E(m;q\delta-\alpha;p-1)\not=0$.

\subsubsection{Case (i)}\label{casii}
Substituting in~(\ref{formp<0}) we get
$m(\gamma'|\gamma')/2-(\rho-s\beta|\gamma')=p$.
The condition
$m>>0$ forces $(\gamma'|\gamma')=0$ and thus 
$(\rho-(p-1)\beta|-\gamma')=p$.
Since $\gamma^{(0)}<0$ the root $-\gamma'$ is positive and isotropic.
From~\Lem{lempri}, $-\gamma'=\alpha$. Hence $b(m;\gamma;s)=p/q$ forces
$(r;s;\gamma)=(m;p-1;l\delta-\alpha)$.

\subsubsection{Case (ii)}
One has $m\gamma+s\beta=m(\gamma+\beta)+(p-1)\beta$ and 
$\gamma+\beta\in\hat{\Delta}$, since 
$\gamma'+\beta=s_{\vareps_0}\beta\in\Delta$
($2\beta^{(0)}\vareps_0$ is a root, so $\Delta$ is invariant
with respect to the reflection $s_{\vareps_0}$).
Clearly, $(\gamma+\beta)^{(0)}=-\beta^{(0)}$. By~\ref{casii},
$b(m;\gamma+\beta;p-1)=p/q$ implies $\gamma+\beta=l\delta-\alpha$,
which contradicts  $\gamma'=-2\beta^{(0)}\vareps_0$.

\subsubsection{Case (iii)}
In this case $\fg=\osp(3,2n),\osp(2n+1,2), n>1$. 
Substituting in~(\ref{formp<0}), we get $p-1+(\rho|\vareps_0)\not=p/2$,
which is impossible since $(\rho|\vareps_0)=n-1/2>1$.

\subsection{Case $M_{-p/q}$} 
We claim that 
\begin{equation}\label{neglen}
b(r;\gamma;s)<0\ \Longrightarrow\ \gamma'\in\Delta^+\setminus\{\beta\}
\text{ or } \gamma'=\mathbb{Z}\vareps_0\cap\Delta.
\end{equation}
Indeed, take $\gamma'\in \Delta^-$.
Since all simple roots, except $\beta$, have positive lengths squared,
both $(\gamma'|\rho), -(\gamma'|\beta)$ are non-positive. From~(\ref{formp<0})
we see that $b(r;\gamma;s)>0$ forces $(\gamma'|\gamma')<0$.
Examining the root systems we see that $(\gamma'|\gamma')<0$ iff
$\gamma'=\mathbb{Z}\vareps_0\cap\Delta$.

\subsection{Proof that $M_{-p/q}=0$ for
 $\fg=\fsl(1,n),\ C(n)=\osp(2,2n-2)$}\label{AC<0}

\subsubsection{}\label{AC}
Take $\gamma'\in \Delta^+_0$ such that $\gamma'+\beta\not\in\Delta$.
Let us show that $E(r;\gamma;s)=0$.

Indeed,  $\gamma'+\beta\not\in\Delta$ 
forces $(\beta|\gamma')\geq 0$. Since
$\beta$ is the only isotropic root in $\Pi$,
$(\alpha|\beta)\leq 0$ for any $\alpha\in\Delta^+$.
Hence $(\gamma'|\beta)=0$. 
Set $\Pi_2:=\{\alpha\in\Pi^{\#}: (\alpha|\beta)=0\}$ and
define $\Delta^+_2$, $\rho_2, W_2$ corresponding to $\Pi_2$. 
It is easy to check that $(\beta|\gamma')=0$ forces
$\gamma'\in\Delta^+_2$.
As a result, $(\rho_2|\gamma')=(\rho|\gamma')$.
Since $b(r;\gamma;s)<0$, (\ref{formp<0}) gives
$(\rho_2-r\gamma'|\rho_2-r\gamma')<(\rho_2|\rho_2)$. 
Then, by~\ref{lemstab}, $\rho_2-r\gamma'$ has a non-trivial stabilizer
in $W_2$. Observe that $\beta,\rho-\rho_2$ are $W_2$-invariant. Hence
$\rho-r\gamma'-s\beta$ has a non-trivial stabilizer
in $W_2\subset W^{\#}$. Therefore $E(r;\gamma;s)=0$, as required.

\subsubsection{}
Retain notation of~\ref{fsl1n}, \ref{Cn}. For $\fsl(1,n),\ C(n)$
(\ref{neglen}) gives 
$b(r;\gamma;s)<0\ \Longrightarrow\ \gamma'\in\Delta^+\setminus\{\beta\}$.
Notice that for $\gamma'\in\Delta^+_1\setminus\{\beta\}$ one has
$\gamma'-\beta\in\Delta$.
Combining~\ref{canc} and~\ref{AC}, we conclude that
$M_{-p/q}$ is the sum of $E(r;\gamma;s)$, where $b(r;\gamma;s)=-p/q$, 
$\gamma',\gamma'+\beta \in \Delta_0^+$ and $r>s\geq 0$.

\subsubsection{$\fg=\fsl(1,n)$}\label{a<0sl}
The conditions $\gamma'\in \Delta_0^+,\gamma'+\beta\in \Delta^+$
mean that $\gamma'=\vareps_1-\vareps_m$ ($1<m\leq n$).
Since $b(r;\gamma;s)<0$, (\ref{formp<0}) gives $m-1+s>r$.
Using the condition $s<r$ we see that 
the permutation $(1; r-s+1)$ stabilizes 
the vector $\rho-r\gamma'-s\beta=(-n/2-s)\vareps_0+(n/2-r+s)\vareps_1
+\sum_{2\leq i\leq n, i\not=m} (n/2+1-i)\vareps_i+(n/2+1-m+r)\vareps_m$.
Hence $M_{-p/q}=0$.

\subsubsection{$\fg=C(n)$}\label{a<0Cn}
In this case  $\rho^{(i)}=n+1-i$ for $i\geq 1$. The conditions 
  $\gamma'\in \Delta_0^+,\gamma'+\beta\in \Delta^+$
mean that $\gamma'=2\vareps_1;\vareps_1\pm\vareps_m$ ($1<m\leq n$).

For $\gamma'=\vareps_1-\vareps_m$ 
the permutation $(1; r-s+1)$ stabilizes $\rho-r\gamma'-s\beta$
as in~\ref{a<0sl}.

Fix $m\in\mathbb{Z}$ such that 
$1<m\leq n$ and set $\gamma':=\vareps_1+\vareps_m$.
One has $b:=b(r;\gamma;s)=\frac{2n+1+s-m-r}{l}$.
$x_1=n+s-r, x_j=n+1-j$ for $j\not=0,1,m$, and $x_m=n+1-m-r$.
Since $b>0$, one has $x_1>-(n+1-m)$; if $x_1\not=n+1-m$
then $(\rho-r\gamma'-s\beta)$ has a non-trivial stabilizer
(since either $x_1=0$ or $x_1=\pm x_j$ for $1<j<m$).
If  $x_1=n+1-m$, then $r-s=m-1$, so $b=\frac{2(n+1-m)}{l}$
and 
\begin{equation}\label{eqCn}
s_{\vareps_1-\vareps_m}(\rho-r\gamma'-s\beta)=(n+1-m-r)\vareps_1+
\sum_{j=2}^n (n+1-j)\vareps_n.
\end{equation}

For $\gamma':=2\vareps_1$ one has $b:=b(r;\gamma;s)=\frac{2(n-r+s)}{l}$,
and so $b>0$ forces $m\leq n$ for $m:=r-s+1$; since $s<r$, we have 
$ 1<m\leq n$. 
One has $\rho-r\gamma'-s\beta=(n+1-m-r)\vareps_1+
\sum_{j=2}^n (n+1-j)\vareps_n$. Using~(\ref{eqCn}) we get
$E(r;l\delta+2\vareps_1;s)+
E(r;l\delta+\vareps_1-\vareps_{r-s+1};s)=0$, and this completes the proof
for $\fg=C(n)$.

\subsection{Proof that $M_{-p/q}\not=0$ for
$\fg\not=\fsl(n,1), C(n)$} Our proof is based on~(\ref{Manot=0}).

\subsubsection{}
$\fg=\osp(2n,2)$. 
One has $b(r;l\delta-2\vareps_0;s)=2\frac{n+s-r-1}{l}$. 
Clearly, $b(r;l\delta-2\vareps_0;s)$ can be any negative rational number.
Fix $(r;s;l)$ such that $b(r;l\delta-2\vareps_0;s)=-p/q$.
The term
$(\rho-r\gamma-s\beta)^{\#}=(n-1+s)\vareps_1+\sum_{i=2}^n(n-i)\vareps_i$
is dominant with respect to $\Pi^{\#}$. It is easy to see that this implies
$E(r;l\delta-2\vareps_0;s)\not=0$ and
$E(r;l\delta-2\vareps_0;s)+E(r';l'\delta-2\vareps_0;s')\not=0$.
The inequality
$b(r;l\delta-2\vareps_0;s)<0$ gives $r\geq n+s$.
Then, by~\ref{gamdel}, $E(r;l\delta-2\vareps_0;s)+
E(r_1;\gamma;s_1)\not=0$ if $\gamma'\in\Delta^+$. Now~(\ref{Manot=0})
follows from~(\ref{neglen}). Hence  $M_{-p/q}\not=0$.

\subsubsection{}
One has
$$b(r;l\delta-\vareps_0;s)=\left\{\begin{array}{ll}
\frac{2n+2s-r-1}{2l} & \text{ for } \fg=\osp(2n+1,2),\osp(3,2n);\\
3\frac{s+3-r}{l} & \text{ for } \fg=F(4);\\
\frac{2s+5-r}{l}& \text{ for } \fg=G(3).
\end{array}\right.$$
Fix $(r;s;l)$ such that
$b(r;l\delta-\vareps_0;s)=-p/q$ and  $r$ is odd.
Then $r\geq s$ and thus, by~\ref{gamdel}, $E(r;l\delta-2\vareps_0;s)+
E(r_1;\gamma;s_1)\not=0$ if $\gamma'\in\Delta^+$.  Using~(\ref{neglen})
we reduce (\ref{Manot=0}) to the formulas 
\begin{equation}\label{nunsha}\begin{array}{ll}
(i) & E(r;l\delta-\vareps_0;s)\not=0,\\
(ii) & E(r;l\delta-\vareps_0;s)+E(r_1;\gamma;s_1)\not=0\ \text{ if }
\ b(r_1;\gamma;s_1)=-p/q \ \&\ 
\gamma'=-t\vareps_0.
\end{array}\end{equation}
Take $\gamma=l_1\delta-t\vareps_0$ ($t=1,2$).
We have: $(\rho-s_1\beta-r_1\gamma)^{\#}
=(\rho-s_1\beta)^{\#}$ is dominant with respect to $\Pi^{\#}$, and this
gives~(\ref{nunsha}, i). To verify~(\ref{nunsha}, ii) assume that
$\rho-s\beta-r(l\delta-\vareps_0)=w(\rho-s_1\beta-r_1\gamma)$ 
for some $w\in W^{\#}$. Then
$w=\id, s_1=s$ and thus $r(l\delta-\vareps_0)=r_1\gamma$, that is
$r=r_1t, rl=r_1l_1$.
Since $r$ is odd, we have $t=1$ and $(r_1,l_1)=(r,l)$.
Hence $E(r;l\delta-\vareps_0;s)+E(r_1;\gamma;s_1)=
2E(r;l\delta-\vareps_0;s)$ and~(\ref{nunsha}, ii) 
follows from~(\ref{nunsha}, i).

\section{Simplicity of minimal $W$-algebras}
\label{Walg}
Let $\fg$ be a simple contragredient finite-dimensional
Lie superalgebra and let $f_{\theta}$ 
be a root vector attached
to the lowest root $-\theta$, which assumed to be even.
Let $(.|.)$ be the invariant bilinear form on $\fg$,
normalized by the condition $(\theta|\theta)=2$.
This normalization may differ in the
super case from the standard normalization (due to inequivalent choices of
$\theta$). The corresponding dual Coxeter numbers $h^{\vee}$ 
are listed in~\cite{KW}. In Section~\ref{rootsy} we list them in the 
standard normalization of $(.|.)$.
For each $k\in\mathbb{C}$, one attaches to the above data 
a vertex algebra $W^k(\fg,f_{\theta})$, as described in~\cite{KWR},\cite{KW},
called the minimal $W$-algebra. We
  denote by $W_k(\fg,f_{\theta})$ its (unique if $k\not=-h^{\vee}$) 
simple quotient.
Our goal is to determine when $W^k(\fg,f_{\theta})$ is simple.
We assume that $k\not=-h^{\vee}$, since in the ``critical'' case, 
when $k=-h^{\vee}$, $W^k(\fg,f_{\theta})$ is never simple.
We shall also exclude the case $\fg=\fsl_2$, since
$W^{k+2}(\fsl_2,f_{\theta})$ is isomorphic to the Virasoro vertex algebra
$V^c$ with $c=1-6\frac{(k-1)^2}{k}$.

\subsection{Main results}\label{funH}
In~\cite{KWR}, \cite{KW} a functor $H$ from the category
of restricted 
$\hat{\fg}$
-modules of level $k$ to the category
of $\mathbb{Z}$-graded  $W^k(\fg,f_{\theta})$-modules is described.
The image of the vacuum $\hat{\fg}$-module $V^k$ is the vertex 
algebra $W^k(\fg,f_{\theta})$,
viewed as a module over itself. The vertex algebra 
$W^k(\fg,f_{\theta})$ is simple  iff $H(V^k)$ is an irreducible module.

\subsubsection{}\label{ar}
According to~\cite{Ar} the functor $H$ is exact and 
$H(L(\lambda))$ is either zero 
or irreducible; one has~\cite{KW},~\cite{Ar}:  $H(L(\lambda))=0$ 
iff $f_{\alpha_0}$ acts locally nilpotently on $L(\lambda)$.

\subsubsection{}
\begin{thm}{Wclaim}
\begin{enumerate}
\item
The vertex algebra $W^k(\fg,f_{\theta})$ 
is simple iff the $\hat{\fg}$-module $V^k$ is irreducible, or
$k\in\mathbb{Z}_{\geq 0}$ and $V^k$ has length two
(i.e., the maximal proper submodule of the 
$\hat{\fg}$-module $V^k$ is irreducible).

\item  If $\fg$ is a simple 
Lie algebra, $\fg\not=\fsl_2$, then the vertex algebra
$W^k(\fg,f_{\theta})$ is simple iff the $\hat{\fg}$-module $V^k$ is 
irreducible.
\end{enumerate}
\end{thm}
\begin{proof}
By~\ref{ar}, $W^k(\fg,f_{\theta})$ is simple if the
$\hat{\fg}$-module $V^k$ is irreducible.
Let $N$ be the maximal proper submodule of $V^k$.
If $k\in\mathbb{Z}_{\geq 0}$ and $N$ is simple 
then, by~\ref{ar}, $H(V^k/N)=H(L(k\Lambda_0))=0$ and $H(N)$ 
is simple. Hence $H(V^k)$ is simple.
Now assume that $W^k(\fg,f_{\theta})$ is simple and
$V^k$ is not irreducible. 
Since $\mathbb{C}[f_{\alpha_0}]$ acts freely on $V^k$,
$f_{\alpha_0}$ does not act locally nilpotently on $N$. Therefore
$H(N)$ is a non-zero submodule of
$W^k(\fg,f_{\theta})$. Hence $H(V^k/N)=H(L(k\Lambda_0))=0$. 
This gives $k\in\mathbb{Z}_{\geq 0}$. It remains to show that
$N$ is simple. 

Recall that $f_{\alpha_0},e_{\alpha_0}$
generate a Lie algebra $\fs$ isomorphic to $\fsl(2)$.
Let $v$ be a singular vector such that $\mathbb{C}[f_{\alpha_0}]v$
is a simple Verma module over $\fs$. Let $N'$ be a $\fhg$-submodule
of $V^k$ generated by $v$ and $N''$ be the maximal proper submodule
of $N'$. Since $\mathbb{C}[f_{\alpha_0}]v$
is a simple Verma module over $\fs$, $N''$ does not meet 
$\mathbb{C}[f_{\alpha_0}]v$ and thus $H(N'/N'')\not=0$ by~\ref{ar}.

Now let $N'$ be any non-zero
submodule of $V^k$ and $v$ be a singular vector in $N'$. 
Note that $\mathbb{C}[f_{\alpha_0}]v$
is a Verma module over $\fs$, which is either simple or has a unique
proper submodule with an $\fs$-singular vector $v'$. Since 
$[f_{\alpha_0},e_{\alpha}]=0$ for any 
$\alpha\in\hat{\Pi}\setminus\{\alpha_0\}$, 
$v'$ is singular. Therefore either $v$ or $v'$
is a singular vector, which generates
a simple Verma module over $\fs$. By above, $H(N')\not=0$.

Let $N'$ be the maximal proper submodule of $N$.
By above, $H(V^k/N)=0$ and $H(N/N')\not=0$.
Since $H(V^k)$ is simple, this gives $N'=0$ and establishes (i).

Finally, (ii) will be proven in~\ref{integr}--\ref{rank2} below.
\end{proof}

Now~\Thm{thm01} gives

\subsubsection{}
\begin{cor}{corWal} Let $\fg$ be a simple 
Lie algebra, $\fg\not=\fsl_2$. Then the vertex algebra
$W^k(\fg,f_{\theta})$ is not simple 
iff $l(k+h^{\vee})$ is a non-negative rational number,
which is not the inverse of an integer 
(here $l$ is the ``lacety'' of $\fg$).
\end{cor}

\subsubsection{}\begin{rem}{}
Recall (see~\cite{KW}) that $W^k(\fg,f_{\theta})$ has central charge
\begin{equation}\label{28} 
c=\frac{k\sdim\fg}{k+h^{\vee}}-6k+h^{\vee}-4,
\end{equation}
if $(\theta|\theta)=2$. 
For example, if $k+n=p/q$, where $p,q\in\mathbb{Z}_{\geq 1}$, 
the vertex algebra $W^k(\fsl_n,f_{\theta})$
has the same central charge for $k_1=-n+p/q$ and
$k_2=-n+\frac{n(n^2-1)}{6}q/p$. 
For $n>2$ we obtain
pairs of non-isomorphic $W$-algebras 
of the same central charge:
if $p=1$ and $q>1$, the vertex algebra 
$W^{k_1}(\fsl_n,f_{\theta})$ is simple, but  the vertex algebra
 $W^{k_2}(\fsl_n,f_{\theta})$ is not simple. (For $n=2$
these $W$-algebras are isomorphic.)
Note, that in contrast to the case of $\fg$ of rank $>1$,  
$W^k(\fsl_2,f_{\theta})$ is simple for 
$k\in\mathbb{Z}_{\geq 0}$.
\end{rem}

\subsubsection{}
The following corollary follows from the above results and the description
of the $N=1,2,3,4$ and big $N=4$ vertex algebras, given in \cite{KW},
in terms of the minimal $W$-algebras.

\begin{cor}{corN} 
\begin{enumerate}
\item The Neveu-Schwarz (N=1) vertex algebra is simple iff its central
charge $c$ is not of the form 
$\frac{3}{2}(1-\frac{2(p-q)^2}{pq})$,
where $p$ and $q$ are relatively prime positive integers such that $p>q$
and $p/q$ is not an odd integer. (The latter set coincides with the set of 
central charges of $N=1$ minimal models, cf. e.g. \cite{KWR}, (6.3).)
\item The $N=2$ vertex algebra is simple iff its central charge $c$ is 
not of the form $3-6p/q$, where $p$ and $q$ are relatively prime
positive integers and $q\geq 2$. (The subset with $p=1$ of the latter set
coincides with the set of central charges of $N=2$ minimal models.)
\item The $N=3$ vertex algebra with central charge $c$ is simple if
$c$ is not a rational number. For all other values of $c$, except, possibly,
for $c=-3b$, where b is a positive odd integer, this vertex algebra is not 
simple.
\item The $N=4$ vertex algebra with central charge $c$ is simple if
$c$ is not a rational number. For all other values of $c$, except, possibly,
for $c=-6b$, where b is a positive integer, this vertex algebra is not simple.
\item The big $N=4$ vertex algebra with central charge $c$ is simple if
$c\not\in\mathbb{Q}_{\geq 0}\cup\mathbb{Q}_{>0}a\cup\mathbb{Q}_{>0}(-1-a)$.
For all other values of $c$, except, possibly,
for $c=-3b$, where b is a positive odd integer, this vertex algebra is 
not simple.
\end{enumerate}
\end{cor}
\begin{proof}
Combining~\Thm{Wclaim}, Theorem~\ref{thm04}, 
and formula (\ref{28}), we obtain (iii)-(v).

(i) follows from~\Thm{thmvacns}. We will give another proof 
by deducing (i) from Theorem~\ref{thm02}.
Indeed,  set $a(k):=2k+3$. Formula (\ref{28}) gives 
$c=\frac{15}{2}-3(a+\frac{1}{a})$. 
By Theorem~\ref{thm02} for $\fg =\osp(1,2)$ with the 
standard normalization $(\theta|\theta)=2$, $V^k$ is simple iff 
$a\not\in\mathbb{Q}\setminus\{\frac{1}{2m+1}\}_{m=0}^{\infty}$. 
By ~\Thm{Wclaim}, 
$W^k(\osp(1,2),f_{\theta})$
is simple for
$a\not\in\mathbb{Q}\setminus\{\frac{1}{2m+1}\}_{m=0}^{\infty}$, 
and is not simple for 
$a\in\mathbb{Q}\setminus\{\frac{1}{2m+1};2m+3\}_{m=0}^{\infty}$.
Since $c(a)=c(1/a)$ and the Neveu-Schwarz vertex algebra 
$W^k(\osp(1,2),f_{\theta})$ is determined by its central charge,
the vertex algebras $W^k(\osp(1,2),f_{\theta})$ and 
$W^{k'}(\osp(1,2),f_{\theta})$ are isomorphic if $a(k)a(k')=1$.
Hence, since $W^k(\osp(1,2),f_{\theta})$ is simple for
$a\in\{\frac{1}{2m+1}\}_{m=0}^{\infty}$,
it is also simple for $a\in\{2m+1\}_{m=0}^{\infty}$.
Hence $W^k(\osp(1,2),f_{\theta})$ is not simple for 
$c=\frac{15}{2}-3(a+\frac{1}{a})$, where
$a\in\mathbb{Q}\setminus\{\frac{1}{2m+1};2m+1\}_{m=0}^{\infty}$.
Since $c(a)=c(1/a)$, we can take 
$a\in\mathbb{Q}\setminus\{\frac{1}{2m+1};2m+1\}_{m=0}^{\infty}$
such that $a>1$ and write $a=p/q$, where $p$ and $q$ are relatively
prime positive integers. This proves (i). 

(ii) The $N=2$ vertex algebra is isomorphic to the minimal $W$-algebra
$W^k(\fsl (2,1),f_{\theta})$ \cite{KW}, and by formula~(\ref{28})
one has $c=-3-6k$.
Combining~\Thm{Wclaim} (i) and Theorem~\ref{thm04}, 
 we see that $W^k(\fsl(2,1),f_{\theta})$ is simple if
$c$ is not of the form $3-6p/q$, where $p$ and $q$ are relatively prime
positive integers, and that for all other values of $c$, except, possibly,
for $c=-3b$, where $b$ is a positive odd integer, this vertex algebra is not 
simple. 

By~\Thm{Wclaim} (i),
it remains to verify that if $k$ is a non-negative integer, then
the vacuum $\hat{\fsl}(2,1)$-module $V^k$
has length two, i.e., the only singular vectors
in $V^k$ have weights $k\Lambda_0$ and $s_{\alpha_0}.k\Lambda_0$.
Consider the natural embedding $\hat{\fsl}(2)$ into $\hat{\fsl}(2,1)$.
We will describe the weights of $\hat{\fsl}(2)$-singular vectors
in $V^k$ and then deduce the required assertion
from the fact that $(k\Lambda_0+\rho,k\Lambda_0+\rho)=(\mu+\rho,\mu+\rho)$
if $\mu$ is the weight of singular vector in $V^k$.

Let $\alpha$ be an even root for ${\fsl}(2,1)$ and $\beta,\alpha+\beta$
be odd roots.
Choose the following set of simple roots for ${\fsl}(2,1)$:
$\{\alpha+\beta,-\beta\}$; then $\theta=\alpha$ and the set of simple roots for
$\hat{\fsl}(2,1)$ is $\{\alpha+\beta,-\beta,\alpha_0:=\delta-\alpha\}$.
Note that $\hat{\fsl}(2,1)$ contains a copy of $\hat{\fsl}(2)$
with a common simple root $\alpha_0$. As a result,
the shifted actions of the reflection  $s_0:=s_{\alpha_0}$
with respect to $\hat{\fsl}(2,1)$ and $\hat{\fsl}(2)$ coincide, i.e.,
$s_0.\mu=s_0(\mu+\hat{\rho})-\hat{\rho}=s_0(\mu+\hat{\rho}')-\hat{\rho}'$,
where $\hat{\rho}$ corresponds to $\hat{\fsl}(2|1)$ and $\hat{\rho}'$
corresponds to $\hat{\fsl}(2)$.

 For $S\subset \hat{\Delta}^+$ set $|S|=\sum_{\gamma\in S}\gamma$.
A Verma module $M(\lambda)$ over $\hat{\fsl}(2|1)$
has a filtration by $\hat{\fsl}(2)$-modules $M'(\lambda-|S|)$: 
$S\subset \hat{\Delta}^+_1\setminus\Delta^+_1$. Let $V^k$ 
be the vacuum module over $\fsl(2,1)$.
Since $\fsl(2)$ acts locally finitely 
on $V^k$, $V^k$ has a filtration
by generalized Verma $\hat{\fsl}(2)$-modules $M'_I(\lambda-|S|)$,
where $I=\{\alpha\}$. One has
$M'_I(\lambda)=M'(\mu)/M'(s_{\alpha}.\mu)$. Since the Weyl group
of $\hat{\fsl}(2)$ is the infinite dihedral group generated by
$s_{\alpha}, s_0=s_{\alpha_0}$,
$M'_I(\mu)$ has at most two $\hat{\fsl}(2)$-singular vectors:
of weight $\mu$ and of weight $s_0.\mu$ if $s_0.\mu<\mu$.
Thus the weight of a $\hat{\fsl}(2)$-singular vector
of $V^k$ is of the form $k\Lambda_0-|S|$ or $s_0.(k\Lambda_0-|S|)$,
where $S\subset \hat{\Delta}^+_1\setminus\Delta^+_1$. 
For $S=\emptyset$
we have $k\Lambda_0$ and $s_0.k\Lambda_0$.
Let us show that there are no other $\hat{\fsl}(2|1)$-singular vectors.

Indeed, if $\mu$ is a weight of $\hat{\fsl}(2)$-singular vector
then $(\lambda+\hat{\rho},\lambda+\hat{\rho})=(\mu+\hat{\rho},\mu+\hat{\rho})$.
Since $(s_0.\mu+\hat{\rho},s_0.\mu+\hat{\rho})=
(\mu+\hat{\rho},\mu+\hat{\rho})$,
it is enough to show that for $S\not=\emptyset$ one has
$(k\Lambda_0+\hat{\rho},k\Lambda_0+\hat{\rho})>
(k\Lambda_0-|S|+\hat{\rho},k\Lambda_0-|S|+\hat{\rho})$, which can be rewritten
as
\begin{equation}\label{N=2}
2(k\Lambda_0+\hat{\rho},|S|)>(|S|,|S|).\end{equation}
For $\mu\in\Delta_1=\{\pm\alpha\pm\beta\}$ set
$S_{\mu}:=S\cap \{k\delta+\mu\}_{k>0},\ \ s_{\mu}:=\# S_{\mu}$.
Then $S=\coprod_{\mu\in\Delta_1} S_{\mu}$, so 
$|S|=\sum_{\mu\in\Delta_1} |S_{\mu}|$. 
Observe that $S_{\mu}=\{r_i\delta+\mu\}_{i=1}^{s_{\mu}}$,
where $1\leq r_1<r_2<\ldots<r_{s_{\mu}}$ and thus
$|S_{\mu}|=r\delta+s_{\mu}\mu$, where $r\geq\frac{s_{\mu}(s_{\mu}+1)}{2}$.
Hence
$$S=m\delta+(s_{\alpha+\beta}-s_{-\alpha-\beta})(\alpha+\beta)
+(s_{\beta}-s_{-\beta})\beta$$
for some $m\geq \sum_{\mu\in\Delta_1}s_{\mu}(s_{\mu}+1)$. 
Taking into account $(\hat{\rho},\beta)=(\hat{\rho},\alpha+\beta)=0,\ 
(\hat{\rho},\delta)=1$ we get
$$2(k\Lambda_0+\hat{\rho},|S|)=2(k+1)m
\geq (k+1)\sum_{\mu\in\{\pm\beta,\pm(\alpha+\beta)\}}s_{\mu}(s_{\mu}+1).$$
On the other hand, 
$$(|S|,|S|)=-(s_{\alpha+\beta}-s_{-\alpha-\beta})(s_{\beta}-s_{-\beta}).$$
For $S\not=\emptyset$ at least one of the quantities
$s_{\beta},s_{-\beta},s_{\alpha+\beta}, s_{-\alpha-\beta}$ is non-zero 
(and all of them are non-negative integers) and thus
$$(k+1)\sum_{\mu\in\{\pm\beta,\pm(\alpha+\beta)\}}s_{\mu}(s_{\mu}+1)>
(s_{-\alpha-\beta}-s_{\alpha+\beta})(s_{\beta}-s_{-\beta}),$$
since $k\geq 0$. This establishes~(\ref{N=2}) and (ii).
\end{proof}

We believe that in all questionable cases in (iii)-(v)
the vertex algebra is not simple,
but we do not know how to prove this.

\subsection{~Outline of the proof of~\Thm{Wclaim} (ii)}\label{integr}
In~\ref{integr}-\ref{rank2} we
assume that $\fg$ is a finite dimensional semisimple {\em Lie algebra}
 and $k$ is a non-negative integer.

\subsubsection{}
In~\ref{cL0} we will show that for $k\in\mathbb{Z}_{\geq 0}$, 
$H(V^k)$ is not simple  
iff $Q_{s_0,w}\not=1$
for some $w\in\hat{W}$, where $Q$ stands for the inverse Kazhdan-Lusztig
polynomial. (This condition does not depend
on the non-negative integer $k$ and thus $H(V^k)$ is simple
iff $H(V^0)$ is simple.)

Remark that for $\fg=\fsl_2$, the Weyl group is the infinite dihedral group;
by~\cite{Hum}, 7.12 one has $P_{x,z}=1$ for $x\leq z$ which implies
$Q_{x,z}=1$ for $x\leq z$. This implies the simplicity of $H(V^k)$ (which is
well-known).

\subsubsection{}
Let $\Theta$ be the set of pairs $(\alpha, X_n)$, where 
$\alpha$ is a node of a  Dynkin diagram $X_n$ satisfying the property:
$$Q_{s_{\alpha},w}\not=1\ \text{ for some } w\in W(X_n),$$
 where $W(X_n)$ is the Coxeter group of type $X_n$.
From~\ref{cL0} we see that $(\alpha, X_n)\not\in\Theta$ iff
for $\fp$ being the maximal parabolic
not containing $\alpha$, the generalized Verma module
$\Ind_{\fp}^{\fg} L'$ has length two, where $L'$ is the trivial 
one-dimensional $\fp$-module.

Geometrically, if $X_n$ is of finite type, $(\alpha, X_n)\in\Theta$
is equivalent to the fact that
a certain codimension one Schubert variety is not rationally smooth.
It is well-known (see, for example,~\cite{Ku}, 12.2.E) that
these Schubert varieties are rationally smooth in rank two cases ($n=2$)
and for the pairs $(\alpha_n, C_n)$ (in enumeration below).
We need to study the case when $X_n$ is an affine diagram. 

\subsubsection{}
We have to show 
that $(\alpha_0, X_n^{(1)})\in \Theta$,
where $X_n^{(1)}$ is the affinization of a finite type diagram $X_n$ ($n>1$)
and $\alpha_0$ is the affine simple root. 

One has
\begin{equation}\label{frakQ}
\begin{array}{ll}
(i) &
(\alpha,X'_n)\in \Theta,\ X'_n \text{ is a subdiagram of } X_m
\ \Longrightarrow\ (\alpha,X_m)\in \Theta;\\
(ii) & \alpha\in X_n  \text{ is connected to } \alpha' \text{ only},\ 
(\alpha',X_n\setminus\{\alpha\})\in \Theta
\ \Longrightarrow\ (\alpha,X_n)\in \Theta
\end{array}
\end{equation}
where ``$X'_n$ is a subdiagram of $X_m$'' in (i) means that the set
of nodes of $X'_n$ is a subset of the nodes of $X_m$ and
the set of edges of $X'_n$ consists of all edges
between these nodes in $X_m$; and
in (ii) the diagram $X_n\setminus\{\alpha\}$ is obtained
from $X_n$ by removing the extremal node $\alpha$ and the edge
between $\alpha$ and $\alpha'$.
We will prove (ii) in~\ref{add0}; (i) follows from the fact that
the inverse Kazhdan-Lusztig polynomials
are the same for a diagram and its subdiagram.

Taking into account~(\ref{frakQ}), 
the verification $(\alpha_0,X_n^{(1)})\in \Theta$ for $n>2$ reduces
to the cases $(\alpha_2,A_3)$, $(\alpha_1,C_3)$. Here
and further we use the following enumeration of the vertices of $X_n$:
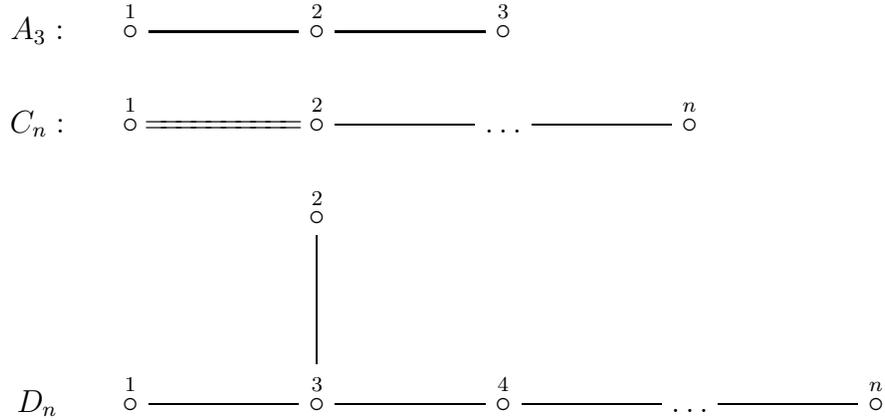
$$\begin{diagram}
A_3: & \overset{\scriptstyle{1}}{\circ} & \rLine & 
\overset{\scriptstyle{2}}{\circ}
& \rLine & \overset{\scriptstyle{3}}{\circ}\\
C_n: &\overset{\scriptstyle{1}}{\circ} 
& \rDline &\overset{\scriptstyle{2}}{\circ} &  \rLine &
\ldots & \rLine & \overset{\scriptstyle{n}}{\circ}\\
&            &  & \overset{\scriptstyle{2}}{\circ}&  & & & & &\\
 &            &  & \uLine &  & & & & & \\
D_n & \overset{\scriptstyle{1}}{\circ} & \rLine & \overset{\scriptstyle{3}}
{\circ}& \rLine & \overset{\scriptstyle{4}}{\circ}& \rLine &
\ldots &\rLine & \overset{\scriptstyle{n}}{\circ}
\end{diagram}$$

Indeed, the pair $(\alpha_0,A_n^{(1)})$ for $n>2$ 
has a subdiagram $(\alpha_2,A_3)$; the pair $(\alpha_0,C_n^{(1)})$ for $n>2$ 
has a subdiagram $(\alpha_1,C_3)$;
applying~(\ref{frakQ}, ii)
to $(\alpha_2,A_3)$ $n-3$ times we obtain the pair $(\alpha_n,D_n)$ 
and now using~(\ref{frakQ}, i) we obtain the pairs
$(\alpha_0,X_n^{(1)})$ for $X=B,D,E$; finally, applying~(\ref{frakQ}, ii)
to $(\alpha_1,C_3)$ twice we get $(\alpha_0,F_4^{(1)})$.

It is easy to verify that for $A_3$ one has $Q_{s_2, s_2s_1s_3s_2}=1+q$,
and that for $C_3$ one has $Q_{s_1, s_1s_2s_1s_3s_2s_1}=1+q^2$.
As a result, $(\alpha_2,A_3), (\alpha_1,C_3)\in \Theta$.
The remaining cases $(\alpha_0,X_2^{(1)})$ are verified in~\ref{rank2}.

\subsection{Multiplicity formula}\label{cL0}
Let $\hat{W}$ be the Weyl group of $\fhg$.
\subsubsection{}
Let $N$ be the maximal proper submodule of $V^k$. By~\Thm{Wclaim} (i),
it is enough to show that $N$ is not simple.
Recall that $k\Lambda_0$ is a dominant integral weight and so
all subquotients of $V^k$ are of the form $L(w.k\Lambda_0)$, $w\in\hat{W}$,
where $\hat{W}$ is the Weyl group of $\fhg$.  The highest weight of $N$ 
is $s_0.k\Lambda_0$ and so $L(s_0.k\Lambda_0)$ is a quotient of
$N$. It remains to verify that
\begin{equation}
\label{NL}
[N: L(w.k\Lambda_0)]\not=0\ \text{ for some } w\not=s_0.
\end{equation}
Let us describe
the multiplicity $[N: L(w.k\Lambda_0)]$ in terms
of Kazhdan-Lusztig polynomials.

\subsubsection{}
One has
$$\ch V^k=\frac{R_I}{R}
 e^{k\Lambda_0},\ \ 
\ch L(k\Lambda_0)=R^{-1}
\sum_{w\in \hat{W}}(-1)^{l(w)}e^{w.k\Lambda_0}$$
where 
$R=\prod_{\alpha\in\hat{\Delta}^+}(1-e^{-\alpha}),\ \
R_I=\prod_{\alpha\in\Delta^+}(1-e^{-\alpha})$.

Using the well-known formula
$R_I=\sum_{w\in W}(-1)^{l(w)}e^{w.0}$
we get
$$\ch V^k=R^{-1}\sum_{w\in W}(-1)^{l(w)}e^{w.k\Lambda_0},$$
that is
$$\ch N=R^{-1}\sum_{w\in \hat{W}\setminus W}(-1)^{l(w)+1}e^{w.k\Lambda_0}=
\sum_{w\in \hat{W}\setminus W}(-1)^{l(w)+1}\ch M(w.k\Lambda_0).$$

\subsubsection{}
In~\cite{KT} the Kazhdan-Lusztig conjecture was established for
the symmetrizable, hence affine, Kac-Moody Lie algebras. This gives
$$\ch M(w.k\Lambda_0)=\sum_{z\in \hat{W}} P_{w,z}(1) \ch L(z.k\Lambda_0),$$
where $P_{w,z}$ are the Kazhdan-Lusztig polynomials defined in~\cite{KL}
(we describe the polynomials in~\ref{KLR}). One has
$P_{w,z}\not=0$ iff $w\leq z$.
We obtain
$$[N:L(z.k\Lambda_0)]=
\sum_{w\in \hat{W}\setminus W}(-1)^{l(w)+1}P_{w,z}(1)=
\sum_{w\in \hat{W}: s_0\leq w\leq z} (-1)^{l(w)+1}P_{w,z}(1).$$

Now the condition~(\ref{NL})  can be rewritten as
\begin{equation}
\label{Pwz}
\sum_{w\in \hat{W}: s_0\leq w\leq z} (-1)^{l(w)+1}P_{w,z}(1)\not=
\delta_{s_0,z}\ \text{ for some } z\in\hat{W}.
\end{equation}

One has $\sum_{w\in \hat{W}: s_0\leq w\leq z} 
(-1)^{l(w)+1}Q_{s_0,w}P_{w,z}=\delta_{s_0,z}$,
where $Q_{s_0,w}$ are the inverse Kazhdan-Lusztig polynomials.
Hence (\ref{Pwz}) is equivalent to
$Q_{s_0,w}(1)\not=1 \text{ for some } w\in\hat{W}$. Using~(\ref{prQ}, i) 
we conclude that
 (\ref{Pwz}) is equivalent to
$$Q_{s_0,w}\not=1 \text{ for some } w\in\hat{W}.
$$

\subsection{Kazhdan-Lusztig polynomials}
\label{KLR}
Let $W$ be a Coxeter group;  denote the unit element
in $W$ by $e$.
For the elements $x,y$ of $W$ set
$$[x,y]:=\{w: x\leq w\leq y].$$

\subsubsection{}
For $x,y\in W$ the Kazhdan-Lusztig polynomials $P_{x,y}(q)$ 
can be computed recursively using the following properties: 
the polynomial $P_{x,y}$ has degree $\leq \frac{l(y)-l(x)-1}{2}$ and
$$P_{x,y}=\left\{\begin{array}{ll}
0, & x\not\leq y,\\
1, & x\leq y \text{ and } l(y)-l(x)\leq 2,\\
\sum_{x\leq w\leq y} (-1)^{l(w)-l(x)} R_{x,w} \ol{P_{w,y}} q^{l(y)-l(w)}, &
\end{array}\right.$$
where  $\ol{P}$ is the image of $P$ under
the algebra involution $q\mapsto q^{-1}$ and 
the polynomials $R_{x,y}(q)$ can be defined
recursively by the formulas
$$R_{x,y}=\left\{\begin{array}{ll}
0, & x\not\leq y,\\
R_{sx,sy}, & sx<x, sy<y,\\
(q-1)R_{sx,y}+q R_{sx,sy}, & sx>x, sy<y,
\end{array}\right.$$
where $s$ is a simple reflection. One has
\begin{equation}
\label{prR}
\begin{array}{ll}
(i)& R_{x,y}=R_{x^{-1},y^{-1}};\\
(ii) & R_{x,y}=(q-1)^{l(y)-l(x)}  \text{ if }x\leq y, l(y)-l(x)\leq 2;\\
(iii) &  \ol{R_{x,y}}=(-q)^{l(x)-l(y)}R_{x,y}.
\end{array}
\end{equation}

By~\cite{KL} 2.3.g, one has:
\begin{equation}
\label{prP}
P_{x,y}=P_{sx,y}\text{ if } sy<y.
\end{equation}

\subsubsection{}
The inverse Kazhdan-Lusztig polynomials $Q_{y,w}(q)$ 
are defined by the formula
\begin{equation}
\label{defQ}
\sum_{w} 
(-1)^{l(w)-l(y)}Q_{y,w}P_{w,z}=\delta_{y,z}.
\end{equation}

A geometric meaning of the inverse Kazhdan-Lusztig polynomials
is discussed in~\cite{KT}. Their results imply that $Q_{x,y}$
have non-negative integer coefficients
in the case of a symmetrizable Kac-Moody Lie algebra.

One has: $Q_{x,z}\not=0$ iff $x\leq z$;
for $x\leq z$ the polynomials $Q_{x,z}$ have the following properties:
\begin{equation}
\label{prQ}
\begin{array}{ll}
(i) & Q_{x,z}=1+a_1q+a_2q^2+\ldots +a_kq^k,\ 
\ a_i\in\mathbb{Z}_{\geq 0},\ k\leq \frac{l(z)-l(x)-1}{2};\\
(ii) & Q_{x,z}=1 \text{ if } l(z)-l(x)\leq 2;\\
(iii) & Q_{x,z}=\sum_{w\in [x,z]} (-1)^{l(z)-l(w)}
q^{l(w)-l(x)} \ol{Q}_{x,w} R_{w,z};\\
(iv) & Q_{e,z}=1, \text{ for all } z.
\end{array}
\end{equation}
The first property follows from~\cite{KT}; (ii) follows from (i).
From~\cite{KT}, Lem. 5.2.1, 5.3 we obtain, using~(\ref{prR}):
$$q^{l(x)} Q_{x,z}=q^{l(z)} \sum_{w\in [x,z]} 
\ol{Q_{x,w} R_{w^{-1},z^{-1}}}=q^{l(z)} \sum_{w\in [x,z]} 
\ol{Q_{x,w}} (-q)^{l(w)-l(z)}R_{w,z},$$
and this gives (iii). 
Finally, combining~(\ref{defQ}) and~(\ref{prP}) we get (iv).

\subsubsection{}
Let 
$$M(x,z):=\sum_{w\in [x,z]} (-1)^{l(z)-l(w)} q^{l(w)-l(x)}R_{w,z}.$$ 
One has
\begin{equation}\label{prQi}
Q_{x,w}=1\ \ \forall w\in [x,z] \ \Longleftrightarrow\ M(x,w)=1\ \ 
\forall w\in [x,z].
\end{equation}
Indeed, assume that $Q_{x,w}=1$ for all  $w\in [x,z]$.
Then for any $y\in [x,z]$ one has $Q_{x,w}=1$ for all  $w\in [x,y]$ 
and~(\ref{prQ}, iii) gives $M(x,y)=1$. For the inverse implication
assume that $M(x,w)=1$ for all  $w\in [x,z]$.
We prove that $Q_{x,w}=1$ by induction on  $w\in [x,z]$ 
with respect to $l(w)$ (note that $l(x)\leq l(w)\leq l(z)$.
If $l(w)=l(x)$, for $w\in [x,z]$, then $w=x$ and
$Q_{x,x}=1$. Suppose that $Q_{x,w}=1$ for all $w\in [x,z]$
with $l(w)<m$. Take $y\in [x,z]$ such that $l(y)=m$.
Then $Q_{x,w}=1$ for all $w\in [x,y]$ and~(\ref{prQ}, iii) gives 
$$Q_{x,y}=M_{x,y}-q^{l(y)-l(x)}R_{y,y}
+q^{l(y)-l(x)}\ol{Q}_{x,y}R_{y,y}=1+q^{l(y)-l(x)}(\ol{Q}_{x,y}-1),$$
that is $Q_{x,y}-1=q^{l(y)-l(x)}\ol{Q_{x,y}-1}$.
If $Q_{x,y}-1=0$,
then $Q_{x,y}-1=b_1 q^{i_1}+b_2q^{i_2}+\ldots b_sq^{i_s}$,
where $i_1<i_2\ldots <i_s$ and $i_1+i_s=l(y)-l(x)$. However,
by~(\ref{prQ}, i) $2i_s<l(y)-l(x)$, a contradiction.

\subsubsection{}\label{prec}
In~\cite{KL} there is the following definition:

\begin{defn}{}
Given $y,w\in W$ we say that $y\prec w$ if the following conditions
are satisfied: $y<w$, $l(w)-l(y)$ is odd and $P_{y,w}$ is a polynomial
in $q$ of degree exactly $\frac{l(w)-l(y)-1}{2}$.
\end{defn}

\begin{lem}{}
Assume that $y\prec z, l(z)-l(y)\geq 3$. Then  $Q_{y,w}\not=1$
for some $w\in [y,z]$.
\end{lem}
\begin{proof}
Suppose that $Q_{y,w}=1$ for all $w\in  [y,z]$.
Then~(\ref{defQ}) gives
$$Q_{y,z}=1+(-1)^{l(z)}\sum_{w\in [y,z]}  (-1)^{l(w)+1}P_{w,z}.$$
The condition $y\prec z$ implies that the degree of $P_{w,z}$ 
is less than the degree of $P_{y,z}$ if $w\in ]y,z]$.
Hence $Q_{y,z}$ has degree $\frac{l(z)-l(y)-1}{2}$
and, in particular, $Q_{y,z}\not=1$, a contradiction.
\end{proof}

\subsubsection{Proof of~(\ref{frakQ}, ii)}
\label{add0}
Let $\alpha_0$ be an extremal node of a Dynkin diagram $X_n$,
$\alpha_1$ be the only node which is connected to $\alpha_0$ and
$X'_n$ be the Dynkin diagram obtained
from $X_n$ by removing the extremal node $\alpha_0$ and the edge
between $\alpha_0$ and $\alpha_1$. Let $W$ (resp., $W'$) be the Coxeter group
of $X_n$ (resp., $X'_n$). Assume that $(\alpha_1, X'_n)\in\Theta$,
that is $Q_{s_1,z}\not=1$ for some
$z\in W'$; let $z$ be a shortest element with this property.

Note that $Q_{s_1,w}=1$ for all $w\in [s_1,z[$.
Formulas~(\ref{prQ}, iv) and~(\ref{prQi}) give 
\begin{equation}\label{muru}
M(s_1,w)=1\ \text{ for 
all }w\in [s_1,z[,\ \ M(s_1,z)\not=1,\ M(e,z)=1.
\end{equation}

Let us show that $Q_{s_0, z'}\not=1$ for some $z'\in [s_0,s_0zs_0]$.
By~(\ref{prQi}) it is enough to verify that $M(s_0,s_0zs_0)\not=1$.
Observe that 
 the elements of $[s_0,s_0zs_0]$ are of the
form $s_0ws_0,s_0w,ws_0$ if $w\in [s_1,z]$ and $s_0w$ if $w\leq z,
w\not\in [s_1,z]$. Since $w$ does not contain $s_0$ (i.e., $w\not\geq s_0$),
one has $l(s_0ws_0)=l(w)+2$ if $w\in [s_1,z]$ and
$l(s_0w)=l(w)+1$ if $w\leq z,
w\not\in [s_1,z]$. The properties of $R_{x,y}$ 
imply $R_{s_0w,s_0zs_0}=R_{ws_0,s_0zs_0}=
(q-1)R_{w,z}$  (since $z\not\geq s_0$) and $R_{s_0ws_0,s_0zs_0}=R_{w,z}$.
We obtain
$$\begin{array}{l}
M(s_0,s_0zs_0)=\sum_{y\in [s_0,s_0zs_0]} 
(-1)^{l(s_0zs_0)-l(y)} q^{l(y)-1}  R_{y,s_0zs_0}\\
=
\sum_{w\in [s_1,z]} (-1)^{l(z)-l(w)}q^{l(w)+1}R_{w,z}+
(q-1)\sum_{w\in [s_1,z]} (-1)^{l(z)+1-l(w)}q^{l(w)}R_{w,z}\\
+
(q-1)\sum_{w\leq z} 
(-1)^{l(z)+1-l(w)}q^{l(w)}R_{w,z}=
q^2M(s_1,z)-q(q-1)M(s_1,z)+(1-q)M(e,z)\\
\ \ \ \ \ \ \  \ \ \ \ \  \ \ \ \ \ \ =1-q+qM(s_1,z)\not=1,\ \text{ 
by~(\ref{muru})}.
\end{array}$$
Hence $Q_{s_0,z'}\not=1$ for some $z'\leq s_0zs_0$.

Using  similar arguments and
the fact that $M(s_1,y)=1$ for all $y\in [s_1,z[$,
we can show that $Q_{s_0,z'}=1$
for $z'<s_0zs_0$ and thus $Q_{s_0, s_0,s_0zs_0}\not=1$.\qed

\subsection{Rank $2$ cases}\label{rank2}

\subsubsection{Case $A_2$}
\label{An}
In this case the  Weyl group $\hat{W}$ is
generated by $s_0,s_1,s_2$, where the relations are
$(s_0s_1)^3=(s_0s_2)^3=(s_0s_2)^2=e$. It is easy to
see that $Q_{s_0,s_0s_1s_2s_0}=1+q$.

\subsubsection{Case $C_2$}\label{C2}
In this case the  Weyl group $\hat{W}$ is
generated by $s_0,s_1,s_2$, where the  non-trivial relations
are $(s_0s_1)^4=(s_1s_2)^4=(s_0s_2)^2=e$.
It is not hard to compute that $Q_{s_0, s_0s_1s_0s_2s_1s_0}=1+q^2$.
We can also check that $Q_{s_0,w}\not=1$ for some $w$
using the tables of Kazhdan-Lusztig polynomials by M.~Goresky:
one has $s_0\prec s_0s_1s_0s_2s_1s_0$ 
(\cite{Gor}, the case $\tilde{B}_2$, No. 57) and then~\Lem{prec}
implies the required assertion.

\subsubsection{Case $G_2$}
Here we use the tables~\cite{Gor}. Take
$z:=s_0(s_1s_2)^2s_0s_1s_2s_1s_0$ 
(No. 133 in the tables~\cite{Gor};
in their notation the affine root is the third one). 
We have $s_0\prec z$
and~\Lem{prec} gives $Q_{s_0,w}\not=1$ for some $w\in [s_0,z]$
(it is easy to see that, in fact, 
$Q_{s_0,w}=1$ for $w<z$ and $Q_{s_0,z}=1+q^4$).

\section{Root systems of defect one}\label{rootsy}
The list of simple Lie superalgebras $\fg$ of defect one consists of
Lie superalgebras $A(0,n)=\fsl(1,n+1), C(n+1)=\osp(2,2n)$, 
$B(1,n)=\osp(3,2n)$, 
$B(n,1)=\osp(2n+1,2)$, $D(n+1,1)=\osp(2n+2,2)$,
where $n\geq 1$, and the exceptional Lie superalgebras
$D(2,1,a), F(4), G(3)$~\cite{Kadv},\cite{KWn}.
In this section we will describe some properties
of the root systems of defect one, which we use in the paper.

\subsection{}\label{piprop}
We choose a set of simple roots of $\fg$ 
which contains a unique isotropic root
$\beta$. We will describe $\Delta^{\#}$ and $W^{\#}$ (see~\ref{kappa2}). 
Let $\Pi^{\#}$ be the system of simple roots for $\Delta^{\#}\cap\Delta^+$.
We have

(i) $\Pi^{\#}=\Pi\setminus\{\beta\}$;

(ii) $W^{\#}\beta\subset \Delta^+$; 

(iii) if $\fg\not=D(2,1,a)$, then
there exists a unique simple root $\alpha_1$ 
such that $(\alpha_1|\beta)\not=0$; for $\fg\not=\osp(3,2)$
one has $(\alpha_1|\beta)=-1,
\ (\alpha_1|\alpha_1)=2$.

The following lemma is used in~\ref{def1p>0}.

\subsubsection{}
\begin{lem}{lempri}
Let $\fg\not=D(2,1,a),\ B(1,1)$.
If $\alpha$ is a positive isotropic root satisfying
$(\rho-t\beta,\alpha)=t+1$ for some $t$ then $\alpha=\beta+\alpha_1$. 
\end{lem}
\begin{proof}
Write $\Pi=\{\beta,\alpha_1,\alpha_2,\ldots,\alpha_m\}$
and $\alpha=m_0\beta+\sum m_i\alpha_i$.
Since $\alpha$ is isotropic, $\alpha\not\in \Delta^{\#}$.
The properties (i), (iii)  imply that  $m_0,m_1\geq 1$. One has
$(\rho|\alpha)\geq m_1\geq 1$ and $-(\beta|\alpha)=m_1\geq 1$. 
The assumption gives $(\rho|\alpha)=-(\beta|\alpha)=1$
and thus $m_1=1$. Then $(\rho|\alpha)=1$ forces
$m_i=0$ for $i>1$. Finally, $(\alpha|\alpha)=2-2m_0=0$
and thus $\alpha=\beta+\alpha_1$ as required.
\end{proof}

\subsubsection{}\label{stnorm}
The standard normalization of the invariant form $B$, 
introduced in~\cite{KWn}, is given by $(\alpha|\alpha)=2$
for an even root $\alpha\in\Delta^{\#}$.
In this normalization the dual Coxeter number $h^{\vee}$ is given by
the following table:

\begin{tabular}{|c||c|c|c|c|c|c|c|c|}
\hline
$\fg$& A(0,n-1)& C(n)& B(1,n) & B(n,1), $n>1$ & D(n+1,1)& F(4)& G(3) &
D(2,1,a)\\
\hline
$h^{\vee}$& n-1& n-1& n-1/2 & 2n-3&              2n-2& 3   & 2    & 0\\
\hline
\end{tabular}


\subsection{Non-exceptional case ($\fg\not=F(4),\ G(3),\ D(2,1,a)$).} 
The root system is described in terms of a basis $\{\vareps_i\}_{i=0,1,...}$.
We use the following bilinear form $(.|.)$, which a multiple of the standard
invariant form: $(\vareps_i|\vareps_j)=0$ if $i\not=j$
and $(\vareps_0|\vareps_0)=-1$,
$(\vareps_i|\vareps_i)=1$ for $i>0$. Then in all cases
$(\gamma_1|\gamma_2)\in\mathbb{Z}$ for all roots $\gamma_1,\gamma_2$.
We choose $\beta:=\vareps_0-\vareps_1$.
One has
$\Delta_0^{\#}=\Delta_0\setminus\mathbb{Z}\vareps_0$ and
$W^{\#}$ is the subgroup of $W$ which stabilizes $\vareps_0$.
One has $\hat{\Delta}^{\#}_0=\Delta_0$
for $A(0,n-1), C(n)$; in all other cases, 
except for $B(1,1)$ and $D(2,1)$,
$\hat{\Delta}^{\#}_0$ corresponds to a simple component of
$\fg_0$ which is not isomorphic to $\fsl(2)$.

\subsection{Case $A(0,n-1)$}\label{fsl1n}
In this case the even part is $\fsl(n)\oplus \mathbb{C}$.
Let $\{\pm(\vareps_i-\vareps_j): 1\leq i<j\leq n\}$ be the root system
for $\fsl(n)$ and $\{\pm(\vareps_0-\vareps_i)\}_{i=1}^n$ be the 
set of odd roots of $\fsl(1,n)$. 
One has $\Delta_0^{\#}=\Delta_0, W^{\#}=W=S_n$.
Take 
$$\begin{array}{l}
\Pi:=\{\vareps_0-\vareps_1,\ 
\vareps_1-\vareps_2,\ \vareps_2-\vareps_3,\ldots,\vareps_{n-1}-
\vareps_n\},\ \Pi^{\#}:=\Pi\cap\Delta_0=
\Pi\setminus\{\beta\},\\
\Delta_1^+=\{\vareps_0-\vareps_i\}_{i=1}^n,\ \ 
\Delta_0^+=\{\vareps_i-\vareps_j\}_{1\leq i<j\leq n}.
\end{array}$$  
The highest root is $\theta=\vareps_0-\vareps_n$ and
$\rho=-\frac{n}{2}\vareps_0+\sum_{i=1}^n (\frac{n}{2}+1-i)\vareps_i$.

\subsection{Case $C(n+1)$}\label{Cn}
In this case the even part is $C_n\oplus \mathbb{C}$.
Let $\{\pm 2\vareps_i;\pm\vareps_i\pm\vareps_j: 1\leq i<j\leq n\}$
 be the root system for the Lie algebra of type $C_n$ and
$\{\pm\vareps_0\pm\vareps_i)\}_{i=1}^n$ be the set of odd roots
of $C(n)$. One has $\Delta_0^{\#}=\Delta_0, W^{\#}=W$.
Take $\Pi:=\{\vareps_0-\vareps_1,\vareps_1-\vareps_2,\ldots,
\vareps_{n-1}-\vareps_n,2\vareps_n\}$.
Then 
$$\Delta_1^+=\{\vareps_0\pm\vareps_i\}_{i=1}^n,\ \ 
\Delta_0^+=\{2\vareps_i,\vareps_i\pm\vareps_j\}_{1\leq i<j\leq n}.$$
One has $\theta=\vareps_0+\vareps_1$ and
$\rho=-n\vareps_0+\sum_{i=1}^n (n+1-i)\vareps_i$.

\subsection{Case $B(1,1), n>1$}\label{B11}
Take $\Pi:=\{\vareps_0-\vareps_1;\vareps_1\}$.  
Then $\Delta_0^+=\{\vareps_0;2\vareps_1\},\
\Delta_1^+=\{\vareps_0\pm\vareps_1;\vareps_1\}$.
One has $\Delta_0^{\#}=\pm 2\vareps_1$, $W^{\#}\cong\mathbb{Z}_2$ is 
the corresponding Weyl group. One has $\theta=\vareps_0+\vareps_1$ and
$2\rho=-\vareps_0+\vareps_1$.

\subsection{Cases $B(n,1), B(1,n): n>1$}\label{Bn1}
Take $\Pi:=\{\vareps_0-\vareps_1;\vareps_1-\vareps_2,\ldots,
\vareps_{n-1}-\vareps_n,\vareps_n\}$. One has
$$\begin{array}{lll}
\Delta_1^+=\{\vareps_0\pm\vareps_i;\vareps_0\}_{i=1}^n, &
\Delta_0^+=\{\vareps_i,\vareps_i\pm\vareps_j;2\vareps_0\}_{1\leq i<j\leq n}
&\text{ if } \fg=B(n,1);\\
\Delta_1^+=\{\vareps_0\pm\vareps_i;\vareps_i\}_{i=1}^n, &
\Delta_0^+=\{2\vareps_i,\vareps_i\pm\vareps_j;\vareps_0\}_{1\leq i<j\leq n}
& \text{ if } \fg=B(1,n)
\end{array}$$

The group $W^{\#}$
is the group of signed permutations of $\{\vareps_i\}_{i=1}^n$ and
$\rho=-(n-\frac{1}{2})\vareps_0+\sum_{i=1}^n
(n-i+\frac{1}{2})\vareps_i$. One has $\theta=2\vareps_0$ if $\fg=B(n,1)$ and
$\theta=\vareps_0+\vareps_1$ if $\fg=B(1,n)$.

\subsection{Case $D(n,1), n>1$}\label{Dn1}
Take $\Pi:=\{\vareps_0-\vareps_1, \vareps_1-\vareps_2,\ldots,
\vareps_{n-1}-\vareps_n,\vareps_{n-1}+\vareps_n\}$.
Then
$$\Delta_1^+=\{\vareps_0\pm\vareps_i\}_{i=1}^n,\ \ 
\Delta_0^+=\{\vareps_i\pm\vareps_j;2\vareps_0\}_{1\leq i<j\leq n};$$
$W^{\#}$
is the group of signed permutations of $\{\vareps_i\}_{i=1}^n$
which change the even number of signs. One has $\theta=2\vareps_0$ and
$\rho=-(n-1)\vareps_0+\sum_{i=1}^n (n-i)\vareps_i$.

\subsection{Case $F(4)$}\label{F4}
The even part of $F(4)$ is $B_3\oplus \fsl(2)$.
Let $\{\pm \vareps_i;\pm\vareps_i\pm\vareps_j: 1\leq i<j\leq 3\}$
 be the root system for the Lie algebra of type $B_3$ and $\vareps_0$
be a root corresponding to $\fsl(2)$. Take
$$\begin{array}{l}
\beta:=\frac{1}{2}(\vareps_0+\vareps_1+\vareps_2+\vareps_3),\ \ 
\Pi:=\{\beta,-\vareps_1,\vareps_1-\vareps_2,\vareps_2-\vareps_3\},\\
\Delta_1^+=\{\frac{1}{2}(\vareps_0\pm \vareps_1\pm\vareps_2\pm\vareps_3)\},
\Delta_0^+=\{\vareps_0;-\vareps_i,-\vareps_i\pm\vareps_j: 1\leq j<i\leq 3\}.
\end{array}$$
Normalize the form in such a way that $(\vareps_0,\vareps_0)=-6$; then
$(\vareps_i,\vareps_j)=2\delta_{i,j}$ if $i\geq 0, j>0$. One has
$\theta=\vareps_0$ and
$\rho=-(3\vareps_0+\vareps_1+3\vareps_2+5\vareps_3)/2$.

\subsection{Case $G(3)$}\label{G3}
The even part of $G(3)$ is $G_2\oplus \fsl(2)$, $\Delta^{\#}$
is the root system for $G_2$ 
and $W^{\#}$ is the Weyl group of $G_2$.
The roots are expressed in terms of $\vareps_1,\vareps_2,\vareps_3:
\vareps_1+\vareps_2+\vareps_3=0$ corresponding to $G_2$ and 
$\vareps_0$ corresponding to $\fsl(2)$. 
We take $\Pi:=\{\vareps_0+\vareps_1,\vareps_2,\vareps_3-\vareps_2\}$,
$\beta:=\vareps_0+\vareps_1$.
Then 
$$\Delta_1^+=\{\vareps_0;\vareps_0\pm\vareps_i: i=1,2,3\},\ \ 
\Delta_0^+=\{2\vareps_0;-\vareps_1,\vareps_2,\vareps_3,\vareps_3-\vareps_2,
\vareps_2-\vareps_1,\vareps_3-\vareps_1\}.$$
Normalize the form in such a way that
$(\vareps_i|\vareps_i)=2$ for $i>0$; then $(\vareps_i|\vareps_j)=-1$
for $0<i<j$ and $(\vareps_0|\vareps_i)=-2\delta_{0,i}$.
One has $\theta=2\vareps_0$ and
$\rho=(-5\vareps_0-3\vareps_1+\vareps_2+3\vareps_3)/2$.

\subsection{Case $D(2,1,a)$}\label{D21alpha}
In this case the even part is $\fsl(2)\oplus \fsl(2)\oplus \fsl(2)$.
We take
$$\Delta_1^+=\{\vareps_0\pm \vareps_1\pm\vareps_2\},
\Delta_0^+=\{2\vareps_0,2\vareps_1,2\vareps_2\}$$
and $\beta:=\vareps_0-\vareps_1-\vareps_2,\ 
\Pi:=\{\beta,2\varesp_1,2\varesp_2\}$. One has  $\theta=2\vareps_0$ and
$\rho=-\beta$.

We take $\Delta^{\#}_0:=\{\pm 2\vareps_1,\pm 2\vareps_2\}$; then 
$W^{\#}\beta=\Delta_1^+$. We  normalize the form as follows:
$$(\vareps_0|\vareps_0)=
\frac{-1-a}{2},\ (\vareps_1|\vareps_1)=a/2, \ 
(\vareps_2|\vareps_2)=1/2, \ (\vareps_i|\vareps_j)=0, \ i \neq j.$$

\section{Appendix}
We will prove two lemmas used in the main text.

Let $\fg$ be a semisimple finite dimensional Lie algebra,
$\Delta^+$  the set of positive roots, $P$  the
weight lattice,  and $W$  the Weyl group of $\fg$. 

\subsection{}
\begin{lem}{lemstab}
If $\lambda\in  P$ is such that
$(\lambda|\lambda)<(\rho|\rho)$, then $E(\lambda)=0$.
\end{lem}
\begin{proof}
First, $\Stab_W \lambda\not=\id\ \Longleftrightarrow\ E(\lambda)=0$,
since $\Stab_W \lambda$ is generated by reflections it
contains (see, for instance,~\cite{jbook}, A.1.1). Hence we may
assume that $\lambda$ has a trivial stabilizer in $W$.
Let $\lambda'$ be the maximal element in the orbit 
$W\lambda$;
then for any simple root $\alpha$ one has $s_{\alpha}\lambda'<\lambda$,
hence $(\lambda'|\alpha)>(\rho|\alpha)$. Therefore $\lambda'=\rho+\xi$,
where $\xi\in P^+$ and we  obtain
$$(\lambda|\lambda)=(\rho+\xi|\rho+\xi)=(\rho|\rho)+(\xi|\xi)+(2\rho|\xi)
\geq(\rho|\rho),$$
since $2\rho\in Q^+$.\end{proof}

\subsection{}
\begin{lem}{lemw}
For each $\alpha\in{\Delta}^+$
and all $r>>0$ one has
$$r\alpha=w.(r'\alpha'),\ \text{ for some }
\alpha'\in{\Delta}\cup\{0\}, r'\geq 1, w\in W\ 
\Longrightarrow\ w=s_{\alpha} \text{ or } w=\id.$$
\end{lem}
\begin{proof}
Since $W$ is a finite group, it is enough to show that
for each $w\in W$, $w\not=\id,s_{\alpha}$ one has
$$r>>0\ \Longrightarrow\ 
r\alpha\not=w.(r'\alpha')\ \text{ for }\alpha'\in{\Delta}\cup\{0\}, 
r'\geq 1.$$
Assume that $r\alpha=w.(r'\alpha')$, that is $\rho-w\rho+r\alpha=r'(w\alpha')$.
Write $\rho-w\rho=:\sum m_i\beta_i, w\alpha'=:\sum k_i\beta_i$,
where  $\{\beta_i\}\subset\Delta$ 
is a set of simple roots such that $\beta_1=\alpha$.
The condition $w\not=\id,s_{\alpha}$
implies that $\rho-w\rho$ is not proportional to $\alpha$
so $m_i\not=0$ for some $i>1$. One has
$\rho-w\rho+r\alpha=(m_1+r)\beta_1+\sum_{i>2}m_i\beta_i$ and thus
$\frac{m_1+r}{m_i}=\frac{k_1}{k_i}$.
Since $w\alpha'$ lies in a finite set $\Delta\cup\{0\}$,
 the set of possible values for
$\frac{m_1+r}{m_i}$ is finite so the set of possible values 
for $r$ is finite as well, which is a contradiction.
\end{proof}


\end{document}